\documentclass[a4paper,11pt]{article}

\usepackage[latin1]{inputenc}
\usepackage[active]{srcltx}
\usepackage{amsfonts}
\usepackage{amsmath}
\usepackage{amssymb}
\usepackage{amsthm}

\usepackage{mathrsfs}
\usepackage{relsize}

\usepackage{chapterbib}
\usepackage{bibunits}
\usepackage[sectionbib]{natbib}
\usepackage{bm}
\usepackage{subfigure}

\usepackage[active]{srcltx}
\usepackage{ifpdf}
\usepackage{setspace}
\usepackage{lscape}
\usepackage{bm}

\usepackage{color}
\definecolor{darkred}{rgb}{0.5,0,0}
\definecolor{darkgreen}{rgb}{0,0.5,0}
\definecolor{darkblue}{rgb}{0,0,0.5}

\ifpdf

\newcommand{\drivy}{pdftex}
\usepackage[pdftex=true,hyperfootnotes=true]{hyperref}
\else

\newcommand{\drivy}{dvips}
\usepackage[hyperfootnotes=true]{hyperref}
\fi
\usepackage[\drivy]{graphicx}
\usepackage{rotating}

\usepackage{setspace}
\usepackage[\drivy]{graphicx}
\usepackage[footnotesize,nooneline,ignoreLTcapwidth,bf]{caption2}
\newcommand{\Qmeas}{\mathbb{Q}}
\newcommand{\Pmeas}{\mathbb{P}}

\newcommand{\mE}{\mathbb{E}}

\theoremstyle{definition}
\newtheorem{assumption}{Assumption}

\newtheorem{remark}{Remark}

\newtheorem{lemma}{Lemma}
\newtheorem{definition}{Definition}

\date{}

\newcommand{\valpha}{\bm \alpha}
\newcommand{\vbeta}{\bm \beta}
\newcommand{\vgamma}{\bm \gamma}

\newcommand{\vzeta}{\bm \zeta}

\newcommand{\vtheta}{\bm \theta}
\newcommand{\vvartheta}{\bm \vartheta}

\newcommand{\vkappa}{\bm \kappa}

\newcommand{\vmu}{\bm \mu}

\newcommand{\vrho}{\bm \rho}

\newcommand{\vtau}{\bm \tau}

\newcommand{\vphi}{\bm \phi}
\newcommand{\vchi}{\bm \chi}

\newcommand{\vPsi}{\mathbf{\Psi}}
\newcommand{\vLambda}{\mathbf{\Lambda}}
\newcommand{\vPhi}{\mathbf{\Phi}}

\makeatletter
\newcommand{\rd}{\@ifnextchar^{\DIfF}{\DIfF^{}}}
\def\DIfF^#1{%
   \mathop{\mathrm{\mathstrut d}}%
   \nolimits^{#1}\gobblespace}
\def\gobblespace{\futurelet\diffarg\opspace}
\def\opspace{%
   \let\DiffSpace\!%
   \ifx\diffarg(%
   \let\DiffSpace\relax
   \else
   \ifx\diffarg[%
   \let\DiffSpace\relax
   \else
   \ifx\diffarg\{%
   \let\DiffSpace\relax
   \fi\fi\fi\DiffSpace}

\title{$GMM$ Estimation of Affine Term Structure Models%
\thanks{The authors thank Eberhard Mayerhofer, Robert Kunst and Paul Schneider as well as the participants of the {\em CFE 2012, 2013 conferences}, the
{\em GPSD 2014} conference and the
{\em COMPSTAT 2014} conference for interesting discussions and comments.
Financial support from the Austrian Central Bank
under Anniversary Grant Nr. 14678 is gratefully acknowledged.} %Moreover, we are grateful to an anonymous referee for helpful comments.}}
}

\author{
Jaroslava Hlouskova
\and
Leopold S\"ogner%
\thanks{
Jaroslava Hlouskova  (jaroslava.hlouskova@ihs.ac.at),
Leopold S\"ogner (soegner@ihs.ac.at),
Department of
Economics and Finance, Institute for Advanced Studies,
Stumpergasse 56, 1060 Vienna, Austria. Leopold S\"ogner has a further affiliation with the Vienna Graduate School of Finance (VGSF) and Jaroslava Hlouskova at Thompson Rivers University, Canada.}}

\date{\today}

\begin{document}

\maketitle
\begin{abstract}
\noindent
This article investigates parameter estimation of affine term structure models by means of the
generalized method of moments. Exact moments of the affine latent process as well as of the yields are obtained by using results derived for ${p-}$polynomial processes. Then the generalized method of moments, combined with Quasi-Bayesian methods, is used to get reliable parameter estimates and to perform inference.
After a simulation study, the estimation procedure is applied to empirical interest rate data.
\\[3pt]
\noindent %polynomial processes.
\end{abstract}

\newpage

\section{Introduction}
\label{sect:intro1}

This article is concerned with parameter estimation and inference in affine term structure models.
%We obtain the exact moments of yields generated from an affine model. 
%In particular,
We use results of \citet{cuchieroetal08} on ${p-}$polynomial processes to obtain the exact conditional moments of a latent affine process driving the term structure. By assuming a stationary affine process, we obtain not only the exact moments of a vector of yields with various maturities but also the first-order auto-covariance matrices of the yields and the squared yields.
Then we estimate the model parameters by means of the {\em Generalized Method of Moments} ($GMM$) introduced in \citet{hansen82}, where  Quasi-Bayesian methods \citep[see][]{chernozhukovhong03} are used to minimize the $GMM$ distance function. A further contribution of this paper is a rigorous study on testing market price of risk specifications discussed in  quantitative finance literature. By considering the Wald test, we observe that test statistics obtained from output provided by Quasi-Bayesian methods strongly outperform test statistics which are obtained by standard procedures with respect to power and size.

Affine term structure models have their origin in the univariate models of \citet{vasicek77} and \citet{cir85}.
The performance of these models and similar univariate setups were already investigated for example in \citet{Ait1996} and
\citet{aitsahalia96}. The articles show that these univariate parametric models inadequately describe the interest rate dynamics. Based on this finding \citet{Ait1996}, \citet{aitsahalia96} as well as \citet{Stanton1997} proposed non-parametric interest rates models. 
As an alternative, \citet{daisingleton00} and \citet{dai03} favored multivariate settings to circumvent the shortcomings of univariate models. This alternative modeling approach has the advantage that a mathematical framework, where bonds and derivatives can be priced in a straightforward way, is available.

Let us briefly discuss some literature on the performance of different estimation approaches:
Regarding parameter estimation, \cite{zhou01} studied the efficient method of moments ($EMM$), the $GMM$, the quasi-maximum likelihood estimation ($QMLE$) and the maximum likelihood estimation ($MLE$) for the \citet{cir85} model. In his study the author assumes that the instantaneous interest rate, driven by a square root process, can be observed. The most efficient results are observed for the $MLE$, which is followed by the $QMLE$ and the $EMM$.%
{\footnote{For stochastic volatility models \citet{Andersenetal1999} have shown that the $EMM$ estimator has almost the same efficiency as the maximum likelihood estimator.}} %
Regarding the $GMM$, this method performs well if the sample size is sufficiently large. In addition, \cite{zhou03} constructed a $GMM$ estimator by deriving moments for univariate latent processes by applying Ito's formula (under the same assumption that the instantaneous interest rate can be observed). This estimator has been compared to the $ML$ estimator. In contrast to \cite{zhou01}, in this setup the $GMM$ estimator performs quite well in the finite sample compared to the maximum likelihood estimation.

More recent literature has proposed different frequentist and Bayesian approaches to estimate the parameters  of multivariate affine term structure models. Bayesian methods have been applied almost recently in \citet{ChibErgashev2008}, an earlier application is e.g.~\citet{fruehwirthgeyer96}. Regarding Bayesian estimation methods, \citet{Jones2003b} pointed out that strong priors are necessary to estimate the parameters in the case of a low degree of mean reversion (i.e., high persistence) of the stochastic process.  MLE has been performed in a three factor Gaussian model \citep[an $\mathbb{A}_{0}(3)$ model in the terminology of][]{daisingleton00} by \citet{HamiltonWu2010}.

Additional articles on parameter estimation for affine models are e.g. \citet{Dieboldetal2006}, \citet{Duffee2011}, \citet{aitsahakiakimmel2009}, \citet{Egorovetal2011} and \citet{Joslinetal2009}. An overview is provided in \cite{piazzesi03}. A further approach is to approximate the transition density of the affine process via approximations of the Chapman/Kolmogorov forward equation. This approach has been explored in series of papers by A\"{i}t-Sahalia \citep[see, e.g.,][]{aitsahalia02,aitsahakiakimmel2009}. \citet{filipovicmayerhoferschneider13} used the moments obtained in \cite{cuchieroetal08} to construct additional likelihood expansions.

In contrast to a lot of other approaches already used in the literature, we use the {\em{exact}} moments of the yields observed, arising from a multivariate affine term structure model. Neither an approximation of the moments (such as an approximation via the solution of the stochastic differential equation) nor an approximation of the likelihood is required.
Since we have to minimize a $GMM$ distance function in more than twenty parameters, $GMM$ estimation is nontrivial. To account for this problem,
we use Quasi-Bayesian methods developed in \citet{chernozhukovhong03}. As standard errors of parameter estimates are byproducts of this estimation routine, we apply them in parameter testing, where we observe rejection rates of the true null hypothesis to be close to the theoretical significance levels. By contrast, when using standard routines to estimate the asymptotic covariance matrix of the unknown parameter vector, the performance of the Wald  test, measured in terms of power and size, is very poor. 

This paper is organized as follows: Section~\ref{sect:affine1} introduces affine term structure models. Section~\ref{sect:polyprocess1} applies results obtained in mathematical finance literature to calculate the moments of the latent process driving the yields and then derives the moments of the yields observed. Section~\ref{sect:mc1} describes the small sample properties of the $GMM$ estimator, while Section~\ref{sect:emp1} applies the estimator to empirical data.  Finally, Section~\ref{sect:conc} offers conclusions.

\section{Affine Models}
\label{sect:affine1}

This section provides a brief description of affine models, which is mainly based on \citet{filipovicbook2009}.
Consider the state space $\mathscr{S}=\mathbb{R}_+^m \times \mathbb{R}^n  \subset \mathbb{R}^d$, where $m,n\geq0$, $m+n=d$,
and the filtered probability space $\left(\Omega,\mathcal{F},(\mathcal{F}_t)_{t \geq 0 }, \mathbb{P}  \right)$. With $\mathbf{X}(t) \in \mathbb{R}^d$, the stochastic process in continuous time $(\mathbf{X}(t))_{t \geq 0}$ is generated by the following affine stochastic differential equation

\begin{equation}
\label{eq:sde3}
d \mathbf{X}(t) = \left( \mathbf{b}^P + \vbeta^P \mathbf{X}(t) \right) dt + \vrho(\mathbf{X}(t)) d \mathbf{W}^P(t) \ ,
\end{equation}
where $\mathbf{b}^P$ is a $d-$dimensional vector and $\vbeta^P$ and $\vrho(\mathbf{x})$ are $d \times d$ matrices. The $d \times d$ diffusion term $\mathbf{a}(\mathbf{x})$ is defined such that $\mathbf{a}(\mathbf{x}) = \vrho(\mathbf{x}) \vrho(\mathbf{x})'=\mathbf{a} + \sum_{i=1}^d x_i \ \valpha_i$, where $\mathbf{a}$ and $\valpha_i$, $i=1,\dots,d$, are $d \times d$ matrices. $\mathbf{W}^P(t)$ is a $d-$dimensional standard Brownian motion. For more details the reader is referred to Appendix~\ref{sect:affine1appa}. In an affine environment the {\em instantaneous interest rate (short rate, $r(t) \in \mathbb{R}$)} follows from
\begin{equation}
\label{eq:short1}
r(t)=\gamma_0 +  \vgamma_x^{\prime} \mathbf{X}(t)  \ ,
\end{equation}
\noindent
where $\gamma_0$ is a scalar and $\vgamma_{x}$ is a $d-$dimensional vector.
\noindent
We consider an arbitrage free market, where $\Pmeas$ is the empirical measure and $\Qmeas$ is an equivalent martingale measure. We
assume that the process $(\mathbf{X}(t))_{t \geq 0}$ is affine also in the measure $\Qmeas$, such that
\begin{equation}
\label{eq:sde5}
d \mathbf{X}(t) = \left( \mathbf{b}^Q + \vbeta^Q \mathbf{X}(t) \right) dt + \vrho(\mathbf{X}(t)) d \mathbf{W}^Q(t) \ ,
\end{equation}
\noindent where $\mathbf{W}^Q(t)$ is a $d-$dimensional standard Brownian motion under $\Qmeas$ measure.

By equations (\ref{eq:sde3}) and (\ref{eq:sde5}), the stochastic process $ \left( \mathbf{X}(t) \right)_{t \geq 0}$ is affine in both measures. While the diffusion parameters ($\mathbf{a}$, $\valpha_i$, $i=1,\dots,d$) remain the same under both measures, we have to consider parameters $\mathbf{b}^P$, $\vbeta^P$, $\mathbf{b}^Q$ and $\vbeta^Q$, in both measures $\Pmeas$ and $\Qmeas$.  This specification, namely equations (\ref{eq:sde3}) and (\ref{eq:sde5}), is called {\em the extended affine market price of risk specification}, and its mathematical foundation is provided in \citet{cheriditofilipovickimmel03}. These authors also show by means of the Girsanov theorem that $\mathbf{W}^Q(t) = \mathbf{W}^P(t) +  \int_0^t \vphi ( \mathbf{X}(s)) ds$. For the affine class
\begin{equation}
\label{eq:sde7}
\vphi ( \mathbf{X}(t)) = \left( \vrho\left(\mathbf{X}(t)\right)\right)^{-1} \left(\mathbf{b}^P-\mathbf{b}^Q + \left(\vbeta^P-\vbeta^Q \right) \mathbf{X}(t) \right)  \ , 
\end{equation}
\noindent
where $ \vphi ( \mathbf{X}(t)) \in \mathbb{R}^d$. The stochastic process $\left( \vphi ( \mathbf{X}(t)) \right)_{t \geq 0}$, is called {\em market price of risk process}.

\begin{remark}
\label{rem:mpr1}
To observe how the market price of risk process $\left( \vphi ( \mathbf{X}(t)) \right)_{t \geq 0}$ is connected to risk premia,
\citet{Cochranebook2005}[p.~339] provides a formal relationship between the process $\left( \vphi ( \mathbf{X}(t)) \right)_{t \geq 0}$ and the (instantaneous) Sharpe ratio.
\end{remark}

We also assume that the process $(\mathbf{X}(t))$ satisfies the admissibility conditions (under both measures), which ensure that the process $(\mathbf{X}(t))$ does not leave the state space $\mathscr{S}$ \citep[see][Theorem~10.2 and Appendix~\ref{appa:restrict1}]{filipovicbook2009}. Next, we define the index sets $I=\{1,\dots,m\}$ and $J=\{m+1,\dots, n \}$, where $m+n=d$. Let $\mathbf{b}_I=(b_1,\dots,b_m)^{\prime}$ and $\vbeta_{II}=\vbeta_{1:m,1:m}$.%
{\footnote{In this article we apply the following notation: For vectors and matrices we use boldface. If not otherwise stated, the vectors considered are column vectors. Given a $r_M \times c_M$ matrix $\mathbf{M}$, the term $\mathbf{M}_{r_a:r_b,c_a:c_b}$ stands for ``from row $r_a$ to row $r_b$ and from column $c_a$ to column $c_b$ of matrix $\mathbf{M}$''. The abbreviation $\mathbf{M}_{r_a:r_b,:}$ stands for ``from row $r_a$ to row $r_b$ of matrix $\mathbf{M}$'', while ``$,:$'' stands for all columns, i.e. columns $1$ to $c_M$. In addition, $\mathbf{M}_{r_a:r_b,c_a}$ extracts the elements $r_a$ to $r_b$ of the column $c_a$. In addition, $\beta_{ij}$ stands for $\left[ \vbeta \right]_{ij}$; $\mathbf{0}_{a \times b}$ and $\mathbf{e}_{a \times b}$ stand for $a \times b$ matrices of zeros and ones; $\mathbf{0}_{a}$ and $\mathbf{e}_{a}$ is used to abbreviate  $\mathbf{0}_{a \times 1}$ and $\mathbf{e}_{a \times 1}$; $\mathbf{I}_a$ is the $a \times a$ identity matrix, while $\mathbb{I}_{(\cdot)}$ stands for an indicator function. Given a vector $\mathbf{x} \in \mathbb{R}^n $, $diag(\mathbf{x})$ transforms $\mathbf{x}$ into a $n \times n$ diagonal matrix. 2 E-3 stands for $2 \cdot 10^{-3}=0.002$.  }}
This notation, the admissibility restrictions (see Appendix~\ref{appa:restrict1}), the short-rate model (\ref{eq:short1}) and the condition $\mathbb{E} \left ( \exp(-\int_0^{\bar \tau} r(z) dz ) \right )< +\infty$, for some $\bar \tau \in \mathbb{R}_+$, imply that there exists a unique solution $\left(\Phi(t,\mathbf{u}) , \vPsi(t,\mathbf{u})^{\prime} \right)^{\prime} \ \in \
  \mathbb{C} \times \mathbb{C}^d$ of the system of Riccati differential equations
\begin{eqnarray}
\label{transformeddk3} %\left.
\begin{array}{lll}
\partial_t \Phi(t,\mathbf{u}) &= \frac{1}{2} \left ( \vPsi_J (t,\mathbf{u})\right)^{\prime}  \mathbf{a}_{JJ} \vPsi_J(t,\mathbf{u}) + \left (\mathbf{b}^Q\right)^{\prime} \vPsi(t,\mathbf{u}) -  \gamma_0;  & \Phi(0,\mathbf{u})=0 \ ,  \\
\partial_t \Psi_i (t,\mathbf{u}) &= \frac{1}{2} \left (\vPsi(t,\mathbf{u})\right)^{\prime} \valpha_i \vPsi(t,\mathbf{u}) + \left (\vbeta_i^Q\right)^{\prime} \vPsi(t,\mathbf{u}) - \gamma_{xi}; \qquad & {\text{ for }} i\in I \ , \\
\partial_t \vPsi_J(t,\mathbf{u}) &= \left (\vbeta_{JJ}^Q\right)^{\prime} \vPsi_J(t,\mathbf{u}) - \vgamma_{xJ}; & \vPsi(0,\mathbf{u}) = \mathbf{u} \ ,
\end{array} %\right\}
\end{eqnarray}
where $t \in [0,\bar \tau]$, $\mathbf{u} \in \imath \mathbb{R}^d$ and $\vbeta=(\vbeta_1, \ldots, \vbeta_d)$, with $\vbeta_i$ being a $d-$dimensional vector, $i=1, \ldots, d$ \citep[see][Theorem~10.4]{filipovicbook2009}.\footnote{Ordinary differential equations similar to (\ref{transformeddk3}) have already been investigated in \citet{duffiekan96} and \citet{duffiepansingleton00}. } %
This system of ordinary differential equations is used to calculate the time $t$ {\em price of a zero coupon bond},
$\pi^0(t,\tau)$, with time to maturity $\tau$. The arbitrage free zero coupon model prices $\pi^{0}(t,\tau)$ and the {\em model yields} $y^{0}(t,\tau)$ follow from \citet{filipovicbook2009}[Corollary~10.2]. That is  %
\begin{eqnarray}
\label{eq:1}
\pi^0(t,\tau) &=& \exp\left(\Phi(\tau,\mathbf{0})+\vPsi(\tau,\mathbf{0})^{\prime} \mathbf{X}(t)\right)  \ {\rm and} \nonumber \\
y^{0}(t,\tau) &=& -\frac{1}{\tau} \log \left ( \pi^0 (t,\tau)\right) =  -\frac{1}{\tau} \left(\Phi(\tau,\mathbf{0})+\vPsi(\tau,\mathbf{0})^{\prime} \mathbf{X}(t) \right)  .
\end{eqnarray}
\noindent The time to maturity, $\tau$, and $\mathbf{u}=\mathbf{0}$ are the arguments of the functions $\Phi(t,\mathbf{u})$ and $\vPsi(t,\mathbf{u})$ described in (\ref{transformeddk3}). The parameters under $\Qmeas$ have to be used to derive $\Phi(\tau,\mathbf{0})$ and $\vPsi(\tau,\mathbf{0})$.

\section{Moments and Polynomial Processes}
\label{sect:polyprocess1}

Since the goal of this paper is to estimate the model parameters by means of the $GMM$, we have to obtain the moments of the yields. Section~\ref{sectsub:polyprocess1} uses a recent theory for polynomial processes to obtain a closed form expression for the moments of the latent process $(\mathbf{X}(t))_{t \geq 0}$. In Section~\ref{sect:model1} we derive the exact moments for the model yields of an affine term structure model with diagonal diffusion term. Finally, Section~\ref{sect:moments} deals with the case of empirical data, when the number of yields observed is larger than the dimension of $(\mathbf{X}(t))_{t \geq 0}$ and thus the yields observed cannot be matched exactly with the model yields derived in (\ref{eq:1}).

\subsection{Polynomial Processes}
\label{sectsub:polyprocess1}

Based on the results of \cite{cuchieroetal08} on ${p-}$polynomial Markov processes, this subsection derives the conditional moments of the latent process $(\mathbf{X}(t))_{t \geq 0}$.
Let us consider a time homogeneous Markov processes $(\mathbf{X}(t))_{t \geq 0}$, started at $\mathbf{X}(0)=\mathbf{x} \in  \mathscr{S}$, where the state space $\mathscr{S}$ is a closed subset of $\mathbb{R}^d$. The \textit{semigroup} $(\mathscr{P}_t)_{t \geq 0}$ described by
\begin{eqnarray}
\mathscr{P}_t f(\mathbf{x})= {\mathbb{E}}(f(\mathbf{X}(t))|\mathbf{X}(0)=\mathbf{x}) = \int_{ \mathscr{S} } f(\vzeta) \nu_t(\mathbf{x}, d \vzeta) \label{semigroup}
\end{eqnarray}
\noindent is defined on all integrable functions $f$: $\mathscr{S} \rightarrow \mathbb{R}$ with respect to the Markov kernels $\nu_t(\mathbf{x},\cdot)$.  For an affine term structure model we need moments of $(\mathbf{X}(t))$ for a process ``started'' at $\mathbf{X}(s)=\mathbf{x}$; $t>s$. Given the filtration $({\mathcal{F}}_t)_{t \geq 0}$ and the assumption that $(\mathbf{X}(t))$ is a homogeneous Markov process, the conditional expectation of $f(\mathbf{X}(t))$, when the process is started at $\mathbf{X}(s)=\mathbf{x}$, is given by ${\mathbb{E}} (f(\mathbf{X}(t))|\mathbf{X}(s)=\mathbf{x}) = \mathscr{P}_{t-s} f(\mathbf{x})$ \citep[see, e.g.,][Theorem~17.9]{Klenke2008}.%

\noindent
Next, let $\mathcal{P}_{ \leq p }(\mathscr{S})$ be the finite dimensional vector space of polynomials on $\mathscr{S}$ up to degree $p \geq 0$, i.e.
\begin{eqnarray}
\label{eq:defdv}
\mathcal{P}_{ \leq p }(\mathscr{S}) &=& \left \{ \sum_{k=0}^{ p } \vkappa_k^{\prime} \mathbf{x}^k| \, \mathbf{x} \in \mathscr{S}, \vkappa_k \in \mathbb{R}^{d_k} \right \} \nonumber  \\ && \text{where }
\mathbf{x}^k =\left(\Pi_{j=1}^d x_j^{l^{(k)}_{1j}}, \Pi_{j=1}^d x_j^{l^{(k)}_{2j}}, \ldots, \Pi_{j=1}^d x_j^{l^{(k)}_{d_kj}} \right)^{\prime} \in \mathbb{R}^{d_k} \text{ and }
d_k =\left(
\begin{array}{c}
k+d-1 \\
k
\end{array}
\right)
 .
\end{eqnarray}
For $i=1, \ldots, d_k$ and $j=1, \ldots, d$ the exponents $l^{(k)}_{ij}$ in the expression for $\mathbf{x}^k$ satisfy
$l_{ij}^{(k)} \in \mathbb{N}_0$ as well as $\sum_{j=1}^d l^{(k)}_{ij} = k$.\footnote{For example, for $d=3$ and $k=2$ we have the following: $\mathbf{x}^2= \left ( x_1^2, x_1 x_2, x_1 x_3, x_2^2, x_2 x_3, x_3^2 \right )^{\prime}$, $d_2=6$ and thus (i) $l_{11}^{(2)}=2, l_{12}^{(2)}=0, l_{13}^{(2)}=0$, (ii) $l_{21}^{(2)}=1, l_{22}^{(2)}=1, l_{23}^{(2)}=0$, (iii) $l_{31}^{(2)}=1, l_{32}^{(2)}=0, l_{33}^{(2)}=1$, (iv) $l_{41}^{(2)}=0, l_{42}^{(2)}=2, l_{43}^{(2)}=0$, (v) $l_{51}^{(2)}=0, l_{52}^{(2)}=1, l_{53}^{(2)}=1$, (vi) $l_{61}^{(2)}=0, l_{62}^{(2)}=0, l_{63}^{(2)}=2$.}
In affine term structure models the basis of $\mathcal{P}_{ \leq p }(\mathscr{S})$ is given by $(1,\mathbf{x}', (\mathbf{x}^2)', \ldots, (\mathbf{x}^p)')'$  and thus its dimension is $N= \sum_{k =0}^p d_{k}$. In addition, the Markov process $(\mathbf{X}(t))_{t\geq s}$ with $\mathbf{X}(s)=\mathbf{x} \in \mathscr{S}$ is called \textit{$p$-polynomial} if for all $f(\mathbf{x}) \in \mathcal{P}_{ \leq p }(\mathscr{S})$ and $t \ge s$
\begin{equation}
\label{eq:defpoly1}
\mathscr{P}_{t-s} f(\mathbf{x})={\mathbb{E}}(f(\mathbf{X}(t))|\mathbf{X}(s)=\mathbf{x}) \in  \mathcal{P}_{ \leq p }(\mathscr{S}) .
\end{equation}
\noindent
That is to say, if $f(\mathbf{x})$ is polynomial, then the ${\mathbb{E}}(f(\mathbf{X}(t))|\mathbf{X}(s)=\mathbf{x})$ is polynomial as well.
\cite{cuchieroetal08}[Theorem 2.7] have shown that a time homogeneous Markov processes $(\mathbf{X}(t))$ is $p-$polynomial if and only if
there exists a linear map $ \mathbf{A}$ on $\mathcal{P}_{\le p}(\mathscr{S})$ such that $\mathscr{P}_{t-s}$ restricted on $\mathcal{P}_{ \leq p }$ can be written as $\mathscr{P}_{t-s}|_{\mathcal{P}_{ \leq p}}= \exp( (t-s) \mathbf{A})$.{\footnote{Note that $\mathscr{P}_{t-s}|_{\mathcal{P}_{ \leq p}}= \exp( (t-s) \mathbf{A})$ also solves the Kolmogorov backward equation
$\frac{\partial \mathtt{u} (t-s,\mathbf{x})}{\partial t}= \mathcal{G} \mathtt{u} (t-s,\mathbf{x})$, where $\mathcal{G}$ is an extended generator as described in \cite{cuchieroetal08}[Definition~2.3]. This follows from the proof of \cite{cuchieroetal08}[Theorem~2.7].}}
Equipped with this mathematical tool and by means of (\ref{semigroup}), the conditional expectation ${\mathbb{E}}(f(\mathbf{X}(t))|\mathbf{X}(s)=\mathbf{x})$, for $t>s$ and $f(\mathbf{x}) \in \mathcal{P}_{ \leq p }(\mathscr{S})$ can be derived
by means of
\begin{equation}
\label{semigroup2}
{\mathbb{E}}(f(\mathbf{X}(t))|\mathbf{X}(s)=\mathbf{x})=\exp ((t-s) \mathbf{A}) f(\mathbf{x})  .
 \end{equation}
\noindent
\noindent
The conditional expectations of $f(\mathbf{X}(t))$ given $\mathbf{X}(s)=\mathbf{x}$, can be derived by obtaining the $N \times N$ matrix $ \mathbf{A}$, where $N= \sum_{k=1}^{p} d_k$, from the generator \citep[see][Theorem~2.9]{cuchieroetal08}
\begin{eqnarray}
\label{eq:gen1}
\mathcal{G} f(\mathbf{x}) &=&  \sum_{i=1}^d \left( b_i^P + \left[\vbeta^P \mathbf{x} \right]_{i} \right) \frac{\partial f(\mathbf{x})}{\partial x_i} + \frac{1}{2} \sum_{i,j=1}^d  \left[\mathbf{a}(\mathbf{x})\right]_{ij} \frac{\partial^2 f(\mathbf{x})}{\partial x_i \partial x_j} \nonumber \\
                      &=&  \sum_{i=1}^d \left( b_i^P + \vbeta^P_{i,1:d} \mathbf{x} \right) \frac{\partial f(\mathbf{x})}{\partial x_i} + \frac{1}{2} \sum_{i,j=1}^d  \left[\mathbf{a}(\mathbf{x})\right]_{ij} \frac{\partial^2 f(\mathbf{x})}{\partial x_i \partial x_j}  .
\end{eqnarray}
\noindent
To obtain the moments of $(\mathbf{X}(t))$ we set $f(\mathbf{X}(t))= \left[ \mathbf{X}(t)^k \right]_{i}$ for {$k=1, \ldots, p$} and $i=1,\dots,d_k$. As already stated above, if the dimension of $\mathbf{X}(t)$ is larger than one, then $\mathbf{X}(t)^k = \left( \Pi_{j=1}^d X(t)_j^{l_{1j}^{(k)}}, \cdots, \Pi_{j=1}^d X_j^{l_{d_k j}^{(k)}} \right)^{\prime}$, where $l_{ij}^{(k)} \in \mathbb{N}_0$, $\sum_{j=1}^d l_{ij}^{(k)}=k \le p $, $i=1,\dots,d_k$ and $j=1, \ldots, d$. In more detail, we consider the basis $(\mathfrak{e}_1,\dots,\mathfrak{e}_N)=(1,\mathbf{x}', (\mathbf{x}^2)', \ldots, (\mathbf{x}^p)')$.
By applying the extended generator $\mathcal{G}$ to the basis element $\mathfrak{e}_i$, we get the $i$-th row of the $N \times N$ matrix $ \mathbf{A}$ by means of
\begin{equation}
\label{semigroup4}
\mathcal{G} \mathfrak{e}_i=\sum_{j=1}^N \mathbf{A}_{ij} \mathfrak{e}_j  .
\end{equation}
\noindent The left hand side has been calculated by applying (\ref{eq:gen1}) to the corresponding basis element. Then $\mathbf{A}_{ij}$ follows from (\ref{semigroup4}) simply by comparing coefficients. This finally results in
\begin{eqnarray}
\label{semigroup7}
{\mathbb E}( \mathbf{X}(t)^k| \, \mathbf{X}(s)=\mathbf{x})
&=& \left (\mathbf{0}_{d_k \times \sum_{j=0}^{k-1} d_j},\mathbf{I}_{d_k},\mathbf{0}_{d_k \times N-\sum_{j=0}^{k} d_j} \right ) \exp((t-s) \mathbf{A}) \left(1,\mathbf{x}', (\mathbf{x}^2)', \ldots, (\mathbf{x}^p)'\right)^{\prime} \  ,
\end{eqnarray}
where $\mathbf{I}_{d_k}$ is the $d_k \times d_k$ identity matrix and $t>s$.

\subsection{\cite{daisingleton00}-Models and Moments of the Latent Process}
\label{sect:model1}

To proceed with an identified model, we work with affine models where the diffusion term can be diagonalized. For this sub-class, \cite{daisingleton00} provided sufficient conditions for identification.%
{\footnote{For example,
the $\mathbb{A}_1(3)$ model, which will be presented in equation (\ref{dk1a13}), has 19 parameter under $\Qmeas$. \cite{daisingleton00} have shown
that the same term structure can be obtained with different parameters. I.e. the model is not identified. Given the \cite{daisingleton00} conditions for identification, only 14 parameters are allowed to be free parameters.
Regarding the diagonal diffusion matrix, \citet{Cheriditoetal2008}[Theorem~2.1] provide conditions where a transformation of a general affine model (\ref{eq:sde3}) to an affine model with diagonal $\mathbf{a}(\mathbf{x})$ exists. For $d \leq 3$ this is always the case.}}
In this case, the affine process $(\mathbf{X}(t))_{t \ge 0}$ follows the stochastic differential equation

\vskip-2em
\begin{eqnarray}
\label{dk1}
d \mathbf{X}(t)&=& (\mathbf{b}^{Q} + \vbeta^{Q} \mathbf{X}(t)) dt + \mathbf{\Sigma}
{\sqrt{ \mathbf{S}(\mathbf{X}(t))}} d \mathbf{W}^{Q} (t),  \ \ \text{ where } \nonumber \\
&& S_{ii} (\mathbf{X}(t))= \mathcal{B}_i^0 + (\mathcal{B}_i^x)^{\prime} \mathbf{X}(t), \   S_{ij}(\mathbf{X}(t))=0 \ , \ \text{for} \  i,j=1,\dots,d, \ i \not = j \ , \\ \nonumber \label{dk2}
&& \text{ and } \mathbf{\Sigma} =  \text{diag} \left( \Sigma_{1},\dots,\Sigma_{d} \right) \text{ such that  } \ \Sigma_i =
\left[ \mathbf{\Sigma} \right]_{ii} > 0   .
\end{eqnarray}
\noindent
Equation (\ref{dk1}) is a special case of (\ref{eq:sde5}). The elements of the $d-$dimensional vector $\mathcal{B}^0$ are $\mathcal{B}_i^0$.
$\mathcal{B}^{x}$ is a $d \times d$ matrix, where $d \times 1$ vector $\mathcal{B}_i^{x}$ is the $i$-th column of this matrix; i.e., $\mathcal{B}^{x}=(\mathcal{B}_1^{x}, \ldots, \mathcal{B}_d^{x})$ with $\mathcal{B}_i^{x}=(\mathcal{B}_{1i}^{x}, \ldots, \mathcal{B}_{di}^{x})'$, $i=1, \ldots, d$.
Since $\mathbf{\Sigma}$ and $\mathbf{S}(\mathbf{X}(t))$ are diagonal matrices we obtain $ \mathbf{a} (\mathbf{X}(t)) = \mathbf{\Sigma}^2 \mathbf{S}(\mathbf{X}(t)) = \mathbf{\Sigma}^2 \text{diag} \left( \mathcal{B}^{0} + (\mathcal{B}^{x})^{\prime} \mathbf{X}(t)  \right)$. The diagonal elements of the $d \times d$ diagonal matrix $ \mathbf{a}$ are given by a$_{ii}= \Sigma_{i}^2 \mathcal{B}_i^0$, $i=1, \ldots, d$ and the diagonal elements of the $d \times d$ diagonal matrices $\valpha_i$, $i=1,\dots,d$, are $\Sigma_1^2 \mathcal{B}_{i1}^{x}$, $\Sigma_2^2 \mathcal{B}_{i2}^{x},\dots,\Sigma_d^2 \mathcal{B}_{id}^{x}$. For $\vbeta^Q$ and $\mathcal{B}_{}^{x}$ \citet{daisingleton00} require
\begin{equation} \label{kappaqpart1}
\vbeta^Q =\left(
\begin{array}{cc}
\vbeta_{II}^Q& {\bf 0}_{m \times n}\\
\vbeta_{JI}^Q \ge 0 & \vbeta_{JJ}^Q  \\
\end{array}
\right) \  \ {\rm and} \ \ \
{\cal B}^{x} =\left(
\begin{array}{cc}
\mathbf{I}_{m} & {\cal B}_{IJ}^{x} \ge 0  \\
{\bf 0}_{n \times m} & {\bf 0}_{n \times n}  \\
\end{array}
\right) \ ,
\end{equation}
\noindent
where $m+n=d$. The matrix $\vbeta_{II}^Q$ is of dimension $m \times m$, $\vbeta_{JI}^Q$ is of dimension $n \times m$, $\vbeta_{JJ}^Q$ is of dimension $n \times n$ and ${\cal B}_{IJ}^{x}$ is of dimension $m \times n$.
As we use findings of \citet{daisingleton00} we need to relate our notation to the notation of \citet{daisingleton00}, where the drift term of the process $(\mathbf{X}(t))_{t \ge 0}$ is considered in the form ${-\vbeta^Q}(\vtheta^{Q} - \mathbf{X}(t)) dt$ and thus $\mathbf{b}^Q= -\vbeta^Q \vtheta^{Q}$. In the following, $\vtheta^Q = -\left(\vbeta^Q \right)^{-1}\mathbf{b}^Q$ is a vector of dimension $d$, partitioned into $\vtheta_I^{Q}$ and $\vtheta_J^{Q}$, where the first term is of dimension $m$ while the second term is of dimension $n$; i.e., $\vtheta^{Q}_I \in \mathbb{R}^{m}$, $\vtheta^{Q}_J \in \mathbb{R}^n$, and thus $\vtheta^{Q}=\left ( \left ( \vtheta^{Q}_I \right )', \left ( \vtheta^{Q}_J \right )' \right )' \in \mathbb{R}^d$. The same partition is applied also to $\mathbf{X}(t)$. This yields to the following.
\begin{definition}[\citet{daisingleton00}-{\em canonical representation} of an $\mathbb{A}_m (d)$ model] \label{def5}
{\it Consider (\ref{dk1}) with diagonal diffusion matrix and the short-rate model (\ref{eq:short1}).
Admissibility and identification require the following:}

\rm{(i)-(a)} {\it For $m>0$ is $\vbeta^Q$ of structure given by (\ref{kappaqpart1}), where in addition $\beta_{ij}^Q \geq 0$ for $1\leq j \leq m$ and $i \not= j$.
Furthermore, $\vtheta^{Q}_I \ge 0$, $\vtheta^{Q}_J= \mathbf{0}$ and $\vbeta_{II}^Q  \vtheta_I^{Q} < \mathbf{0}$.}

\rm{(i)-(b)} {\it For $m=0$ is $\vbeta^{Q}$ a lower (or upper) triangular matrix \citep[][p.~1948]{daisingleton00}.}

\rm{(ii)}
$\mathbf{\Sigma}=\mathbf{I}_d$.

\rm{(iii)} {\it $\gamma_0$ and $\gamma_{xi}$ are unrestricted for $i \in I$, while $\gamma_{xj} \geq 0$ for $j \in J$.}

\rm{(iv)} {\it ${\cal B}^{0}=\left( \mathbf{0}_{1\times m}, \mathbf{e}_{1 \times n} \right)^{\prime}$ and ${\cal B}^{x}$ is of structure provided by (\ref{kappaqpart1}).}

\noindent {\it
If the admissibility conditions {\rm (i)-(iv)} for the affine process $(\mathbf{X}(t))_{t \ge 0}$ are satisfied, then model (\ref{dk1}) with diagonal diffusion term  will be called $\mathbb{A}_m(d)$ model.}
\end{definition}

Definition~\ref{def5}(i)-(a) implies that $b_i^Q=-\sum_{j=1}^m \beta_{ij}^Q \theta_j^{Q}>0$, for $i=1,\dots,m$, and thus the first $m$ elements of $\mathbf{b}^Q$ are strictly positive and the last $n$ elements of $\mathbf{b}^Q$ are negative. Namely
\begin{eqnarray}
\label{eq:bQ}
\mathbf{b}^Q = \left (
\begin{array}{c}
\mathbf{b}_I^Q \\
\mathbf{b}_J^Q
\end{array}
\right )= \left (
\begin{array}{lr}
-\vbeta_{II}^Q \vtheta_I^{Q} & > \mathbf{0} \\
-\vbeta_{JI}^Q\vtheta_I^{Q} & \le \mathbf{0}
\end{array}
\right )  .
\end{eqnarray}
This implies that the diagonal elements of $\vbeta_{II}$ are negative.
We slightly deviate from the canonical representation in Definition~\ref{def5} by assuming $\mathbf{\Sigma}$ to be a diagonal matrix with entries $\Sigma_{i}>0$ and $\vgamma_{x}=\mathbf{e}_{d}$.%
\footnote{Note that the canonical representation of \citet{daisingleton00} is one of many representations where the admissibility and identification conditions are met. The Appendix of \citet{daisingleton00} presents affine linear transformations $\Lambda_A \mathbf{X}(t) = \mathbf{L}_A \mathbf{X}(t) + \mathbf{l}_A$ where the model is still admissible and identified. } %
Since $\vtheta_J^Q$ is restricted to zero, not all elements of $\vbeta^Q$ and $\mathbf{b}^Q$ can be unrestricted. In the estimation procedure we account for this fact by using $\vtheta^Q$ as a parameter. Then $\mathbf{b}^Q=-\vbeta^{Q} \vtheta^{Q}$.

Now we apply the tools developed in Section~\ref{sectsub:polyprocess1} to $\mathbb{A}_m(d)$ models. To observe how this works we first derive matrix $ \mathbf{A}$ for the
\citet{vasicek77} and the \citet{cir85} model. Then we calculate $ \mathbf{A}$ for an $\mathbb{A}_m(d)$ model for arbitrary $0\leq m\leq d$ and $d \leq 3$. Matrix $\mathbf{A}$, for $d=3$, is presented in Appendix~\ref{appa:matrixAmd}, as for $p=4$ moments its dimension becomes large ($35 \times 35$).%

Let us start with the \citet{vasicek77} model, where  $d=1$ and $m=0$ such that $(X(t))$ follows an Ornstein-Uhlbeck process
$d X(t)= (b^P + \beta^P X(t)) dt + \Sigma \, d W^P(t)$.
For this model the generator of Markov-transition probabilities $\mathcal{G}$ is given by
\begin{equation}
\label{eq:gen1mod}
\mathcal{G} f(x)= \left(b^P + \beta^P x \right) \frac{d f(x)}{d x} + \frac{1}{2} \Sigma^2 \frac{d^2 f(x)}{d x^2}  .
\end{equation}
\noindent
Consider the basis $1,x,x^2,\dots,x^p$. The linear map $ \mathbf{A}$ used to derive the moments $\leq p$ (under $\Pmeas$) is given by
the $(p+1) \times (p+1) $ matrix
$$ \mathbf{A}=\left(
\begin{array}{cccccccc}
0& \dots &&&& \\
b^P& \beta^P &0 \dots&&&&&\\
\Sigma^2 & 2 b^P &2 \beta^P &0&\dots&&&\\
0& 3 \Sigma^2 & 3 b^P &3 \beta^P &0&\dots&&\\
&&&&\ddots&&& \\
0 &\dots&0&\frac{k(k-1)}{2}\Sigma^2& k b^P & k \beta^P&& \\
&&&&\ddots&\ddots&\ddots& \\
0 &\dots&\dots&\dots&0&\frac{p(p-1)}{2}\Sigma^2& p b^P & p \beta^P\\
\end{array} \right)  .
$$
\noindent
For the \citet{cir85} model, where $d=1$ and $m=1$, $(X(t))$ follows a square-root process
$d X(t)= (b^P + \beta^P X(t)) dt + \Sigma {\sqrt{X(t)}} d W^P(t)$.
The generator of Markov-transition probabilities $\mathcal{G}$ is given by
$$\mathcal{G} f(x)= (b^P + \beta^P x) \frac{d f(x)}{d x} + \frac{1}{2} \Sigma^2 x \frac{d^2 f(x)}{d x^2} \ , $$
\noindent
such that the linear map $ \mathbf{A}$ is given by the $(p+1) \times (p+1) $ matrix
$$ \mathbf{A}=\left(
\begin{array}{cccccccc}
0& \dots &&&& \\
b^P& \beta^P &0 \dots&&&&&\\
0 & 2 b^P + \Sigma^2 &2 \beta^P &0&\dots&&&\\
0& 0 & 3 b^P +3 \Sigma^2  &3 \beta^P &0&\dots&&\\
&&&&\ddots&&& \\
0 &\dots&\dots&0& k b^P +\frac{k(k-1)}{2}\Sigma^2 & k \beta^P&& \\
&&&&&\ddots&\ddots& \\
0 &\dots&\dots&\dots&\dots&0& p b^P+\frac{p(p-1)}{2}\Sigma^2 & p \beta^P\\
\end{array} \right)  ,
$$
\noindent
where $1 \le k \leq p$. For an $\mathbb{A}_1(3)$ model, where  $d=3$ and $m=1$, $(\mathbf{X}(t))$ follows a stochastic process
 containing one square root component. Let us start with the model under $\Qmeas$
\begin{eqnarray}
\label{dk1a13}
d \mathbf{X}(t)&=&
\left( \left(
\begin{array}{c}
b_1^Q=-\beta^Q_{11} \theta_1^{Q} > 0\\
b_2^Q=-\beta^Q_{21} \theta_1^{Q} \le 0\\
b_3^Q=-\beta^Q_{31} \theta_1^{Q} \le 0\\
\end{array}
\right)
 +
 \left(
\begin{array}{ccc}
\beta_{11}^Q<0&0&0\\
\beta_{21}^Q \ge  0&\beta_{22}^Q&\beta_{23}^Q\\
\beta_{31}^Q \ge 0&\beta_{32}^Q&\beta_{33}^Q\\
\end{array} \right)
   \mathbf{X}(t) \right) dt \nonumber \\
&&    +
\left(
\begin{array}{ccc}
\Sigma_{1} {\sqrt{X_1(t)}} & &\\
&\Sigma_{2} {\sqrt{1 + \mathcal{B}^x_{12} X_1(t) }}&\\
& &\Sigma_{3} {\sqrt{1 + \mathcal{B}^x_{13} X_1(t)}} \\
\end{array}
\right)
d \mathbf{W}^{Q} (t)  .
\end{eqnarray}
\noindent
The \citet{daisingleton00} restrictions discussed above yield:
$\theta_1^{Q} > 0$ and $\beta^Q_{11}<0$, $\mathcal{B}^x_{12}, \mathcal{B}^x_{13} \ge 0$, and $\Sigma_{1}, \Sigma_{2}, \Sigma_{3}>0$. Note that (\ref{dk1a13}) has 13 parameters while under $\Qmeas$ we can identify 14 parameters. These parameters are the thirteen parameters in (\ref{dk1a13}) and $\gamma_0$ arising in (\ref{eq:short1}).%
\footnote{In more detail:
 $\vbeta^Q$ (7 parameters), $\theta^{Q}_1$ (1 parameter;  which is $\theta_1^{Q} \geq 0$ while $\theta_2^{Q}=\theta_3^{Q}=0$, and thus $\mathbf{b}^Q= - \vbeta^Q \vtheta^{Q}=-[\beta^Q_{11}, \beta^Q_{21}, \beta^Q_{31}]' \theta^{Q}_1$), $\mathbf{\Sigma}$ (3 parameters, only the elements in the main diagonal are positive, the other parameters are zero), $\mathcal{B}^x_{12}\geq 0 $ and $\mathcal{B}^x_{13}\geq 0$.} %
The same structure is assumed under $\Pmeas$.
Based on \citet{cheriditofilipovickimmel03} this {\em extended affine market price of risk} specification is mathematically well defined given that $b_I^P=b_1^P \geq 0$, $b_J^P =(b_2^P,b_3^P)^{\prime} \leq 0$, and eight additional parameters $\beta_{11}^P \le 0$, $\beta_{21}^P \ge 0$, $\beta_{31}^P \ge 0$, $\beta_{22}^P$, $\beta_{32}^P$, $\beta_{23}^P$, $\beta_{33}^P$, contained in $\vbeta^P$, and $\theta_1^P \geq 0$ contained in $\vtheta^P$, where $\theta_2^P = \theta_3^P = 0$. Then $\mathbf{b}^P= - \vbeta^P \vtheta^{P}$. Since $\vtheta_{2:3}^P= \vtheta_{2:3}^Q= \mathbf{0}_{2}$ for the $\mathbb{A}_1(3)$ model considered, we write $\theta^Q$ and $\theta^P$ instead for $\theta_1^Q$ and $\theta_1^P$ in the following.  By collecting these parameters (not subject to an equality restriction), we obtain the vector of model parameters $\vvartheta_{\mathbb{A}_{1(3)}} \in
\mathbb{R}^{22}$.

By means of (\ref{dk1a13}) and the extended affine market price of risk assumption the generator becomes
\begin{equation}
\label{eq:A1_A13}
\mathcal{G} f(\mathbf{x}) = \sum_{i=1}^3 \left(b_i^P + \vbeta_i^P \mathbf{x} \right) \frac{\partial f(\mathbf{x})}{\partial x_i} +
\frac{1}{2} \sum_{i=1}^3 \Sigma^2_{i} \left ( \mathcal{B}_i^0+ \mathcal{B}_{1i}^x x_1 \right ) \frac{\partial^2 f(x)}{\partial x_i^2}.
\end{equation}
\noindent
The conditional expectation ${\mathbb{E}}(f(\mathbf{X}(t))|\mathbf{X}(s)=\mathbf{x})$
for $f(\mathbf{x}) \in \mathcal{P}_{ \leq p }(\mathscr{S})$ follows from Section~\ref{sectsub:polyprocess1}.
In particular, the conditional moments ${\mathbb{E}}(\mathbf{X}(t)^k|\mathbf{X}(s)=\mathbf{x})$, $t>s,$ can be derived by means of (\ref{semigroup7}), %
where $ \mathbf{A}$ is a matrix of dimension $N \times N$. We shall consider the first four moments, i.e., $p=4$. The number of moments, $N$, follows from the multinomial coefficients.
Regarding the basis elements $\mathfrak{e}_j$, $j=1, \ldots, N$, of our polynomial, we choose the basis $\left(1 \, | \, x_1, x_2, x_3 \, | \, x_1^2, \dots, x_3^2 \, | \, x_1^3, \dots, x_3^3 \, | \, x_1^4, \dots, x_3^4 \, \right)$.
In this expression we have separated the terms of different power by means of $|$.
Matrix $ \mathbf{A}$ is derived by comparing coefficients, such that $\mathcal{G} \mathfrak{e}_j = \sum_{l=1}^N  A_{jl} \mathfrak{e}_l$, for $j=1, \ldots, N$, where $A_{jl} = \left[\mathbf{A} \right]_{jl}$. With $(\mathbf{X}(t))$ of dimension $3$, we get one term for $k=0$, three terms for $k=1$, six for $k=2$, ten for $k=3$ and fifteen for $k=4$. Therefore $N=35$.
%Supplementary Material~\ref{appa:matrixAmd} derives matrix $\mathbf{A}$ for an unrestricted $\mathbb{A}_m(3)$ model.
Restricting the corresponding model parameters provides us with the matrix $\mathbf{A}$ for an $\mathbb{A}_1(3)$ model.

In the remaining part of this article we stick to following assumption.
\begin{assumption}
\label{ass:stat}
{\it The background driving process $(\mathbf{X}(t))$ is stationary.}
\end{assumption}

\noindent Sufficient conditions for a  stationary process $(\mathbf{X}(t))$ are provided in \citet{GlassermanKim2009}. For $\mathbb{A}_m(d)$ models, when $d\leq 3$, sufficient conditions for a stationary process are also reported in \citet{aitsahakiakimmel2009} and in Appendix~\ref{appa:restrict1}.

For a stationary $(\mathbf{X}(t))$, we get $\mathbb{E}  \left( \mathbf{X}(t) \right) = \vtheta^P$. In addition, to obtain higher order moments, we use the following abbreviations:  $\tilde{\mathbf{x}} %_{0:p}
 =(1,(\mathbf{x}^1)',(\mathbf{x}^2)',\dots,(\mathbf{x}^p)')^{\prime}$, which is of dimension $N$, while  $\tilde{\mathbf{x}}_{2:N}= ((\mathbf{x}^1)',(\mathbf{x}^2)',\dots,(\mathbf{x}^p)')^{\prime}$ is a $(N-1)-$dimensional vector. $\tilde{\mathbf{X}}(t)_{}$ and $\tilde{\mathbf{X}}(t)_{2:N}$ are defined in the same way. Since ${\mathbb{E}}\left( \tilde{\mathbf{X}}(t)_{}\right) = {\mathbb{E}}\left( {\mathbb{E}}(\tilde{\mathbf{X}}(t)_{}| \mathbf{X}(s)) \right)$, for $0 \leq s < t$, by the tower, rule we obtain
\begin{eqnarray}
\label{eq:matrix1}
{\mathbb{E}}\left( \tilde{\mathbf{X}}(t)_{}\right) &=& \left(\begin{array}{c}
1 \\
{\mathbb{E}} \left( \tilde{\mathbf{X}}(t)_{2:N}\right)
\end{array}
\right) =
\mathbb{E} \left( \left[\exp((t-s) \mathbf{A})\right] \tilde{\mathbf{X}}(t)_{} \right)
=  [\exp((t-s) \mathbf{A})] \mathbb{E} \left( \tilde{\mathbf{X}}(t)_{} \right) \nonumber \\
&=& \left(
\begin{array}{cc}
1 & \mathbf{0}_{1 \times N-1} \\
\left[ \exp((t-s) \mathbf{A}) \right]_{2:N,1} & \left[ \exp((t-s) \mathbf{A}) \right]_{2:N,2:N}
\end{array}
\right) \left(\begin{array}{c}
1 \\
{\mathbb{E}} \left( \tilde{\mathbf{X}}(t)_{2:N}\right)
\end{array}
\right),
\end{eqnarray}
\noindent where the $N \times N$ matrix $\exp((t-s) \mathbf{A})$ can be partitioned into four blocks: (i) north-western $\left[\exp((t-s) \mathbf{A}) \right]_{11}=1$, {(ii)} north-eastern  $\left[\exp((t-s) \mathbf{A}) \right]_{1,2:N}=\mathbf{0}_{1\times N-1}$, {(iii)} south-western $ \left[\exp((t-s) \mathbf{A}) \right]_{2:N,1}$, and {(iv)} south-eastern $ \left[\exp((t-s) \mathbf{A}) \right]_{2:N,2:N}$.%
\footnote{Note that $\exp((t-s) \mathbf{A})$ and $\mathbf{A}$ are of the same structure. This follows from the power series representation of the matrix exponential $\exp((t-s) \mathbf{A}) = \sum_{v=0}^{\infty} \frac{1}{v} \left((t-s) \mathbf{A}\right)^{v}$. In addition, the existence of $ \left( \mathbf{I}_{N-1} - \left[ \exp((t-s) \mathbf{A}) \right]_{2:N,2:N} \right)^{-1}$ follows from the properties of the matrix exponential.}
Hence, the (unconditional) moments of order $1$ to $p$ follow from
\begin{eqnarray}
\label{eq:matrix3}
{\mathbb{E}} \left( \tilde{\mathbf{X}}(t)_{2:N} \right) &=& \left( \mathbf{I}_{N-1} - \left[ \exp((t-s) \mathbf{A}) \right]_{2:N,2:N} \right)^{-1}
\left[ \exp((t-s)  \mathbf{A}) \right]_{2:N,1}  .
\end{eqnarray}

\subsection{Moments of the Observed Yields}
\label{sect:moments}

The previous Section~\ref{sect:model1} provided us with the moments of the latent process $(\mathbf{X}(t))$. By means of (\ref{eq:1}) the {\em model yields} are
\begin{eqnarray}
\label{eq:1aa}
y^{0}(t , \tau) &=&
-\frac{1}{\tau} \left(\Phi(\tau,\mathbf{0})+\vPsi(\tau,\mathbf{0})^{\prime} \mathbf{X}(t) \right)  . \nonumber
\end{eqnarray}
\noindent
Now we have to account for the fact that real world data cannot be observed on a continuous time scale, but only on a discrete grid $\Delta,2\Delta,  \dots,{\mathtt{t}}\Delta, \dots, T \Delta$, where $T$ is the time series dimension and $\Delta$ is the step-width. We set $\Delta=1$ and assume that $\mathbf{X}_{\mathtt{t}}$ stands for $\mathbf{X}({\mathtt{t}} \Delta)$. Additionally, the maturities $\tau$ available are given by $\vtau=(\tau_1,\dots,\tau_M)^{\prime}$, where $M$ is the number of maturities observed. For model yields with a maturity $\tau_i \in \{ \tau_1,\dots,\tau_M \}$ observed at $t={\mathtt{t}} \Delta$ we use the notation
$y^{0}_{\mathtt{t}i}$, $i=1, \ldots, M$. Since $M$ yields cannot be matched exactly by $d<M$ factors, we add the noise term $\varepsilon_{\mathtt{t}i}$ and arrive at the
{\em yields observed}
\begin{eqnarray}
\label{eq:3aa}
y_{\mathtt{t}i}&=&y_{\mathtt{t}i}^{0}+   \varepsilon_{\mathtt{t}i} = -\frac{1}{\tau_i} \left(\Phi(\tau_i,\mathbf{0} )+\vPsi(\tau_i,\mathbf{0} )^{\prime} \mathbf{X}_{\mathtt{t}} \right) +  \varepsilon_{\mathtt{t}i}, \ \ i=1, \ldots, M, \ \ \mathtt{t} =1, \ldots, T.  \nonumber
\end{eqnarray}

\noindent With $M$ maturities $\vtau=(\tau_1,\dots,\tau_M)$ we define
{\small
\begin{eqnarray}
\tilde{\vPhi} &=&
\left( \begin{array}{c}  -\Phi(\tau_1,\mathbf{0})/\tau_1 \\
\vdots \\
-\Phi(\tau_M,\mathbf{0})/\tau_M
\end{array} \right) \ \in \mathbb{R}^M
 \ \ \text{,} \ \
\tilde{\vPsi}  = \left(
\begin{array}{c}
-\vPsi(\tau_1,\mathbf{0})^{\prime}/\tau_1 \\
\cdots \\
-\vPsi(\tau_M,\mathbf{0})^{\prime}/\tau_M
\end{array}
\right) \ \in \mathbb{R}^{M \times d} \  \ \text{ and } \ \
{\bm \varepsilon}_{{\mathtt{t}}}  = \left(
\begin{array}{c}
\varepsilon_{\mathtt{t}1} \\
\vdots \\
\varepsilon_{\mathtt{t}M}
\end{array}
\right) \ \in \mathbb{R}^{M} \ , \nonumber
\end{eqnarray} }
such that the $M-$dimensional vector of yields, $\mathbf{y}_{\mathtt{t}} =(y_{\mathtt{t}1}, \ldots, y_{\mathtt{t}M})' $, is given by
\begin{eqnarray}
\label{eq:5aa}
\mathbf{y}_{{\mathtt{t}}} &=& \tilde{\vPhi}+ \tilde{\vPsi} \mathbf{X}_{\mathtt{t}}  +  {\bm \varepsilon}_{{\mathtt{t}}} \in \mathbb{R}^M  .
\end{eqnarray}
\noindent
Based on (\ref{eq:5aa}) we observe that the moments of $y_{{\mathtt{t}}i}$ have to follow from the moments of $\mathbf{X}_{\mathtt{t}}$.
For the noise term $\varepsilon_{{\mathtt{t}}i}$ we apply the following assumption.
\begin{assumption}
\label{ass:noise}
{\it Let $\varepsilon_{{\mathtt{t}}i}$, $\mathtt{t}=1, \ldots, T$, $i=1, \ldots, M$, be independent with zero mean, variance $0 < \sigma_i^2 < +\infty$ and  $\mathbb{E}(\varepsilon_{{\mathtt{t}}i}^4) < +\infty$. In addition $|\mathbb{E}(\varepsilon_{{\mathtt{t}}i}^p)|<+\infty$ for $i=1,\dots,M$ and $\mathbb{E}(\varepsilon_{{\mathtt{t}}i}^{2 \iota - 1} )=0$ for $\iota =1,\dots, \lfloor p/2 \rfloor $, where $\lfloor p/2 \rfloor$ is the largest integer smaller or equal to $p/2$. }
\end{assumption}
\noindent
Note that by Assumption~\ref{ass:noise} all maturities are assumed to be observed with noise. In addition, $\mathbb{E}(\varepsilon_{{\mathtt{t}}i}^{} \varepsilon_{{\mathtt{t}}j} )=0$ for $i \not = j$, $i,j=1,\dots,M$ and $\mathbb{E}(\varepsilon_{{\mathtt{t}}i}^4)<+\infty$.
By means of equation (\ref{eq:5aa}) and Assumption~\ref{ass:noise} we derive the moments of the empirical yields $\mathbb{E}(y^k_{{\mathtt{t}}i} y^l_{{\mathtt{t}}j} ) = \mathbb{E}(([\tilde{\vPhi} + \tilde{\vPsi} \mathbf{X}_{\mathtt{t}}  +  {\bm \varepsilon}_{{\mathtt{t}}}]_i)^k ([\tilde{\vPhi} + \tilde{\vPsi} \mathbf{X}_{\mathtt{t}}  +  {\bm \varepsilon}_{{\mathtt{t}}}]_j)^l )$, where {$0\leq k+l \leq p$ and} $[\cdot]_i$ extracts the $i$-th element of a vector. Hence, we derive the first four moments of the yields observed, i.e. $\mathbb{E}(y^k_{ \mathtt{t} i} )$, $k=1,\dots,4$. In addition,
  applications in finance often take the auto-covariance of the yields, $\mathbb{E}(y_{{\mathtt{t}}i} y_{{\mathtt{t}}-1 i} ) $, and the auto-covariance of the squared yields, $\mathbb{E}(y_{{\mathtt{t}}i}^2 y_{{\mathtt{t}}-1 i}^2 ) $, into consideration
(``indicator for volatility clustering" - see, e.g., the discussion in \citet{piazzesi03}[p.~649]). Therefore also the terms $\mathbb{E}(y_{{\mathtt{t}}i} y_{{\mathtt{t}}-1 i} ) $ and $\mathbb{E}\left(y^2_{\mathtt{t} i} y^2_{{\mathtt{t}}-1 i} \right) $ {are} calculated. Since this part is straightforward, but tedious algebraic manipulations were necessary to obtain all these moments, we present the results in Appendix~\ref{appa:momentsyields1}. We put the noise parameters necessary to obtain the moments of the observed yields into the parameter vector $\vvartheta_{\sigma}$. The dimension of $\vvartheta_{\sigma}$ depends on how $\sigma_i^2$ is specified and on the moments used in the estimation. If $\sigma_i^2$ is different for each maturity, we have $M$ parameters for the second moments of the noise. If, in addition, the fourth moments of the yields are calculated, the fourth moments of the noise enter into the calculations as well, i.e. we get another $M$ parameters for the moments of the noise. In this case the dimension of $\vvartheta_{\sigma}$ is $2M$. Since the dimension of the model parameter $\vvartheta_{\mathbb{A}_{1(3)}}$ is already over twenty, we continue with a more parsimonious specification of the noise, where $\sigma_i^2=\sigma^2$ and $\mathbb{E}(\varepsilon_{{\mathtt{t}}i}^4)=
\tilde{\sigma}^4$ for all $i=1,\dots,M$. Hence, the dimension of $\vvartheta_{\sigma}$ is two if fourth moments are required in the calculation of the yields observed, otherwise it is one.
This results in the model parameter vector $\vvartheta=(\vvartheta_{\mathbb{A}_{1(3)} }^{\prime},\vvartheta_{\sigma}^{\prime})^{\prime}$ of dimension $\mathfrak{p}$, which is contained in the parameter space $\Theta \in \mathbb{R}^{\mathfrak{p}}$, where due to the \citet{daisingleton00} and stationarity restrictions, $\Theta$ is proper subset of  $\mathbb{R}^{\mathfrak{p}}$. The components of $\vvartheta$ are introduced by the first column of Table~\ref{tab:res4}.

The calculation of the moments also requires to solve the Riccati equations (\ref{transformeddk1}). For the Vasicek and the Cox-Ingersol-Ross model closed form solutions are available, as e.g. presented  in \citet{filipovicbook2009}[Chapter~10.3.2]. For $\mathbb{A}_m (d)$ models, however, $\Phi$ and $\vPsi$ have to be derived by means of numerical tools in general.\footnote{See also \citet{duffiekan96,daisingleton00,chenjoslin11}.}
In this paper we follow a computationally efficient way proposed by \citet{GrasselliTebaldi2008}, to obtain an (almost) closed form solution for
$\Phi(t,\mathbf{u})$ and $\vPsi(t,\mathbf{u})$. This methodology requires the matrix $\vbeta_{II}^Q$ to be diagonal. Given \citet{daisingleton00} setup, this implies no further restrictions for $m \leq 1$, while for $m \geq 2$ the off-diagonal parts of $\vbeta_{II}^Q$ have to be set to zero.
Appendix~\ref{appa:psiphi1} shows how $\Phi$ and $\vPsi$ could be derived for an $\mathbb{A}_m (d)$ model with diagonal $\vbeta_{II}$ in a numerically parsimonious way.

\section{Parameter Estimation and Finite Sample Properties}
\label{sect:mc1}

\subsection{Parameter Estimation}

By observing yields for maturities $\tau_i$, $i=1,\dots,M$, in periods $\mathtt{t}=1,\dots,T$,
we obtain $M-$variate vectors $\mathbf{y}_{\mathtt{t}} = \left(y_{\mathtt{t}1},\dots, y_{\mathtt{t}M} \right)^{\prime}$, $\mathtt{t}=1,\dots,T$, the observations of $M-$variate time series
$\mathbf{y}_{1:T}= \left(\mathbf{y}_1^{\prime},\dots, \mathbf{y}_T^{\prime} \right)^{\prime}$, as well as
$\tilde{\mathfrak{q}}-$dimensional vectors
$\tilde{\mathbf{m}}_{({\mathtt{t}})} \left( \mathbf{y}_{1:T} \right) =
\left( y_{\mathtt{t}1}, \dots, y_{ \mathtt{t}M}^p,y_{ \mathtt{t}1} y_{ \mathtt{t}-1,1},  \dots ,y_{\mathtt{t}M}^2y_{ \mathtt{t}-1,M}^2  \right)^{\prime}$ %for ${{\mathtt{t}}}=2,\dots,T$,
 and  $\tilde{\mathbf{m}}_{T} \left( \mathbf{y}_{1:T} \right) = \left( \frac{1}{T} \sum_{{\mathtt{t}}=1}^T y_{\mathtt{t}1}, \dots, \frac{1}{T} \sum_{{\mathtt{t}}=1}^T y_{ \mathtt{t}M}^p,
 \frac{1}{T-1}\sum_{{\mathtt{t}}=2}^T y_{ \mathtt{t}1} y_{ \mathtt{t}-1,1},  \dots , \frac{1}{T-1} \sum_{{\mathtt{t}}=2}^T y_{\mathtt{t}M}^2y_{ \mathtt{t}-1,M}^2  \right)^{\prime}$.

Let $\tilde{\vmu}(\vvartheta)$$ =\left( \mathbb{E} (y_{\mathtt{t} 1}),\dots,\mathbb{E} (y_{\mathtt{t} M}^p),\mathbb{E} (y_{\mathtt{t} 1} y_{\mathtt{t}-1, 1}),\dots,
\mathbb{E} (y_{\mathtt{t} M}^2 y_{\mathtt{t}-1, M}^2) \right)^{\prime} $ stand for the corresponding vector of moments as a function of the unknown parameter vector $\vvartheta \in \Theta \subset \mathbb{R}^\mathfrak{p}$. The components of the vector $\tilde{\vmu}(\vvartheta)$ are provided in Appendix~\ref{appa:momentsyields1} (see equations (\ref{eq:momy1}), (\ref{mm2}), (\ref{m3i2j}), (\ref{m3ij2}), (\ref{eq:momy4}), (\ref{mst21}) and (\ref{mst22})).

The generalized method of moments demands for $\mathfrak{q} \geq \mathfrak{p}$ moments to be selected. By means of
a $\mathfrak{q} \times \tilde{\mathfrak{q}}$ selector matrix $\mathcal{M}$, where $\left[ \mathcal{M} \right]_{ij}=1 $ if the corresponding moment is used and zero otherwise, we obtain
$\vmu(\vvartheta) = \mathcal{M} \, \tilde{\vmu}(\vvartheta) \in \mathbb{R}^{\mathfrak{q}}$,
${\mathbf{m}}_{(\mathtt{t})} \left( \mathbf{y}_{1:T} \right) = \mathcal{M} \, \tilde{\mathbf{m}}_{(\mathtt{t})} \left( \mathbf{y}_{1:T} \right) \in \mathbb{R}^{\mathfrak{q}}$
 and ${\mathbf{m}}_{T} \left( \mathbf{y}_{1:T} \right) = \mathcal{M} \, \tilde{\mathbf{m}}_{T} \left( \mathbf{y}_{1:T} \right) \in \mathbb{R}^{\mathfrak{q}}$. Next we define $\mathbf{h}_{(\mathtt{t})}  \left( {\vvartheta} ; \mathbf{y}_{1:T} \right)  =  \mathbf{m}_{(\mathtt{t})} (\mathbf{y}_{1:T}) - \vmu(\vvartheta)$ and  $\mathbf{h}_T \left( {\vvartheta};\mathbf{y}_{1:T} \right) = \mathbf{m}_T (\mathbf{y}_{1:T}) - \vmu(\vvartheta)$ as well as the $GMM$ distance function
\begin{equation}
\label{def:Qtgmm1}
Q_T( \vvartheta; \mathbf{y}_{1:T}) = \mathbf{h}_T ({\vvartheta};\mathbf{y}_{1:T} )^{\prime} \, \mathbf{C}_T \, \mathbf{h}_T({\vvartheta}; \mathbf{y}_{1:T})  .
\end{equation}

\noindent
The $GMM$ estimate $\widehat{\vvartheta}$ (of $\vvartheta$) minimizes $Q_T( \cdot)$ in (\ref{def:Qtgmm1}), where $\mathbf{C}_T$ is a $\mathfrak{q} \times \mathfrak{q}$ symmetric positive semi-definite weighting matrix \citep[see, e.g.,][Chapters~21-22]{Ruudbook2000}. In particular, the {\em continuous updating estimator} ($CUE$) is used to obtain an efficient $GMM$ estimate. That is, we run an iterative procedure
with iteration steps $\mathtt{m}=1,\dots,\mathtt{M}$, where we commute between (i) augmenting the parameter-estimate to ${\vvartheta}^{(\mathtt{m})}$
based on $Q_T( \cdot)$ given $C_T$ and (ii) updating $C_T$ given ${\vvartheta}^{(\mathtt{m}-1)}$ from the previous iteration step $\mathtt{m}-1$. The weighting matrix applied is
%$\mathbf{C}_T = \mathbf{I}_{\mathfrak{q}}$ in the initial step $\mathtt{m}=1$, %of this iterative procedure,
%while 
$\mathbf{C}_T  = \left(\hat{\vLambda}_T({\vvartheta}^{(\mathtt{m}-1)} )\right)^{-1}$, with $\hat{\vLambda}_T \left( \vvartheta^{(\mathtt{m}-1)} \right)$$ = \frac{1}{T-1} \sum_{ \mathtt{t}=2}^T \mathbf{h}_{(\mathtt{t})} ( {\vvartheta}^{(\mathtt{m}-1)} ;\mathbf{y}_{1:T} ) \, \mathbf{h}_{(\mathtt{t})} \, ( {\vvartheta}^{(\mathtt{m}-1)} ;\mathbf{y}_{1:T} )^{\prime}$. 
For regularity conditions and further issues on $GMM$ estimation see, e.g., \citet{hansen82,AltonjiSegal1996,PoetscherPrucha1997,Windmeijer2005,Guggenberger2005,NeweyWindmeijer2009}.

To satisfy the order condition, the inequality {``}$\mathfrak{q} \geq \mathfrak{p}${''} has to be fulfilled. For the $\mathbb{A}_1(3)$ model considered in Section~\ref{sect:polyprocess1}, the dimension of the parameter vector $\vvartheta$ is $23$ ($\mathfrak{p}=23$), if moments of order smaller than four are used.
Including fourth order moments of the yields results in $\mathfrak{p}=24$. The number of maturities $M$ available is around ten. Therefore, by using the moments $\mathbb{E}(y_{ \mathtt{t} i} )$, $\mathbb{E}(y_{ \mathtt{t} i}^2)$ and $\mathbb{E}(y_{ \mathtt{t} i} y_{ \mathtt{t-1}, i} )$ for $i=1,\dots,M$, we are already equipped with $3M$ moment conditions. Hence, for $M \geq 8$ the order condition $\mathfrak{q} \geq \mathfrak{p}$ is already met.
By using the first four moments ($p=4$) and the auto-covariances (for $M=10$), the number of moments is much larger than the number of parameters.

To obtain parameter estimates, a high-dimensional nonlinear minimization problem has to be solved and
$\mathfrak{q}$ moment conditions have to be selected from the set of moments available. Regarding the latter issue,
it turned out that the instability of the parameter estimates is amplified if higher order moments are added. Due to this instability, using the Wald and the distance difference tests to test for redundant moment conditions \citep[testing for over-identifying restrictions; see, e.g.,][Chapter 22.2]{Ruudbook2000} provide us with very ambiguous results. Hence, the selection of these moments was performed by means of simulation experiments. Based on the simulation results, we work with $\mathfrak{q}=27$ moment conditions, namely, $\mathbb{E}(y_{ \mathtt{t} i})$, $\mathbb{E}(y_{ \mathtt{t} i} y_{ \mathtt{t}-1, i})$ , $i=1,\dots,M=10$, and $\left[ \mathbb{E} \left( \mathbf{y}_{ \mathtt{t}} \mathbf{y}_{\mathtt{t}}^{\prime} \right) \right]_{ij}$, for ${(i,j)} = (1,1)$, $(2,2)$, $(3,2)$, $(5,5)$, $(7,7)$, $(9,10)$ and $(10,10)$.

Regarding the minimization of the $GMM$ distance function, we observe that standard minimization procedures designed to find local minima do not result in reliable parameter estimates. 
In more detail, to investigate the properties of our estimation routine we performed Monte Carlo experiments with simulated yields where $M=10$, $T=500$ and the number of simulation runs is $1,000$.  The
parameter vector $\vvartheta$ used to generate the yields is presented in the second column of Table~\ref{tab:res4}. The initial values for the $GMM$ estimation, $\vvartheta^{(\mathtt{m}_0)}$, are generated as follows:
$[\vvartheta^{(\mathtt{m}_0)}]_j = [\vvartheta]_j + c_{\vartheta} [ | \vvartheta|]_j \zeta_j$ for coordinate $j$, when the support is the real axis, while
$[\vvartheta^{(\mathtt{m}_0)}]_j = \exp\left( \log [|\vvartheta |]_j  + c_{\vartheta} \zeta_j \right) \text{sgn} \left(  [\vvartheta]_j  \right)$ is used for the elements $j$ living on the non-positive or non-negative part of the real axis. $\zeta_j$ is $iid$ standard normal and distortion parameter $c_{\vartheta}$ is set to $0$, $0.1$, $0.25$, $0.5$ and $1$. Then, parameter estimates are obtained by means of the MATLAB minimization routine $\mathtt{fminsearch}$ based on the Nelder-Mead algorithm.\footnote{See $\mathtt{http://www.mathworks.de/de/help/matlab/ref/fminsearch.html}$}
With this algorithm an estimate $\widehat{\vvartheta}$ is provided by ${\vvartheta}^{(\mathtt{M})}$, where -- in this case --- $\mathtt{M}$ is the last iteration step. 
We observe that the parameters can be estimated easily by means of this standard minimization tool when $c_{\vartheta} \leq 0.25$; i.e., when the optimization is started sufficiently close to the true parameter $\vvartheta$. 
{However,} the parameter estimation {with} $c_{\vartheta}=0.5$ or $c_{\vartheta}=1$ becomes a difficult problem.%
{\footnote{By combining multistart random search methods \citep[see, e.g.,][]{ToernZilinskas1989} with  the Nelder-Mead algorithm, we observe that the parameter estimates improve. However, performing inference still remains a difficult problem. For more details see Appendix~\ref{sect:gmmest1}.}

To cope with this problem, we combine multistart random search methods with Quasi-Bayesian methods
\citep[see, e.g.,][]{ToernZilinskas1989,chernozhukovhong03}.
For each Monte Carlo run $\ell$, where $\ell=1,\dots,\mathtt{L}=200$, we proceed as follows:
First, parameter estimation is started with the random draws $\vvartheta^{(\mathtt{n})}$, where $\mathtt{n} = 1,\dots, \mathtt{N} = 2,000$.
The samples $\vvartheta^{(\mathtt{n})}$ are generated in the same way as
$\vvartheta^{(\mathtt{m}_{0})}$ in the above paragraph with distortion parameter $c_{\vartheta}=1$.
Then $\vvartheta^{(\mathtt{n})}$ with the smallest GMM distance function is used as the starting value
of the Quasi-Bayesian sampler. Appendix~\ref{sect:gmmest1} describes how the draws, ${\vvartheta}^{(\mathtt{m})}$, from an ergodic Markov Chain are obtained.\footnote{In our analysis $\mathtt{m}=1,2,\dots,\mathtt{M}=20,000$.} Finally, parameter estimates $\widehat{\vvartheta}_{\ell}$ as well as the estimates of the variance $\hat{\mathbb{V}}_{BM} \left( \left[ \widehat{\vvartheta}_{\ell} \right]_{\iota \iota} \right)$, $\iota =1,\dots,\mathfrak{p}$, are derived from these draws, where the latter are obtained by applying a batch mean estimator \citep[see][in particular, Equation~(6)]{FlegalJones2010}.

Tables~\ref{tab:res4} and \ref{tab:res5} present results from our Monte Carlo experiments. In both tables the true parameter vector $\vvartheta$ is provided in the second column. In Table~\ref{tab:res4} the data are generated such that $\theta^P=1.5 \not= 10=\theta^Q$, while {$\theta^P=\theta^Q=1.5$} in Table~\ref{tab:res5}. In all Monte Carlo experiments an unrestricted model is estimated.
That is, we obtain separate estimates for $\theta^P$ and $\theta^Q$, respectively. We force our multistart random search routine to generate samples {such that} $\left(\theta^P\right)^{(\mathtt{n})} =  \left(\theta^Q \right)^{(\mathtt{n})}$ as well as $\left(\theta^P\right)^{(\mathtt{n})} \not=  \left(\theta^Q \right)^{(\mathtt{n})}$ {(for both experiments presented in Tables~\ref{tab:res4} and ~\ref{tab:res5}, respectively)}.
 In addition, a reversible jump move, based on \citet{Green1995} and \citet{RichardsonGreen1997} is included in the Bayesian sampler. The reversible jump move turned out to be useful in the case when $\theta^P=\theta^Q$  (see Appendix~\ref{sect:gmmest1} for more details).

From estimates $\widehat{\vvartheta}_{\ell}$, $\ell=1,\dots,\mathtt{L}=200$, we obtain the sample $mean$, $median$, minimum ($min$), maximum ($max$), standard deviation ($std$), skewness ($skew$) and kurtosis ($kurt$).
These descriptive statistics are reported in columns three to nine of Tables~\ref{tab:res4} and \ref{tab:res5}. The last column presents the absolute difference between the sample mean of the estimates and the true parameter value.
 
Comparing results based on Quasi-Bayesian methods (see Table~\ref{tab:res4}) for the case when $\theta^P \not=  \theta^Q$ to results based on a standard minimization procedure (see~Table~\ref{tab:res1} in Supplementary Material~\ref{sect:gmmest1}), we see that the Quasi-Bayesian approach
reduces the standard deviations of the point estimates for most parameters.
For example, the standard deviation of the point estimate of $\theta^Q$ is reduced from $6.05$ (see~Table~\ref{tab:res1}) to approximately $3.05$ (see Table~\ref{tab:res4}). Similar effects are observed for the estimates of the terms driving volatility, i.e., $\Sigma_{1}$, $\Sigma_{2}$, $\Sigma_{3}$ and $\sigma_{\varepsilon}^2$, {which are difficult to estimate}. By considering the smallest and the largest point estimates ($min$ and $max$ in the corresponding tables), we observe a substantially smaller dispersion in the point estimates of $\vvartheta$ for the
Quasi-Bayesian approach. Note that an estimate of $\theta^P$ is an estimate of the expected value of the first component of the process $\left( \mathbf{X}(t) \right)_{t \geq 0}$.
Since the serial correlation of $\left( \mathbf{X}_{\mathtt{t}} \right)_{\mathtt{t} \in \mathbb{N}_0}$ is quite high, we know from estimating means of an autoregressive process, that the standard error of the estimator of the mean becomes large (e.g., when the Fisher-information matrix of an $AR(1)$ process is calculated).
Similar results are presented in Table~\ref{tab:res5} for the $\theta^P=\theta^Q$ case.

\begin{table}[h]
\begin{center}
\begin{tabular}{cr|rrrrrrrr}%
\hline \hline
\multicolumn{2}{c|}{$\vvartheta$}	&	$mean$	&	$median$	&	$min$	&	$max$	&	$std$	& $skew$	&	$kurt$ &	$|\vvartheta- \widehat{\vvartheta}|$		 \\
&& \multicolumn{1}{|c}{$\widehat{\vvartheta}$}  &&&&&&& \\
\hline
\hline
$\theta^Q$	&	10	&	8.8660	&	8.5486	&	0.1534	&	19.2247	&	3.5008	&	0.6071	&	4.2157	&	1.1340	\\
$\theta^P$	&	1.5	&	1.6610	&	1.4883	&	0.0042	&	2.5920	&	1.0643	&	0.4911	&	-0.3916	&	0.1610	\\
$\beta^Q_{11}$	&	-1	&	-1.6418	&	-1.2797	&	-9.2212	&	-0.4173	&	1.5798	&	-3.2923	&	11.8217	&	0.6418	\\
$\beta^Q_{21}$	&	0.2	&	0.1817	&	0.1524	&	0.0025	&	0.3591	&	0.1299	&	1.8759	&	6.5029	&	0.0183	\\
$\beta^Q_{31}$	&	0.02	&	0.0350	&	0.0214	&	1.86E-5	&	0.3473	&	0.0457	&	3.1329	&	14.2229	&	0.0150	\\
$\beta^Q_{22}$	&	-1	&	-1.4731	&	-1.0671	&	-8.1154	&	-0.4823	&	1.1478	&	-2.7690	&	10.1519	&	0.4731	\\
$\beta^Q_{32}$	&	0.04	&	0.0373	&	0.0219	&	-0.0662	&	0.2711	&	0.0606	&	2.3781	&	10.2813	&	0.0027	\\
$\beta^Q_{23}$	&	0	&	0.0006	&	-0.0003	&	-0.0840	&	0.0266	&	0.0176	&	1.5436	&	16.5725	&	0.0006	\\
$\beta^Q_{33}$	&	-0.8	&	-1.5327	&	-1.2070	&	-7.8466	&	-0.6308	&	1.2389	&	-2.5704	&	8.1375	&	0.7327	\\
$\beta^P_{11}$	&	-1	&	-1.5069	&	-0.9650	&	-7.0168	&	-0.1670	&	1.4929	&	-1.5812	&	1.8702	&	0.5069	\\
$\beta^P_{21}$	&	0.02	&	0.0288	&	0.0037	&	3.67E-6	&	0.0170	&	0.0778	&	5.4759	&	35.4115	&	0.0088	\\
$\beta^P_{31}$	&	0.01	&	0.0099	&	0.0032	&	4.44E-7	&	0.0006	&	0.0206	&	5.0431	&	32.4652	&	0.0001	\\
$\beta^P_{22}$	&	-0.7	&	-1.1194	&	-0.6085	&	-7.5792	&	-0.1400	&	1.2938	&	-2.2389	&	6.0933	&	0.4194	\\
$\beta^P_{32}$	&	0.01	&	-1.1194	&	-0.6085	&	-7.5792	&	-0.1400	&	1.2938	&	-2.2389	&	6.0933	&	1.1294	\\
$\beta^P_{23}$	&	0	&	-0.0015	&	0.0000	&	-0.0551	&	0.0017	&	0.0104	&	-1.1382	&	8.4369	&	0.0015	\\
$\beta^P_{33}$	&	-0.7	&	-0.9059	&	-0.4692	&	-6.5051	&	-0.1844	&	1.1881	&	-2.7918	&	7.8669	&	0.2059	\\
$\mathcal{B}^{x}_{12}$	&	0.1	&	0.0623	&	0.0123	&	2.43E-6	&	0.0493	&	0.1652	&	5.6178	&	35.3695	&	0.0377	\\
$\mathcal{B}^{x}_{13}$	&	0.01	&	0.1045	&	0.0352	&	8.08E-7	&	0.8676	&	0.1856	&	3.3315	&	13.2906	&	0.0945	\\
$\gamma_0$	&	2	&	1.7855	&	1.8939	&	-0.0070	&	3.2115	&	0.8411	&	-0.0384	&	-0.0520	&	0.2145	\\
$\Sigma_{1}$	&	0.7	&	0.5921	&	0.5238	&	0.2002	&	1.3639	&	0.3121	&	0.9334	&	0.4182	&	0.1079	\\
$\Sigma_{2}$	&	1	&	0.4704	&	0.3714	&	0.1060	&	0.9983	&	0.3336	&	1.3636	&	1.4000	&	0.5296	\\
$\Sigma_{3}$	&	0.8	&	0.4563	&	0.3447	&	0.1071	&	1.0514	&	0.3451	&	1.4573	&	1.8141	&	0.3437	\\
$\sigma_{\varepsilon}^2$	&	0.0067	&	0.0113	&	0.0096	&	0.0053	&	0.0176	&	0.0047	&	0.7672	&	-0.5302	&	0.0046	\\
\hline																	
\hline												
\end{tabular}
\caption{Parameter estimates for the $\mathbb{A}_{1}(3)$ based on Quasi-Bayesian methods. Data simulated with $M=10$, $T=500$ and $\theta^Q \not= \theta^P$. $c_{\vartheta}=1$ is controlling for the noise in the generation of the starting value of the optimization routine.  Statistics are obtained from $\mathtt{L}=200$ simulation runs. $mean$, $median$, $min$, $max$, $std$, $skew$ and $kurt$ stand for the sample mean, median, minimum, maximum, standard deviation, skewness and kurtosis of the point estimates $\widehat{\vvartheta}_{\ell}$, $\ell=1,\dots,\mathtt{L}$. $|\vvartheta- \widehat{\vvartheta}|$ stands for absolute value of the mean deviation from the true parameter. The true parameter values  $\vvartheta$ are reported in the second column.}
\label{tab:res4}
\end{center}
\end{table}

\begin{table}[h]
\begin{center}
\begin{tabular}{cr|rrrrrrrr}%
%\begin{tabular}{cr|cccccccc}%
\hline \hline
\multicolumn{2}{c|}{$\vvartheta$}	&	$mean$	&	$median$	&	$min$	&	$max$	&	$std$	& $skew$	&	$kurt$ &	$|\vvartheta- \widehat{\vvartheta}|$		 \\
&& \multicolumn{1}{|c}{{$\widehat{\vvartheta}$}}  &&&&&&& \\
\hline
$\theta^Q$	&	1.5	&	1.7127	&	1.2500	&	0.0148	&	5.4034	&	1.5225	&	2.4322	&	6.6231	&	0.2127	\\
$\theta^P$	&	1.5	&	1.4298	&	1.4745	&	0.0218	&	2.1810	&	0.5370	&	-0.2753	&	0.6087	&	0.0702	\\
$\beta^Q_{11}$	&	-1	&	-0.9482	&	-0.7216	&	-9.3936	&	-0.2657	&	1.1017	&	-5.9892	&	42.4434	&	0.0518	\\
$\beta^Q_{21}$	&	0.2	&	0.2760	&	0.1745	&	0.0082	&	0.5801	&	0.3184	&	2.7465	&	8.6887	&	0.0760	\\
$\beta^Q_{31}$	&	0.02	&	0.0365	&	0.0188	&	0.0001	&	0.0271	&	0.0501	&	3.5544	&	16.0667	&	0.0165	\\
$\beta^Q_{22}$	&	-1	&	-1.4434	&	-1.1180	&	-8.5167	&	-0.6810	&	1.1585	&	-2.6154	&	10.2604	&	0.4434	\\
$\beta^Q_{32}$	&	0.04	&	0.0391	&	0.0280	&	-0.0514	&	0.0828	&	0.0483	&	1.8007	&	6.8699	&	0.0009	\\
$\beta^Q_{23}$	&	0	&	-0.0013	&	-0.0001	&	-0.0562	&	0.0295	&	0.0108	&	-0.9656	&	8.6647	&	0.0013	\\
$\beta^Q_{33}$	&	-0.8	&	-1.3134	&	-1.0069	&	-6.6218	&	-0.5230	&	1.0095	&	-2.1866	&	6.7837	&	0.5134	\\
$\beta^P_{11}$	&	-1	&	-1.8616	&	-1.4688	&	-6.8225	&	-0.7239	&	1.5857	&	-0.8374	&	0.4486	&	0.8616	\\
$\beta^P_{21}$	&	0.02	&	0.2610	&	0.1233	&	0.0017	&	0.8445	&	0.4000	&	4.3370	&	26.3265	&	0.2410	\\
$\beta^P_{31}$	&	0.01	&	0.0314	&	0.0127	&	0.0001	&	0.0602	&	0.0489	&	3.4009	&	15.1149	&	0.0214	\\
$\beta^P_{22}$	&	-0.7	&	-1.1592	&	-0.8226	&	-6.6295	&	-0.1769	&	1.0613	&	-1.5312	&	4.0173	&	0.4592	\\
$\beta^P_{32}$	&	0.01	&	0.0383	&	0.0207	&	-0.1791	&	0.0872	&	0.0606	&	1.8303	&	6.8339	&	0.0283	\\
$\beta^P_{23}$	&	0	&	-0.0010	&	0.0002	&	-0.2496	&	0.0425	&	0.0231	&	-4.7594	&	70.9923	&	0.0010	\\
$\beta^P_{33}$	&	-0.7	&	-1.3493	&	-1.0871	&	-6.3858	&	-0.1818	&	1.2570	&	-1.5028	&	3.0873	&	0.6493	\\
$\mathcal{B}^{x}_{12}$	&	0.1	&	0.0769	&	0.0326	&	0.0007	&	0.0456	&	0.1328	&	3.8884	&	18.7903	&	0.0231	\\
$\mathcal{B}^{x}_{13}$	&	0.01	&	0.1262	&	0.0677	&	0.0019	&	0.2241	&	0.1816	&	4.0900	&	22.9321	&	0.1162	\\
$\gamma_0$	&	2	&	1.9495	&	1.9564	&	0.0111	&	2.1332	&	0.5018	&	-2.0313	&	7.2117	&	0.0505	\\
$\Sigma_{1}$	&	0.7	&	0.8427	&	0.9478	&	0.0186	&	1.1315	&	0.3418	&	-0.7662	&	-0.4319	&	0.1427	\\
$\Sigma_{2}$	&	1	&	0.6263	&	0.5797	&	0.0225	&	1.0547	&	0.3573	&	0.4775	&	-0.4588	&	0.3737	\\
$\Sigma_{3}$	&	0.8	&	0.5591	&	0.4891	&	0.0182	&	1.2413	&	0.3631	&	0.6926	&	-0.4573	&	0.2409	\\
$\sigma_{\varepsilon}^2$	&	0.0067	&	0.0106	&	0.0093	&	0.0009	&	0.0215	&	0.0049	&	0.6131	&	-0.2135	&	0.0039	\\
\hline																	
\hline												
\end{tabular}
\caption{Parameter estimates for the $\mathbb{A}_{1}(3)$ based on Quasi-Bayesian methods. Data simulated with $M=10$, $T=500$ and $\theta^Q=\theta^P$. $c_{\vartheta}=1$ is controlling for the noise in the generation of the starting value of the optimization routine.  Statistics are obtained from $\mathtt{L}=200$ simulation runs. $mean$, $median$, $min$, $max$, $std$, $skew$ and $kurt$ stand for the sample mean, median, minimum, maximum, standard deviation, skewness and kurtosis of the point estimates $\widehat{\vvartheta}_{\ell}$, $\ell=1,\dots,\mathtt{L}$. $|\vvartheta- \widehat{\vvartheta}|$ stands for absolute value of the mean deviation from the true parameter. The true parameter values  $\vvartheta$ are reported in the second column.}
\label{tab:res5}
\end{center}
\end{table}

\subsection{Inference}
%{\em Inference:}
The asymptotic distribution of $\sqrt{T} \left( \widehat{\vvartheta} - {\vvartheta} \right)$ is a normal distribution with mean vector $\mathbf{0}_{\mathfrak{p}}$ and the asymptotic covariance matrix $\mathbf{V}$ \citep[for more details and regularity conditions see, e.g.,][]{hansen82,PoetscherPrucha1997,neweyMcfadden94,Ruudbook2000}.
As our test statistics rely on asymptotic results, we have to investigate the finite sample properties of our tests. Since a lot of parameters are considered and various restrictions can be constructed, we focus now on the restriction $\theta^P=\theta^Q$, which is often discussed in finance literature.

To test for parameter restrictions, we assume that the null hypothesis consists of $\mathfrak{r}_{\mathfrak{p}}$ restrictions. %
Suppose that these restrictions are described by a twice continuously differential function $r(\vvartheta): \mathbb{R}^\mathfrak{p} \rightarrow \mathbb{R}^{\mathfrak{r}_{\mathfrak{p}}}$ and the $\mathfrak{r}_{\mathfrak{p}} \times \mathfrak{p}$ matrix of partial derivatives
\begin{eqnarray}
\mathbf{R} = \mathbf{D}_{\vvartheta} \mathbf{r}( \widehat \vvartheta) = \left (
\begin{array}{ccc}
\frac{\partial r_1 (\widehat \vvartheta)}{\partial \vartheta_1} & \cdots &  \frac{\partial r_1 (\widehat \vvartheta)}{\partial \vartheta_{\mathfrak{p}}} \\
\cdots & \cdots & \cdots \\
\frac{\partial r_{\mathfrak{r}_{\mathfrak{p}}} (\widehat \vvartheta)}{\partial \vartheta_1} & \cdots &  \frac{\partial r_{\mathfrak{r}_{\mathfrak{p}}} (\widehat \vvartheta)}{\partial \vartheta_{\mathfrak{p}}}
\end{array}
\right ) ,
\end{eqnarray}
\noindent
which has rank $\mathfrak{r}_{\mathfrak{p}}$. Under the null hypothesis we have $ \mathbf{r}(\vvartheta)=\mathbf{0}_{\mathfrak{r}_{\mathfrak{p}}}$ and thus the Wald-statistic becomes
\begin{eqnarray}
\label{eq:wald1}
\mathscr{W}&=& T \mathbf{r}(\widehat \vvartheta )^{\prime} \left(  \mathbf{R} \hat{\mathbf{V}}_T  \mathbf{R}^{\prime} \right)^{-1}  \mathbf{r}(\widehat \vvartheta)  \ ,
\end{eqnarray}
\noindent
where $\hat{\mathbf{V}}_T$ is an estimate of the asymptotic covariance {matrix} of $\sqrt{T}(\widehat{\vvartheta}-\vvartheta)$. Under the null hypothesis the Wald-statistic $\mathscr{W}$ follows a $\chi^2$-distribution with $\mathfrak{r}_{\mathfrak{p}}$ degrees of freedom. The null hypothesis is rejected if $\mathscr{W} > \chi^2_{\mathfrak{r}_{\mathfrak{p}},1-{\mathit{\alpha_S}}}$, where ${\mathit{\alpha_S}}$ is the significance level and $\chi^2_{\mathfrak{r}_{\mathfrak{p}},1-{\mathit{\alpha_S}}}$ is the $1-\mathit{\alpha_S}$ percentile of a $\chi^2$-distribution with $\mathfrak{r}_{\mathfrak{p}}$ degrees of freedom. In particular, if the goal is to test the null hypothesis $\theta^P = \theta^Q$ against the alternative $\theta^P \not= \theta^Q$, then $\mathfrak{r}_{\mathfrak{p}}=1$, $r(\vvartheta)= \left(1,-1,0,\dots,0 \right) \vvartheta = \theta^Q-\theta^P$ and $\mathbf{R} = \left(1,-1,0,\dots,0 \right)$.%
\footnote{The components of the parameter vector $\vvartheta$ are presented in the first column of Table~\ref{tab:res4}.}

Appendix~\ref{sect:gmmest1} demonstrates that the performance of the Wald test implemented in a standard way (as well as the distance difference test) is poor. 
In particular with $c_{\vartheta}=1$, substantial undersizing is observed for the Wald test while the power is very low. 
With the distance difference test we observe 
only minor oversizing, and even if it's power is already better than the power of the Wald test, is is still low (approximately 55\% rejection rate on a 5\% significance level).%
\footnote{We used here the same simulation designs as in Tables~\ref{tab:res4} and ~\ref{tab:res5}.} %
 To implement a {``}standard{''} Wald or distance difference test, 
the $\mathfrak{p} \times \mathfrak{p}$ covariance matrix $\mathbf{V}$ is estimated by means of the {``}standard $GMM$ covariance matrix estimate{''} \citep[see, e.g.,][Chapters~21 and 22, for a  {``}standard{''} implementation of the Wald and the distance difference test]{Ruudbook2000}. That is, when the {following estimate is applied}
\begin{eqnarray}
\label{eq:VNstand}
\hat{\mathbf{V}}_T &=& \left(\hat{\mathbf{H}}_T ^{\prime} \hat{\vLambda}_T^{-1}  \hat{\mathbf{H}}_T  \right)^{-1}  \ , \ \text{where }  \nonumber \\
 \hat{\mathbf{H}}_T &=& \frac{1}{T-1} \sum_{\mathtt{t}=2}^{T}
 \mathbf{D}_{\vvartheta} \mathbf{h}_{(\mathtt{t})} \left( \widehat{\vvartheta}; {\mathbf{y}}_{1:T} \right)
\in \mathbb{R}^{\mathfrak{q} \times \mathfrak{p}}
  \ \text{and } \nonumber \\
\hat{\mathbf{\vLambda}}_T &=& \frac{1}{T-1} \sum_{ \mathtt{t}=2}^T \mathbf{h}_{(\mathtt{t})} \left( \widehat{\vvartheta}; {\mathbf{y}}_{1:T} \right)  \mathbf{h}_{(\mathtt{t})} \left( \widehat{\vvartheta}; {\mathbf{y}}_{1:T} \right)^{\prime} \in \mathbb{R}^{\mathfrak{q} \times \mathfrak{q}}  .
\end{eqnarray}
\noindent
Note that in (\ref{eq:VNstand}) matrices of dimension $\mathfrak{p} \times \mathfrak{p}$ (with $\mathfrak{p} \geq 23$) have to be inverted and partial derivatives in matrix $\mathbf{D}_{\vvartheta} \mathbf{h}_{(\mathtt{t})} \left( \widehat{\vvartheta}; {\mathbf{y}}_{1:T} \right)$ have to be derived numerically. Hence, estimating the covariance matrix $\mathbf{V}$ by means of (\ref{eq:VNstand}) is numerically demanding. Additionally, $\hat{\mathbf{H}}_T$ as well as $\hat{\mathbf{\vLambda}}_T$ also depend on ${\mathbf{y}}_{1:T}$, and therefore are subject to the variation of the finite samples.

To cope with this problem, we use the output of the Bayesian sampler to perform inference. Based on \citet{chernozhukovhong03}, asymptotic normality still holds and the draws from an ergodic Markov Chain, ${\vvartheta}^{(\mathtt{m})}$, can be used to estimate the covariance matrix $\mathbf{V}$.
In particular, to estimate the asymptotic variance of $\hat \theta^P- \hat \theta^Q = \left(1, -1, 0,\dots,0 \right) \widehat{\vvartheta}$, we use Markov-Chain Monte Carlo output and the batch mean estimator \citep[see][Equation~(6)]{FlegalJones2010}. For the Wald test, rejection rates of the true and the false null-hypothesis are provided in Table~\ref{tab:res7}. We observe that the rejection rates of the true null-hypothesis $\theta^Q=\theta^P$ are quite close to their theoretical values $\alpha_S$.

\begin{table}[h]
\begin{center}
\begin{tabular}{c|c|c}%
\hline \hline
$\alpha_{S}$ & \multicolumn{1}{c}{$\theta^Q=10 \not=1.5= \theta^P$} & \multicolumn{1}{c}{$\theta^Q=\theta^P=1.5$}\\
 \hline
0.01	&	1.0000	&	0.0286 \\
0.05	&	1.0000	&	0.0476	\\
0.10	& 1.0000	& 0.0857		\\
\hline																	
\hline												
\end{tabular}
\caption{Parameter tests based on the Wald test~(\ref{eq:wald1}): Data simulated with $M=10$ and $T=500$; $\alpha_{S}$ stands for the significance level; $c_{\theta} =1$ controls for the noise in the generation of the starting value of the optimization routine. The null hypothesis is $\theta^Q=\theta^P$, which is tested against the two sided alternative $\theta^Q \not= \theta^P$.
The draws of the Quasi-Bayesian sampler are used to estimate $\theta^Q$, $\theta^P$ as well as the asymptotic variance of $\hat{\theta}^Q - \hat{\theta}^P$.
The quantities presented are rejection rates of the null hypothesis given the significance level $\alpha_{S}$. Statistics are obtained from $\mathtt{L}=200$ simulation runs.
}
\label{tab:res7}
\end{center}
\end{table}

\section{Parameter Estimation in Empirical Data}
\label{sect:emp1}

This section applies the estimator developed in the previous sections to empirical data. We downloaded H-15 interest rate data from the Federal Reserve.\footnote{\texttt{http://federalreserve.gov/releases/h15/data.htm}} In particular, we used weekly data (measured every Friday) of {``}Treasury constant maturity{''} yields.
The time period considered is August 3, 2001 to August 30, 2013.  An almost full panel of maturities from one month to thirty years is available for these periods. Since the thirty year maturity time series exhibits a lot of missing values this maturity has been excluded. Thus, we have $M=10$ maturities such that ${\bm \tau}=\{1/12,1/4,1/2,1,2,3,5,7,10,20 \}$ and $T=631$ observations per yield. Although the H-15 data set can only be seen as a proxy for the risk-free term structure, we follow the related literature \citep[see, e.g.,][]{ChibErgashev2008} and work with this dataset.

In contrast to the analysis in Section~\ref{sect:mc1}, where $\mathtt{L}$ draws from the data generating process were considered, this section investigates one panel of interest rate data.
The purpose of running the $GMM$ estimation procedure $\mathtt{L}-$times with the same data, is to check for the stability of our estimation routine in the empirical data.%
\footnote{For the mulitstart random search, the vector of parameters presented in the second row of Table~\ref{tab:res4} is used.}
 By doing this, we observe that in all simulation runs, $\ell = 1,\dots,\mathtt{L}=5$,  the intervals $\left[ \widehat{\vvartheta}_{\ell} \right]_{\iota} \pm \hat{\mathbb{V}}_{BM} \left( \left[ \widehat{\vvartheta}_{\ell} \right]_{\iota \iota}
 \right)^{0.5}$, $\iota = 1,\dots,\mathfrak{p}$, overlap. Without the Quasi-Bayesian algorithm, this stability result would not have been attained. In addition, in all simulation runs the p-values for the test $\theta^Q = \theta^P$ against the two-sided alternative $\theta^Q  \not= \theta^P$ are smaller than $0.05$. Hence, we reject the null hypothesis $\theta^Q = \theta^P$ at the significance level $\alpha_S=$ $0.05$.

To obtain parameter estimates, the draws of the Bayesian sampler ${\vvartheta}^{(\mathtt{m})}$, $\mathtt{m}=5,001, \ldots, 20,000 $, are used from which we obtain the sample mean $\widehat{\vvartheta}$ and the vector of sample standard deviations $
\left( \left[ \hat{\mathbb{V}}_{BM} \left(  \widehat{\vvartheta} \right) \right] _{{11}}^{0.5}, \ldots, \left[ \hat{\mathbb{V}}_{BM} \left(  \widehat{\vvartheta} \right) \right]^{0.5}_{{ \mathfrak{p} \mathfrak{p}}} \right)^{\prime}$, where again the batch mean estimator \citet{FlegalJones2010}[Equation~6] is applied. In contrast to Tables~\ref{tab:res4} and \ref{tab:res5}, where the descriptive statistics based on the various point estimates $\widehat{\vvartheta}_{\ell}$ are presented, we now obtain the
$\widehat{median}_{\vvartheta}$, sample minimum, $\widehat{min}_{\vvartheta}$, sample maximum, $\widehat{max}_{\vvartheta}$, sample standard deviation, $\widehat{std}_{\vvartheta}$,
 sample skewness, $\widehat{skew}_{\vvartheta}$ and sample kurtosis, $\widehat{kurt}_{\vvartheta}$ from the draws of one particular chain ${\vvartheta}^{(\mathtt{m})}$,  $\mathtt{m}=5,001, \ldots, 20,000$.
These descriptive statistics are presented in Table~\ref{tab:res9}.

Following mathematical finance literature
\citep[see, e.g.,][]{cheriditofilipovickimmel03,Cochranebook2005}, a usual way to investigate how the market demands for a compensation ({\em risk premium}) for the risk generated by $\mathbf{W}^P(t)$, is to consider the
market price of risk process $ \left( \vphi ( \mathbf{X}(t)) \right)_{t \geq 0}$ described in (\ref{eq:sde7}). This process depends on the model parameters
$\vvartheta$. If $\mathbf{b}^P= \mathbf{b}^Q$  and $\vbeta^P =\vbeta^Q$, then $\vphi ( \mathbf{X}(t)) = \mathbf{0}_d$.
In terms of the parametrization used in this article, $\vphi ( \mathbf{X}(t)) = \mathbf{0}_d$ if $\theta^P=\theta^Q$ and
$\vbeta^P =\vbeta^Q$, while if $\theta^P \not= \theta^Q$  or $\vbeta^P \not= \vbeta^Q$, then $\vphi ( \mathbf{X}(t)) \not= \mathbf{0}_d$ (almost surely).
 In the following we test whether this is the case.

By considering the estimates $\hat{\theta}^Q=12.0667$ and $\hat{\theta}^P=0.0682$ and their estimated standard deviations
$\hat{\mathbb{V}}_{BM} \left(\hat{\theta}^Q \right)^{0.5} = 2.0573$ and $\hat{\mathbb{V}}_{BM} \left(\hat{\theta}^P \right)^{0.5} = 0.0728$, respectively, we observe that the difference in the parameter estimates is relatively large, compared to their estimated standard deviations. We obtained the Wald statistic $\mathscr{W} =4.32546$ with p-value being $0.03056$. Based on this, the null hypothesis $\theta^Q = \theta^P$ is rejected at the significance level
$\alpha_S = 0.05$ for this empirical data set.

Next, we perform the test $\vbeta^P$ $=$ $\vbeta^Q$ against the alternative $\vbeta^P$ $\not=$ $\vbeta^Q$, where $\vbeta^{\cdot}$ contains seven parameters.
In more details, we test the null hypothesis $\left( \beta^Q_{11},\beta^Q_{21},\beta^Q_{31},\beta^Q_{22},\beta^Q_{32},\beta^Q_{23},\beta^Q_{33}\right)$
$=$ $\left( \beta^P_{11},\beta^P_{21},\beta^P_{31},\beta^P_{22},\beta^P_{32},\beta^P_{23},\beta^P_{33}\right)$ against the two sided alternative
$\left( \beta^Q_{11},\beta^Q_{21},\beta^Q_{31},\beta^Q_{22},\beta^Q_{32},\beta^Q_{23},\beta^Q_{33}\right)$ $\not=$ $\left( \beta^P_{11},\beta^P_{21},\beta^P_{31},\beta^P_{22},\beta^P_{32},\beta^P_{23},\beta^P_{33}\right)$.
By estimating $\left( \beta^Q_{11},\beta^Q_{21},\beta^Q_{31},\beta^Q_{22},\beta^Q_{32},\beta^Q_{23},\beta^Q_{33}\right)^{\prime}
-\left( \beta^P_{11},\beta^P_{21}, \beta^P_{31}, \beta^P_{22}, \beta^P_{32}, \beta^P_{23},\beta^P_{33}\right)^{\prime}$ and its covariance matrix from Monte Carlo output, we obtain the Wald statistic
$\mathscr{W} = 38.7047$ with a corresponding p-value of $2.223 $ E-6. That is, also the null hypothesis
$\left( \beta^Q_{11},\beta^Q_{21},\beta^Q_{31},\beta^Q_{22},\beta^Q_{32},\beta^Q_{23},\beta^Q_{33}\right)$ $=$ $\left( \beta^P_{11},\beta^P_{21},\beta^P_{31},\beta^P_{22},\beta^P_{32},\beta^P_{23},\beta^P_{33}\right)$ is rejected on significance levels $\alpha_S \geq 0.01$. Summing up, since the null hypothesis
$\theta^P = \theta^Q$ and $\vbeta^P = \vbeta^Q$ are rejected, the market price of risk process is significantly different from zero.

\begin{table}[h]
\begin{center}
\begin{tabular}{c|rrrrrrr}%
\hline \hline \\[-9pt]
$\vvartheta$ & \multicolumn{1}{|c}{$\widehat{\vvartheta}$} &  $\widehat{median}_{\vvartheta}$ & $\widehat{min}_{\vvartheta}$ & $\widehat{max}_{\vvartheta}$ & $\widehat{std}_{\vvartheta}$ & $\widehat{skew}_{\vvartheta}$& $\widehat{kurt}_{\vvartheta}$ \\
\hline
$\theta^Q$	&	12.0667	&	11.6886	&	6.1634	&	15.2185	&	2.0573	&	-0.5395	&	3.0115	\\
$\theta^P$	&	0.0682	&	0.0460	&	0.0174	&	0.3855	&	0.0728	&	2.7690	&	10.3944	\\
$\beta^Q_{11}$	&	-0.1036	&	-0.1005	&	-0.1754	&	-0.1000	&	0.0108	&	-3.8346	&	17.1485	\\
$\beta^Q_{21}$	&	0.0331	&	0.0245	&	0.0128	&	0.1169	&	0.0245	&	1.8071	&	5.5307	\\
$\beta^Q_{31}$	&	0.0108	&	0.0088	&	0.0060	&	0.0283	&	0.0052	&	1.6536	&	4.6572	\\
$\beta^Q_{22}$	&	-1.4100	&	-1.3316	&	-2.3193	&	-0.8051	&	0.5064	&	-0.3689	&	1.6129	\\
$\beta^Q_{23}$	&	0.0925	&	0.0915	&	0.0782	&	0.1203	&	0.0073	&	0.5622	&	3.3120	\\
$\beta^Q_{32}$	&	-0.0096	&	-0.0092	&	-0.0143	&	-0.0060	&	0.0020	&	-0.4681	&	2.4442	\\
$\beta^Q_{33}$	&	-0.8124	&	-0.7957	&	-1.1401	&	-0.7108	&	0.0720	&	-2.4422	&	9.4258	\\
$\beta^P_{11}$	&	-0.7390	&	-0.5119	&	-2.1933	&	-0.1430	&	0.5474	&	-0.8072	&	2.3157	\\
$\beta^P_{21}$	&	0.0542	&	0.0475	&	0.0191	&	0.1164	&	0.0231	&	0.6122	&	2.4151	\\
$\beta^P_{31}$	&	0.0196	&	0.0196	&	0.0083	&	0.0379	&	0.0053	&	0.2360	&	2.9440	\\
$\beta^P_{22}$	&	-2.9191	&	-2.9900	&	-5.5775	&	-1.1761	&	1.0701	&	-0.2561	&	2.1254	\\
$\beta^P_{23}$	&	0.0047	&	0.0049	&	0.0017	&	0.0088	&	0.0019	&	0.0118	&	1.4428	\\
$\beta^P_{32}$	&	-0.0019	&	-0.0020	&	-0.0030	&	-0.0010	&	0.0005	&	0.2345	&	2.0680	\\
$\beta^P_{33}$	&	-0.4352	&	-0.4247	&	-0.8304	&	-0.3137	&	0.0704	&	-2.0444	&	9.6661	\\
$\mathcal{B}^{x}_{12}$	&	0.0324	&	0.0295	&	0.0155	&	0.0570	&	0.0101	&	0.4526	&	2.0998	\\
$\mathcal{B}^{x}_{13}$	&	0.0871	&	0.0842	&	0.0508	&	0.1412	&	0.0215	&	0.4104	&	2.1585	\\
$\gamma_0$	&	1.8155	&	1.8188	&	1.5172	&	1.9797	&	0.0662	&	-0.7029	&	4.5135	\\
$\Sigma_{1}$	&	0.2004	&	0.2002	&	0.2000	&	0.2024	&	0.0004	&	2.2614	&	8.7845	\\
$\Sigma_{2}$	&	1.1715	&	1.1476	&	0.7876	&	1.4995	&	0.2554	&	0.0554	&	1.4457	\\
$\Sigma_{3}$	&	1.4936	&	1.4957	&	1.4258	&	1.4999	&	0.0074	&	-3.3418	&	21.3889	\\
$\sigma_{\varepsilon}^2$	&	0.0119	&	0.0119	&	0.0106	&	0.0137	&	0.0006	&	-0.0545	&	3.4937	\\
\hline																	
\hline												
\end{tabular}
\caption{Parameter estimates for empirical H-15 interest rate data for the  $\mathbb{A}_{1}(3)$ model.
%based on Quasi-Bayesian methods for .
Statistics are obtained from $\mathtt{M}=20,000$ draws with $\mathtt{M}_b=5,000$ burn-in steps. $\widehat{\vvartheta}$ stands for sample mean, $\widehat{median}_{\vvartheta}$ for sample median, $\widehat{min}_{\vvartheta}$ for sample minimum, $\widehat{max}_{\vvartheta}$ for sample maximum, $\widehat{std}_{\vvartheta}$ for sample standard deviation, $\widehat{skew}_{\vvartheta}$ for sample skewness and $\widehat{kurt}_{\vvartheta}$ for sample kurtosis obtained from the draws of the chain  $\left( {\vvartheta}^{(\mathtt{m})}: \ \mathtt{m}=\mathtt{M}_b+1, \ldots, \mathtt{M} \right)$.}
\label{tab:res9}
\end{center}
\end{table}

\section{Conclusions}
\label{sect:conc}

In this article we developed a new method allowing for parameter estimation based on the exact moments of the yields for affine term structure models.
By applying the results of \citet{cuchieroetal08} on ${p-}$polynomial processes the conditional moments are derived. By assuming a stationary process, we obtain the exact moments of the yields as well as the first order auto-covariance of the yields and the squared yields.
By means of these moments, the model parameters can be estimated by the generalized method of moments.

Since the number of parameters is relatively large and the moments are non-linear in the model parameters, the implementation of the generalized method of moments becomes a non-trivial problem. We observe that standard minimization  routines perform poorly. To cope with this problem, we use random search methods combined with Quasi-Bayesian methods to minimize the $GMM$ distance function as proposed in \citet{chernozhukovhong03}. By these techniques parameter estimation becomes more stable. The standard deviations as well as the dispersions of the point estimates decrease for most parameters, compared to parameter estimation based on a standard minimization of the $GMM$ distance function. For some parameters this decline is substantial.

Another main contribution of this article is a rigorous investigation of the testing problem, whether parameters controlling for the mean of the latent affine process in the empirical and in the equivalent martingale measure are different. We observe substantial undersizing, when implementing a Wald test based on standard estimates of the covariance matrix of the unknown parameter. By applying methods developed by \citet{chernozhukovhong03}, the standard errors of the corresponding components of the parameter vector can be obtained from the draws provided by a Bayesian sampler. We observe that in this case the rejection rates of the true null hypothesis are close to theoretically correct levels.

In a final step, our estimation methodology is applied to empirical term structure data. By applying the testing procedure developed in this article, the null hypothesis of equal parameters controlling for the mean of the latent affine process, in the empirical as well as in the equivalent martingale measure, is rejected. Our estimates support a significant market price of risk.

\clearpage

\appendix

\section{Affine Models}
\label{sect:affine1appa}
The following paragraphs - based on \citet{filipovicbook2009} -  describe {\em affine processes}. Let us assume the following: The state space is given by $\mathscr{S} \subset \mathbb{R}^d$, $\mathbf{W}(t)$ stands for $d-$dimensional standard Brownian motion on a filtered probability space $\left(\Omega,\mathcal{F},(\mathcal{F}_t)_{t \geq 0 }, \mathbb{Q}  \right)$ and for any initial value $\mathbf{X}(0)=\mathbf{x}$, $\mathbf{x} \in \mathscr{S} $, there exists a unique solution $(\mathbf{X}(t))$ for the  stochastic differential equation
\begin{equation}
\label{eq:sde1}
d \mathbf{X}(t) = \tilde \vbeta^Q(\mathbf{X}(t)) dt + \vrho(\mathbf{X}(t)) d \mathbf{W}(t), \  {\text{ where  $\tilde \vbeta^Q(\mathbf{x}) \in \mathbb{R}^d$ and $\vrho(\mathbf{x}) \in \mathbb{R}^{d \times d }$.}}
\end{equation}

\noindent An affine stochastic process is defined as follows:
\begin{definition}[Affine Process] \label{def1}{\it
Consider $\mathbf{X}(t) \in \mathbb{R}^d$. $\left(\mathbf{X}(t) \right)_{t \geq 0}$ described by the stochastic differential equation
(\ref{eq:sde1}) is called \textit{affine} stochastic process if the $\mathcal{F}_s$ conditional characteristic function of $\mathbf{X}(t)$ is exponentially affine in $\mathbf{X}(s)$, $0\leq s \leq t$. Thus, there exist functions $\Phi(t,\mathbf{u}) \in \mathbb{C}$ and $\vPsi(t,\mathbf{u}) \in \mathbb{C}^d$, with jointly continuous $t$-derivatives, such that
\begin{eqnarray}
\mathbb{E} \left( \exp(\mathbf{u}^{\prime} \mathbf{X}(t)) |\mathcal{F}_s  \right) = \exp\left(\Phi(t-s,\mathbf{u})+\vPsi(t-s,\mathbf{u})^{\prime} \mathbf{X}(s)\right) \label{charact}
\end{eqnarray}
for all $\mathbf{u} \in \imath \mathbb{R}^d$ and $s \le t$. }
\end{definition}
As the conditional characteristic function is bounded by one, the real part of the exponent $\Phi(t-s,\mathbf{u})+\vPsi(t-s,\mathbf{u})^{\prime} \mathbf{X}(s)$ is negative. The functions $\Phi(t,\mathbf{u})$ and $\vPsi(t,\mathbf{u})$ are uniquely determined by (\ref{charact}) for $t \ge 0$ and $\mathbf{u} \in \imath \mathbb{R}^d$ and satisfy the initial conditions $\Phi(0,\mathbf{u})=0$ and $\vPsi(0,\mathbf{u})=\mathbf{u}$.

If $\left( \mathbf{X}(t) \right)_{t \in \mathbb{R}_{+}}$ is affine, then the \textit{drift term} $\tilde \vbeta^Q(\mathbf{X}(t))$ and the (positive definite) \textit{diffusion matrix} $ \mathbf{a}(\mathbf{X}(t))=\vrho(\mathbf{X}(t))\vrho(\mathbf{X}(t))^{\prime}$  are affine functions in $\mathbf{X}(t)$ (see \citet{filipovicbook2009}[Definition 10.1 and Theorem 10.1]); i.e., $\tilde \vbeta^Q (\mathbf{x})=\mathbf{b}^Q + \sum_{i=1}^d x_i \vbeta_i^Q$ and $ \mathbf{a}(\mathbf{x})= \mathbf{a}+ \sum_{i=1}^d x_i \valpha_i$ where $\mathbf{b}^Q$, $\vbeta_i^Q$ and $\mathbf{x}$ are vectors of dimension $d$ and $\mathbf{a}(\mathbf{x})$, $ \mathbf{a}$ and $\valpha_i$ are $d \times d $ matrices. $\vbeta^Q =(\vbeta_1^Q,\dots,\vbeta_d^Q)$ is a $d \times d $ matrix. In addition, $\Phi(t,\mathbf{u}) $ and $\vPsi(t,\mathbf{u}) $ solve the following system of Riccati equations; see \citet{filipovicbook2009}[Eq.~10.4]\footnote{Extensions with jumps are possible - for some theory see \citet{KellerMayerhofer2011}, \citet{mayerhoferetal10}, \citet{duffiepansingleton00}, \citet{duffiefilipovicschachermayer03}.}
\begin{eqnarray}
\label{transformeddk1}
\partial_t \Phi(t,\mathbf{u}) &=& \frac{1}{2} \vPsi(t,\mathbf{u})^{\prime} \mathbf{a} \vPsi(t,\mathbf{u}) + (\mathbf{b}^Q)^{\prime} \vPsi(t,\mathbf{u}),  \qquad \Phi(0,\mathbf{u})=0 \ , \nonumber \\
\partial_t \vPsi_i(t,\mathbf{u}) &=& \frac{1}{2} \vPsi(t,\mathbf{u})^{\prime} \valpha_i \vPsi(t,\mathbf{u}) + (\vbeta_i^Q)^{\prime} \vPsi(t,\mathbf{u}), \ \ \ \ \vPsi(0,\mathbf{u})=\mathbf{u}\ ,
\end{eqnarray}
$i=1,\dots,d$ and $\mathbf{u} \in \imath \mathbb{R}^d$.
\section{Matrix $\mathbf{A}$ for $\mathbb{A}_m(3)$ Models}
\label{appa:matrixAmd}

This section derives the matrix $ \mathbf{A}$ for an arbitrary $\mathbb{A}_m(3)$ setting; where $0 \leq m \leq 3$. In the first step we ignore all the restrictions arising from admissibility, the boundary conditions, stationarity and identification, and calculate  $ \mathbf{A}$ for a model with diagonal diffusion, where all elements in $\mathbf{b}^P$, $\vbeta^P$, $\mathbf{\Sigma}$, $\mathcal{B}^{{0}}$ and $\mathcal{B}^{{x}}$ are free parameters. To obtain $ \mathbf{A}$ for a particular $\mathbb{A}_m(3)$ model, the corresponding parameter restrictions have to be taken into consideration. Moreover, restrictions like $\beta_{ij}^Q=\beta_{ij}^P$ for some $ij$ can also be included. This allows a joint treatment of all models.

For the first four moments $\mathbf{x}^k$, $k=1, \ldots, p=4$, we choose the basis $\left(1 \, | \, x_1, x_2, x_3 \, | \, x_1^2, \dots,	x_3^2 \, | \, x_1^3, \dots, x_3^3 \, | \, x_1^4	, \dots, x_3^4 \, \right)$. In this expression we have separated the terms of different power by means of $|$. I.e., with $d=3$, we get one term for $k=0$, three for $k=1$, six for $k
=2$, ten for $k=3$ and fifteen for $k=4$. Therefore $N=35$. The elements of matrix $\mathbf{A}$ not presented are zero by the model assumptions.

\noindent In the following we use (\ref{eq:A1_A13}) and start with $k=0$: Here we immediately observe that the first row of $ \mathbf{A}$ is $ \mathbf{A}_{1,:}=\mathbf{0}_{ 1\times N }$.
With $k=1$ we obtain the rows $2$ to $d+1$ of the matrix $ \mathbf{A}$ as follows: With $f(\mathbf{x})=x_i$ we get
$\frac{\partial x_i}{\partial x_i}=1$, $\frac{\partial x_j}{\partial x_i}=0$ and $\frac{\partial^2 x_i}{\partial x_i^2}=0$. Hence,
$\mathcal{G}(x_i) = b_i^P + \vbeta_i^P \mathbf{x} $, $i=1,\dots,d$. This yields
$$ \mathbf{A}_{2:4,:}=\left(
\begin{array}{c|ccc|cc}
b_1^P & \beta_{11}^P  &\beta_{12}^P & \beta_{13}^P & 0   \dots\\
b_2^P  & \beta_{21}^P & \beta_{22}^P & \beta_{23}^P & 0  \dots\\
b_3^P  & \beta_{31}^P & \beta_{32}^P & \beta_{33}^P & 0  \dots\\
\end{array}
\right)  .
$$

Next, for $k=2$ we have to consider $d(d+1)/2 = d_2$ basis elements, corresponding to rows $d+2$ to $d+1 + \frac{d(d+1)}{2}$ of
$ \mathbf{A}$. We arrange the basis elements as follows $\mathbf{x}^2=\left(x_1^2 \right.$, $x_1 x_2$, $x_1 x_3$, $x_2^2$, $x_2 x_3$, $ \left. x_3^2 \right)$. Since the diffusion matrix is diagonal we have only non-zero elements in the generator for $i=j$.

The first partial derivatives with respect to $x_1$ are $2 x_1$, $x_2$, $x_3$, 0, 0, 0 for these basis elements. The second partial derivatives with respect to $x_1$ are
2, 0 ,0, 0, 0, 0, 0, etc. For $x_2$ and $x_3$ we proceed in the same way.

For example, consider $f(\mathbf{x})=x_1^2$ and thus equation (\ref{eq:A1_A13}) yields
$\mathcal{G}(x_1^2) = \left( b_i^P + \vbeta_i^P \mathbf{x} \right) 2 x_1 +
\frac{1}{2} \sum_{j=1}^{d} \left(   \Sigma^2_{1} \left ( \mathcal{B}_1^0+ \mathcal{B}_{j1}^x x_j \right ) \right) 2$.
For $f(\mathbf{x})=x_1 x_2$, where $\frac{\partial (x_1 x_2)}{\partial x_1}=x_2$, $\frac{\partial (x_1 x_2)}{\partial x_2}=x_1$ and  $\frac{ \partial^2 (x_1 x_2)}{\partial x_1 x_2}=1$,  (\ref{eq:A1_A13}) and the fact that $\mathbf{S}(\mathbf{X}(t))$ and $\mathbf{\Sigma}$ are diagonal matrices,\footnote{Where $S_{ii} (\mathbf{X}(t))= \mathcal{B}_i^0 + \left (\mathcal{B}_i^x \right )^{\prime} \mathbf{X}(t)$ and $S_{ij}(\mathbf{X}(t))=0, \  i,j=1,\dots,d$.}
result in $\mathcal{G}(x_1 x_2) = \left( b_1^P + \vbeta_1^P \mathbf{x} \right) x_2 + \left( b_2^P + \vbeta_2^P \mathbf{x} \right) x_1 +
\frac{1}{2} \left[ \mathbf{\Sigma} \mathbf{S} \right]_{12} \cdot 1 +  \frac{1}{2} \left[ \mathbf{\Sigma} \mathbf{S} \right]_{21} \cdot 1 $.
%$=\left( b_1^P + \vbeta_1^P \mathbf{x} \right) x_2 + \left( b_2^P + \vbeta_2^P \mathbf{x} \right) x_1$.
With $x_1x_3,\dots,x_3^2$ we proceed in the same way. This results in
{\tiny
\begin{eqnarray}
&&\mathbf{A}_{5:10,1:10} \nonumber \\
&& =\left(
\begin{array}{c|ccc|cccccc|cc}
\Sigma_{1}^2 \mathcal{B}_1^0 & 2 b_{1}^P  + \Sigma_{1}^2\mathcal{B}_{11}^x & \Sigma_{1}^2\mathcal{B}_{21}^x & \Sigma_{1}^2\mathcal{B}_{31}^x & 2 \beta_{11}^P & 2 \beta_{12}^P & 2 \beta_{13}^P&0&0& 0& 0&\dots\\
0  & b_{2}^P & b_{1}^P & 0 & \beta_{21}^P & \beta_{11}^P+\beta_{22}^P &  \beta_{23}^P &\beta_{12}^P& \beta_{13}^P&0 &0&\dots\\
0  & b_{3}^P& 0 & b_{1}^P & \beta_{31}^P &  \beta_{32}^P & \beta_{11}^P+\beta_{33}^P &0& \beta_{12}^P& \beta_{13}^P&0&\dots\\
\Sigma_{2}^2 \mathcal{B}_2^0 & \Sigma_{2}^2  \mathcal{B}_{12}^x & 2 b_{2}^P+\Sigma_{2}^2  \mathcal{B}_{22}^x & \Sigma_{2}^2  \mathcal{B}_{32}^x & 0 & 2 \beta_{21}^P & 0& 2 \beta_{22}^P& 2 \beta_{23}^P& 0& 0&\dots\\
0  & 0 & b_{3}^P &  b_{2}^P &  0  &  \beta_{31}^P &\beta_{21}^P & \beta_{32}^P & \beta_{22}^P+\beta_{33}^P& \beta_{23}^P&0&\dots\\
\Sigma_{3}^2 \mathcal{B}_3^0 &  \Sigma_{3}^2  \mathcal{B}_{13}^x &  \Sigma_{3}^2  \mathcal{B}_{23}^x &2 b_{3}^P +\Sigma_{3}^2  \mathcal{B}_{33}^x& 0 & 0 & 2 \beta_{31}^P & 0 &  2 \beta_{32}^P& 2 \beta_{33}^P& 0&\dots\\
\end{array}
\right)  . \nonumber
\end{eqnarray}
}
For $k=3$ we obtain ${5 \choose 3}=10= d_3$  elements. Therefore we consider the rows $11$ to $20$. The basis elements are
$\mathbf{x}^3=\left(x_1^3 \right. $, $x_1^2 x_2 $, $x_1^2 x_3$, $x_1 x_2^2$, $x_1 x_2 x_3$, $x_1 x_3^2$, $x_2^3$, $x_2^2 x_3$, $x_2 x_3^2$, $\left. x_3^3 \right)$. Then,
{\tiny
\begin{eqnarray}
&& \mathbf{A}_{11:20,1:10}
=  \nonumber \\
&&\left(
\begin{array}{c|ccc|cccccc}
0	&	3 \Sigma_{1}^2 \mathcal{B}_1^0	&	0	&	0	&	3 b_1^P + 3 \Sigma_{1}^2 \mathcal{B}_{11}^x	&	3 \Sigma_{1}^2 \mathcal{B}_{21}^x	&	3 \Sigma_{1}^2 \mathcal{B}_{31}^x	&	0	&	0	&	0	\\
0	&	0	&	\Sigma_{1}^2 \mathcal{B}_1^0	&	0	&	b_2^P	&	2 b_1^P + \Sigma_{1}^2 \mathcal{B}_{11}^x	&	0	&	\Sigma_{1}^2 \mathcal{B}_{21}^x	 &	\Sigma_{1}^2 \mathcal{B}_{31}^x	&	0	\\
0	&	0	&	0	&	\Sigma_{1}^2 \mathcal{B}_1^0	&	b_3^P	&	0	&	2 b_1^P + \Sigma_{1}^2	\mathcal{B}_{11}^x &	0	&	\Sigma_{1}^2	 \mathcal{B}_{21}^x	&	\Sigma_{1}^2	\mathcal{B}_{31}^x	\\
0	&	\Sigma_{2}^2 \mathcal{B}_2^0	&	0	&	0	&	\Sigma_{2}^2 \mathcal{B}_{12}^x	&	2 b_2^P+ \Sigma_{2}^2 \mathcal{B}_{22}^x	&	\Sigma_{2}^2 \mathcal{B}_{32}^x	&	b_1^P	 &	 0	&	0	\\
0	&	0	&	0	&	0	&	0	&	b_3^P	&	b_2^P	&	0	&	b_1^P	&	0	\\
0	&	\Sigma_{3}^2 \mathcal{B}_3^0	&	0	&	0	&	\Sigma_{3}^2 \mathcal{B}_{13}^x	&	\Sigma_{3}^2 \mathcal{B}_{23}^x	&	2 b_3^P	+ \Sigma_{3}^2 \mathcal{B}_{33}^x &	0	&	0	&	b_1^P	\\
0	&	0	&	3 \Sigma_{2}^2 \mathcal{B}_2^0	&	0	&	0	&	3 \Sigma_{2}^2 \mathcal{B}_{12}^x	&	0 &	3 b_2^P	+ 3 \Sigma_{2}^2 \mathcal{B}_{22}^x &	 3 \Sigma_{2}^2 \mathcal{B}_{32}^x	&	0	\\
0	&	0	&	0	&	\Sigma_{2}^2 \mathcal{B}_2^0	&	0	&	0	&	\Sigma_{2}^2 \mathcal{B}_{12}^x	&	b_3^P	&	2 b_2^P	+ \Sigma_{2}^2 \mathcal{B}_{22}^x &	\Sigma_{2}^2 \mathcal{B}_{32}^x	\\
0	&	0	&	\Sigma_{3}^2 \mathcal{B}_3^0	&	0	&	0	&	\Sigma_{3}^2 \mathcal{B}_{13}^x	&	0	&	\Sigma_{3}^2 \mathcal{B}_{23}^x	&	2 b_3^P+\Sigma_{3}^2 \mathcal{B}_{33}^x	&	b_2^P	\\
0	&	0	&	0	&	3 \Sigma_{3}^2 \mathcal{B}_3^0	&	0	&	0	&	3 \Sigma_{3}^2 \mathcal{B}_{13}^x	&	0	&	3 \Sigma_{3}^2 \mathcal{B}_{23}^x	 &	3 b_3^P+3 \Sigma_{3}^2 \mathcal{B}_{33}^x	\\
\end{array}
\right)  . \nonumber
\end{eqnarray}
}
\noindent and
{\tiny
\begin{eqnarray}
&& \mathbf{A}_{11:20,11:20}
= \nonumber \\
&&\left(
\begin{array}{cccccccccc|ccc}
3 \beta_{11}^P	&	3 \beta_{12}^P	&	3 \beta_{13}^P	&	0	&	0	&	0	&	0	&	0	&	0	&	0	&	0	&	\dots	\\
\beta_{21}^P	&	2 \beta_{11}^P + \beta_{22}^P	&	\beta_{23}^P 	&	2 \beta_{12}^P	&	2 \beta_{13}^P	&	0 	&	0	&	0	&	0	&	0	&	0	 &	\dots	\\
\beta_{31}^P	&	\beta_{32}^P	&	2 \beta_{11}^P+\beta_{33}^P	&	0	&	2 \beta_{12}	&	2 \beta_{13}	&	0	&	0	&	0	&	0	&	0	&	 \dots	\\
0	&	2 \beta_{21}^P	&	0	&	\beta_{11}^P+2\beta_{22}^P	&	2 \beta_{23}^P	&	0	&	\beta_{12}^P	&	\beta_{13}^P	&	0	&	0	&	0	&	 \dots	\\
0	&	\beta_{31}^P	&	\beta_{21}^P	&	\beta_{32}^P	&	\beta_{11}^P+\beta_{22}^P+\beta_{33}^P	&	\beta_{23}^P	&	0	&	\beta_{12}^P	&	 \beta_{13}^P	&	0	 &	0	&	\dots	\\
0	&	0	&	2 \beta_{31}^P	&	0	&	2 \beta_{32}^P	&	\beta_{11}^P+2\beta_{33}^P	&	0	&	0	&	\beta_{12}^P	&	\beta_{13}^P	&	0	&	 \dots	\\
0	&	0	&	0	&	3 \beta_{21}^P	&	0	&	0	&	3 \beta_{22}^P	&	3 \beta_{23}^P	&	0	&	0	&	0	&	\dots	\\
0	&	0	&	0	&	\beta_{31}^P	&	2 \beta_{21}^P	&	0	&	\beta_{32}^P	&	2 \beta_{22}^P + \beta_{33}^P	&	2 \beta_{23}^P	&	0	&	0	 &	\dots	\\
0	&	0	&	0	&	0	&	2 \beta_{31}^P	&	\beta_{21}^P	&	0	&	2\beta_{32}^P	&	2 \beta_{33}^P + \beta_{22}^P	&	\beta_{23}^P	&	0	 &	\dots	\\
0	&	0	&	0	&	0	&	0	&	3 \beta_{31}^P	&	0	&	0	&	3 \beta_{32}^P	&	3 \beta_{33}^P	&	0	&	\dots	\\
\end{array}
\right)  . \nonumber
\end{eqnarray}
}
\noindent
Last but not least, with $k=4$ we have $d_4 = 15$ basis elements $\\\mathbf{x}^4=\left(x_1^4 \right.$,	$x_1^3 x_2$, $x_1^3 x_3$, $x_1^2 x_2^2$, $x_1^2 x_2 x_3$, $x_1^2 x_3^2$, $x_1 x_2^3$, $x_1 x_2^2 x_3$, $x_1 x_2 x_3^2$, $x_1 x_3^3$, $x_2^4$, $x_2^3 x_3$, $x_2^2 x_3^2$, $x_2 x_3^3$, $\left. x_3^4\right)$.
Then we obtain:

\begin{eqnarray}
&& \mathbf{A}_{21:35,1:10}
=  \nonumber \\
&&\left(
\begin{array}{c|ccc|cccccc}
0	&	0	&	0	&	0	&	6	\Sigma_{1}^2 \mathcal{B}_1^0 &	0	&	0	&	0	&	0	&	0	\\
0	&	0	&	0	&	0	&	0	&	3 \Sigma_{1}^2 \mathcal{B}_1^0	&	0	&	0	&	0	&	0	\\
0	&	0	&	0	&	0	&	0	&	0	&	3 \Sigma_{1}^2 \mathcal{B}_1^0	&	0	&	0	&	0	\\
0	&	0	&	0	&	0	&	\Sigma_{2}^2 \mathcal{B}_2^0&	0	&	0	&	\Sigma_{1}^2 \mathcal{B}_1^0	&	0	&	0	\\
0	&	0	&	0	&	0	&	0	&	0	&	0	&	0	&	\Sigma_{1}^2 \mathcal{B}_1^0	&	0	\\
0	&	0	&	0	&	0	&	\Sigma_{3}^2 \mathcal{B}_3^0	&	0	&	0	&	0	&	0	&	\Sigma_{1}^2 \mathcal{B}_1^0	\\
0	&	0	&	0	&	0	&	0	& 3 \Sigma_{2}^2 \mathcal{B}_2^0	&	0	&	0	&	0	&	0	\\
0	&	0	&	0	&	0	&	0	&	0	&	\Sigma_{2}^2 \mathcal{B}^0_{2}	&	0	&	0	&	0	\\
0	&	0	&	0	&	0	&	0	&	\Sigma_{3}^2 \mathcal{B}^0_{3}	&	0	&	0	&	0	&	0	\\
0	&	0	&	0	&	0	&	0	&	0	&	3 \Sigma_{3}^2 \mathcal{B}_3^0	&	0	&	0	&	0	\\
0	&	0	&	0	&	0	&	0	&	0	&	0	&	6 \Sigma_{2}^2 \mathcal{B}_2^0	&	0	&	0	\\
0	&	0	&	0	&	0	&	0	&	0	&	0	&	0	&	3 \Sigma_{2}^2 \mathcal{B}_2^0	&	0	\\
0	&	0	&	0	&	0	&	0	&	0	&	0	&	\Sigma_{3}^2 \mathcal{B}_3^0	&	0	&	\Sigma_{2}^2 \mathcal{B}_2^0	\\
0	&	0	&	0	&	0	&	0	&	0	&	0	&	0	&	3\Sigma_{3}^2 \mathcal{B}_3^0	&	0	\\
0	&	0	&	0	&	0	&	0	&	0	&	0	&	0	&	0	&	6 \Sigma_{3}^2 \mathcal{B}_3^0	\\
\end{array}
\right)
\end{eqnarray}

\begin{landscape}
\begin{tiny}{
\begin{eqnarray}
&& \mathbf{A}_{21:35,11:20}
= \nonumber \\
&&\left(
\begin{array}{ccccccccccc}
%1
 4 b_1^P+6 \Sigma_{1}^2	\mathcal{B}^x_{11}&	6 \Sigma_{1}^2	\mathcal{B}^x_{21}	&	6 \Sigma_{1}^2	\mathcal{B}^x_{31}	&	0	&	0	&	0	&	0	&	0	 &	0	&	0	\\
%2
b_2^P	&	3 b_1^P + 3 \Sigma_{1}^2 \mathcal{B}^x_{11} 	&	0	&	3 \Sigma_{1}^2 \mathcal{B}^x_{21}	&	3 \Sigma_{1}^2 \mathcal{B}^x_{31}	&	0	&	 0	&	0	&	0	&	0	\\
%3
b_3^P	&	0	&	3 b_1^P + 3\Sigma_{1}^2 \mathcal{B}^x_{11}	&	0	&	3 \Sigma_{1}^2 \mathcal{B}^x_{21}	&	3 \Sigma_{1}^2 \mathcal{B}^x_{31}	&	0	 &	0	&	0	&	0	 \\
%4
\Sigma_{2}^2 \mathcal{B}_{12}^x	&	2 b_2^P	+ \Sigma_{2}^2 \mathcal{B}_{22}^x  & \Sigma_{2}^2 \mathcal{B}_{32}^x  	&	2 b_1^P +\Sigma_{1}^2 \mathcal{B}_{11}^x	 &	0	&	0	&	\Sigma_{1}^2 \mathcal{B}_{21}^x	 &	 \Sigma_{1}^2 \mathcal{B}_{31}^x	&	0	&	0	\\
%5
0	&	b_3^P	&	b_2^P	&	0	&	2 b_1^P + \Sigma_{1}^2 \mathcal{B}_{11}^x	&	0	&	0	&	\Sigma_{1}^2 \mathcal{B}_{21}^x	&	\Sigma_{1}^2 \mathcal{B}_{31}^x	&	0	\\
%6
\Sigma_{3}^2 \mathcal{B}_{13}^x	&	\Sigma_{3}^2 \mathcal{B}_{23}^x	&	2 b_3^P+\Sigma_{3}^2 \mathcal{B}_{33}^x	&	0	&	0	&	2 b_1^P + \Sigma_{1}^2 \mathcal{B}_{11}^x	&	0	&	0	&	 \Sigma_{1}^2 \mathcal{B}_{21}^x	 &	\Sigma_{1}^2 \mathcal{B}_{31}^x	\\
%7
0	& 3 \Sigma_{2}^2 \mathcal{B}_{12}^x	&	0	&	3 b_2^P+3 \Sigma_{2}^2 \mathcal{B}_{22}^x	&	3 \Sigma_{2}^2 \mathcal{B}_{32}^x& 0		&	b_1^P	&	 0	&	0	&	0	\\
%8
0	&	0	&	\Sigma_{2}^2 \mathcal{B}^x_{12}	&	b_3^P &	2 b_2^P	+\Sigma_{2}^2 \mathcal{B}^x_{22}	&	\Sigma_{2}^2 \mathcal{B}^x_{32} &	0	&	b_1^P	 &	0	&	0	\\
%9
0	&	\Sigma_{3}^2 \mathcal{B}^x_{13}	&	0	&	\Sigma_{3}^2 \mathcal{B}^x_{23}	&	2 b_3^P +\Sigma_{3}^2 \mathcal{B}^x_{33}	&	b_2^P 	&	0	&	0	 &	b_1^P	&	0	\\
%10
0	&	0	&	3 \Sigma_{3}^2 \mathcal{B}_{13}^x	&	0	&	3\Sigma_{3}^2 \mathcal{B}_{23}^x	&	 3 b_3^P + 3 \Sigma_{3}^2 \mathcal{B}_{33}^x 	&	0	 &	0	&	0	&	b_1^P	 \\
%11
0	&	0	&	0	&	6 \Sigma_{2}^2 \mathcal{B}^x_{12}	&	0	&	0	&	4 b_2^P + 6 \Sigma_{2}^2 \mathcal{B}^x_{22}	&	6 \Sigma_{2}^2 \mathcal{B}^x_{32}	&	0	&	0	\\
%12
0	&	0	&	0	&	0	&	3 \Sigma_{2}^2 \mathcal{B}_{12}^x	&	0	&	b_3^P	&	3 b_2^P + 3 \Sigma_{2}^2 \mathcal{B}_{22}^x	&	3 \Sigma_{2}^2 \mathcal{B}_{32}^x	&	0	\\
%13
0	&	0	&	0	&	\Sigma_{3}^2 \mathcal{B}_{13}^x	&	0	&	\Sigma_{2}^2 \mathcal{B}_{12}^x	&	\Sigma_{3}^2 \mathcal{B}_{23}^x	&	2 b_3^P	 +\Sigma_{3}^2 \mathcal{B}_{33}^x &	 2 b_2^P+\Sigma_{2}^2 \mathcal{B}_{22}^x	&	 \Sigma_{2}^2 \mathcal{B}_{32}^x	\\
%14
0	&	0	&	0	&	0	&	3 \Sigma_{3}^2 \mathcal{B}_{13}^x	&	0	&	0	&	3 \Sigma_{3}^2 \mathcal{B}_{23}^x	&	3 b_3^P	+3 \Sigma_{3}^2 \mathcal{B}_{33}^x &	b_2^P	\\
%15
0	&	0	&	0	&	0	&	0	&	6 \Sigma_{3}^2 \mathcal{B}^x_{13}	&	0	&	0	&	6 \Sigma_{3}^2 \mathcal{B}^x_{23}	&	4 b_3^P	 + 6 \Sigma_{3}^2 \mathcal{B}^x_{33} \\
\end{array}
\right)  . \nonumber
\end{eqnarray}
}
\end{tiny}
\end{landscape}

\begin{landscape}

\tiny{
\begin{eqnarray}
&&\mathbf{A}_{21:35,21:35} = \nonumber \\[2pt]
&&  \left(
\begin{array}{ccccccccccccccc}
4 \beta_{11}	& 4 \beta_{12} &			4 \beta_{13}	&	0	&	0	&	0	&	0	&	0	&	0	&	0	&	0	&	0	&	0	&	0	&	0		 \\
\beta_{21}	&	3 \beta_{11}	&	\beta_{23}	&	3\beta_{12}	&	3 \beta_{13}	&	0	&	0	&	0	&	0	&	0	&	0	&	0	&	0	&	 0 &0			 \\
          	&	 +\beta_{22}	&		&		&		&		&		&		&		&		&		&		&		&	  &			\\
\beta_{31}	&	\beta_{32}	&	3 \beta_{11}	&	0	&	3 \beta_{12}	&	 3 \beta_{13} 	&	0	&	0	&	0	&	0	&	0	&	0	&	0	 &	0	 &	0			 \\
          	&	            &	+\beta_{33}	&	 	&		&	  	&		&		&		&		&		&		&		 &		&				 \\
0	&	2 \beta_{21}	&	0	&	2 \beta_{11}	&	2 \beta_{23}	&	0	&	2 \beta_{12}	&	2 \beta_{13}	&	0	&	0	&	0	&	 0	&	0	 &	0	&	 0			 \\
	&	            	&	 	&	2 + 2\beta_{22}	&		&		&		&		&		&		&		&	 	&		&		&	 			 \\
0	&	\beta_{31}	&	\beta_{21}	&	\beta_{32}	&	2 \beta_{11}	          &	\beta_{23}	&	0	&	2 \beta_{12}	&	2 \beta_{13}	&	 0	&	0	 &	0	&	 0	&	0	&	 0		\\
	&	 	&	       	&	                         &	+\beta_{22}+\beta_{33}	&		&		&		&		&	 	&		&		&	 	&		&	 		 \\
	0	&	0	&	2 \beta_{31}	&	0	&	2 \beta_{32}	&	2 \beta_{11}	&	0	&	0	&	2 \beta_{12}	&	2 \beta_{13}	&	0	&	 0	&	 0	&	0	&	 0		\\
	&		&	              	&		&		&	+ 2\beta_{33}	&		&		&	&	&		&	 	&		&		&	 		\\
0	&	0	&	0	&	3 \beta_{21}	&	0	&	0	&	\beta_{11}    &	3 \beta_{23}	&	0	&	0	&	 \beta_{12}	&	 \beta_{13}	&	0	 &	0	&	 0		 \\
	&		&		&	              &		&		&	+3 \beta_{22}	&	3            	&		&		&	 	&	 	&		 &	&			 \\
0	&	0	&	0	&	\beta_{31}	&	2 \beta_{21}	&	0	&	\beta_{32}	&	 \beta_{11}+2\beta_{22}+\beta_{33}	&	2\beta_{23}	&	0	&	0	&	  \beta_{12}	 &	 \beta_{13}	&	0	&	0	\\
0	&	0	&	0	&	0	&	2 \beta_{31}	&	\beta_{21}	&	0	&	2 \beta_{32}	&	\beta_{11}+\beta_{22}              	&	\beta_{23}	&	0	&	 0	&	 \beta_{12}	&	  \beta_{13}	&	0		 \\
	&	 	&	 	&	 	&	2 \beta_{31}	&	\beta_{21}	&	0	&	2 \beta_{32}	&	                     +2 \beta_{33}	&		&		&	 	&	 	 &	  	&			 \\
0	&	0	&	0	&	0	&	0	&	3 \beta_{31}	&	0	&	0	&	 3\beta_{32}	&	\beta_{11}	&	0	&	0	&	0	&	 \beta_{12}	 &	  \beta_{13}		 \\
	&		&		&		&		&	             	&		&		&	             	&	+3 \beta_{33}	&		&		&		&		 &	  		 \\
0	&	0	&	0	&	0	&	0	&	0	&	4 \beta_{21}	&	0	&	0	&	0	&	4 \beta_{22}	&	4 \beta_{23}	&	0	&	0	&	0	\\
0	&	0	&	0	&	0	&	0	&	0	&	\beta_{31}	&	3 \beta_{21}	&	0	&	0	&	\beta_{32}	&	3 \beta_{22}	&	3 \beta_{23}	 &	0	 &	 0	\\
	&		&		&		&		&		&	           	&	             	&		&		&	          	&	+\beta_{33}	&	 &	&	 	\\
0	&	0	&	0	&	0	&	0	&	0	&	0	&	2 \beta_{31}	&	2 \beta_{21}	&	0	&	0	&	2 \beta_{32}	&	2 \beta_{22}	 &	2 \beta_{23}	&	 0		 \\
	&		&		&		&		&		&		&	            	&	             	&		&		&	              &	+ 2\beta_{33}	 &		&	 		 \\
0	&	0	&	0	&	0	&	0	&	0	&	0	&	0	&	3 \beta_{31}	&	\beta_{21}	&	0	&	0	&	3 \beta_{32}	&	\beta_{22}	 &	 \beta_{23}	\\
	&		&		&		&		&		&		&		&	              &	           	&		&		&	             	&	+3 \beta_{33}	 &		\\
0	&	0	&	0	&	0	&	0	&	0	&	0	&	0	&	0	&	4 \beta_{31}	&	0	&	0	&	0	&	4 \beta_{32}	&	4 \beta_{33}		\\
\end{array}
   \right)  . \nonumber
\end{eqnarray}
 }
\end{landscape}

\section{Moments of the Observed Yields}
\label{appa:momentsyields1}

The following paragraphs obtain the first four moments of the yields observed, i.e. $\mathbb{E}\left(y^k_{{\mathtt{t}}i} \right)$, $k=1,\dots,4$, the auto-covariance of the yields, $\mathbb{E}\left(y_{{\mathtt{t}}i} y_{{\mathtt{t}}-1, i} \right) $, and the auto-covariance of the squared yields, $\mathbb{E}\left(y^2_{{\mathtt{t}}i} y^2_{{\mathtt{t}}-1, i} \right)$.
 Assumption~\ref{ass:noise} specifies the moments $\mathbb{E}\left(\varepsilon_{{\mathtt{t}}i}^k \varepsilon_{{\mathtt{t}} i}^l \right)$.%
{\footnote{Here we derive  the 1st moments with $k=1$, $l=0$, where $\mathbb{E}(\varepsilon_{{\mathtt{t}}i})=0$ for $i=1,\dots,M$. For the 2nd moments: $k=2,$ $l=0$, such that $\mathbb{E}\left(\varepsilon_{{\mathtt{t}}i}^2 \right)=\sigma_{i}^2$ for all $i$; with $k=l=1,$ $i \not = j$ we get  $\mathbb{E}(\varepsilon_{{\mathtt{t}}i} \varepsilon_{{\mathtt{t}}j})=0$, $i \not = j$. For the 3rd moments: $k=3,$ $l=0,$ all $i$, $k=2,$ $l=1,$ $i \not = j$, and $k=1,$ $l=2,$ $i\not = j$. All these terms are zero by assumption, i.e. $\mathbb{E}\left(\varepsilon_{{\mathtt{t}}i}^2 \varepsilon_{tj}\right)=0$, and $\mathbb{E}\left(\varepsilon_{{\mathtt{t}}i} \varepsilon_{{\mathtt{t}}j}^2 \right)=0,$ $i \not = j$. For the 4th moments: $k=4,$ $l=0,$ all $i$, $k=3,$ $l=1,$ $i\not = j$, $k=l=2,$ $i\not = j$, $k=1,$ $l=3,$ $i\not = j$, $\mathbb{E}\left(\varepsilon_{{\mathtt{t}}i}^3 \varepsilon_{{\mathtt{t}}j}\right)=0$, $\mathbb{E}\left(\varepsilon_{{\mathtt{t}}i}^2 \varepsilon_{{\mathtt{t}}j}^2 \right)=0$, and $\mathbb{E}\left(\varepsilon_{{\mathtt{t}}i} \varepsilon_{{\mathtt{t}}j}^3 \right)=0$, $i\not = j$, $\mathbb{E}\left(\varepsilon_{{\mathtt{t}}i}^4 \right)= \sigma_i^4$. Note that $\sigma_i^4$ stands for the fourth moment of $\varepsilon_{{\mathtt{t}}i}$, where in general $(\sigma^2_i)^2 \not= \sigma_i^4$. }}
If $p$ moments of $\mathbf{y}_{\mathtt{t}}$ should be considered, we get the number of moments
by summing over the multinomial coefficients, i.e.
\begin{equation}
\label{eq:nummuom1}
N_y=
\sum_{j=1}^p \left(
\begin{array}{c}
j+M-1 \\
j
\end{array}
\right)  .
\end{equation}
\noindent Powers of sums can be obtained by means
of the  {\em multinomial formula}. With $k= \sum_{i=1}^d l_i$, $l_i \geq 0$, we get
\begin{equation}
\label{eq:pasc2}
   \left(x_1 + x_2 + \cdots + x_d \right)^k = \sum_{l_1+l_2+\cdots+l_d=k} {k \choose l_1, l_2, \ldots, l_d} \prod_{1 \le i \le d}x_{i}^{l_{i}} \ ,
\end{equation}

\noindent where
${k \choose l_1, l_2, \ldots, l_d} = \frac{k!}{l_1!\, l_2! \cdots l_d!}$. %\footnote{Pascal's triangle is a special case of (\ref{eq:pasc2}).}
Let
\begin{equation}
\label{eq:defdv2}
d_{(i,K)} = \left(
\begin{array}{c}
i+K-1 \\
i
\end{array}
\right)
\end{equation}
\noindent
for $K \in \mathbb{N}$ and $i \le p$. In accordance with equation (\ref{eq:defdv}), we write $d_{i} \equiv d_{(i,d)} ;$ when $K=d$. That is, the notation is simplified when $K=d$. Note that $d_i$ calculates the dimension of conditional moments $\mathbb{E}(\mathbf{X}(t)^i|\mathbf{X}(s)=\mathbf{x} )$. In addition, $N_i=\sum_{j=0}^i d_{j}$ corresponds to the sum of the conditional moments smaller or equal to $i$.

We  shall derive the first four moments, which implies that $p=4$ in the following.  From (\ref{eq:5aa}) we get the first moments by means of
\begin{eqnarray}
\label{eq:momy1}
\mathbb{E}(\mathbf{y}_{{\mathtt{t}}}) &=& \tilde{\vPhi}+ \tilde{\vPsi} \mathbb{E} (\mathbf{X}_{\mathtt{t}}) \ ,  \nonumber \\
\mathbb{E}(\mathbf{y}_{{\mathtt{t}}i}) &=&  \Phi_i+ \vPsi_i'\mathbb{E}(\mathbf{X}_{\mathtt{t}}) = \Phi_i+\vPsi_i'\mathbb{E}(\tilde{\mathbf{X}}_{{\mathtt{t}},1:3}) \ ,
\end{eqnarray}
\noindent
where $\Phi_i =-\frac{1}{\tau_i} \Phi(\tau_i, \mathbf{0}) \in \mathbb{R}$, $\vPsi_i = -\frac{1}{\tau_i} \vPsi(\tau_i, \mathbf{0}) \in \mathbb{R}^d $,
$\tilde{\vPhi} \in \mathbb{R}^{M}$, $\mathbf{X}_{\mathtt{t}} \in \mathbb{R}^d$, $\mathbf{y}_{\mathtt{t}} \in \mathbb{R}^M$
and $\tilde{\vPsi} \in \mathbb{R}^{M \times d} $.

The second moments of the yields are given by:
\begin{eqnarray}
\mathbb{E}(y_{{\mathtt{t}}i}y_{{\mathtt{t}}j})&=& \Phi_i \Phi_j + \left (\Phi_i \vPsi_j' + \Phi_j \vPsi_i' \right ) \mathbb{E}(\mathbf{X}_{\mathtt{t}}) + \vPsi_i' \mathbb{E}(\mathbf{X}_{\mathtt{t}} \mathbf{X}_{\mathtt{t}}')\vPsi_j  + \mathbb{E} (\varepsilon_{{\mathtt{t}}i} \varepsilon_{{\mathtt{t}}j}) \nonumber \\
&=&  \Phi_i \Phi_j + \left (\Phi_i \vPsi_j' + \Phi_j \vPsi_i' \right ) \mathbb{E}(\tilde{\mathbf{X}}_{{\mathtt{t}},1:d}) + \vPsi_i' \mathbb{E}\left(vech^{-1}(\tilde{\mathbf{X}}_{{\mathtt{t}},d+1:d+d_2}) \right)\vPsi_j + \mathbb{E} (\varepsilon_{{\mathtt{t}}i} \varepsilon_{{\mathtt{t}}j}) \label{m2} \ ,
\end{eqnarray}
\noindent for $i,j=1,\dots,M$. In (\ref{m2}) we need the function $vech^{-1}$. The purpose of this function is to transform the $d(d+1)/2 \times 1$ vector $\tilde{\mathbf{X}}_{{\mathtt{t}},d+1:d+d_2}$ into the symmetric $d \times d$  matrix $\mathbf{X}_{\mathtt{t}} \mathbf{X}_{\mathtt{t}}'$. In more details,
$\tilde{\mathbf{X}}_{{\mathtt{t}},d+1:d+d_2} = vech \left( \mathbf{X}_{\mathtt{t}} \mathbf{X}_{\mathtt{t}}' \right) $, where $vec \left( \mathbf{X}_{\mathtt{t}} \mathbf{X}_{\mathtt{t}}' \right)$ vectorizes the $d \times d$ matrix
$\mathbf{X}_{\mathtt{t}} \mathbf{X}_{\mathtt{t}}'$ and $vech \left( \mathbf{X}_{\mathtt{t}} \mathbf{X}_{\mathtt{t}}' \right)$ eliminates the supra-diagonal elements from the $d^2 \times 1$ vector $vec \left( \mathbf{X}_{\mathtt{t}} \mathbf{X}_{\mathtt{t}}' \right)$
\citep[see, e.g.,][page~646]{poirier1995}. Hence $vech \left( \mathbf{X}_{\mathtt{t}} \mathbf{X}_{\mathtt{t}}' \right)$ is a $d(d+1)/2 \times 1$ vector.
The function $vech^{-1}$ takes us back to $\mathbf{X}_{\mathtt{t}} \mathbf{X}_{\mathtt{t}}'$, i.e. $vech^{-1}$ maps the $d(d+1)/2 \times 1$ vector $\mathbb{E}(\tilde{\mathbf{X}}_{{\mathtt{t}},d+1:d+d_2})$ to the symmetric $d \times d$ matrix $\mathbb{E}(\mathbf{X}_{\mathtt{t}} \mathbf{X}_{\mathtt{t}}')$. For $d=3$ this works as follows:
\begin{eqnarray}
&&
vech^{-1}\left(
\begin{array}{c}
a_1 \\
\cdots \\
a_6
\end{array}
\right) = \left [
\begin{array}{ccc}
a_1 & a_2 & a_3 \\
a_2 & a_4 & a_5 \\
a_3 & a_5 & a_6
\end{array}
\right ]
\end{eqnarray}
\noindent
and thus $vech^{-1}(\tilde{\mathbf{X}}_{{\mathtt{t}},d+1:d+d_2}) = \mathbf{X}_{\mathtt{t}} \mathbf{X}_{\mathtt{t}}'$.
By Assumption~\ref{ass:noise} we obtain $\mathbb{E}(X_{{\mathtt{t}}l}\varepsilon_{{\mathtt{t}}i})=0$ for $l=1, \ldots, d$,  $i=1, \ldots, M$ and
\begin{eqnarray}
\mathbb{E}(\varepsilon_{{\mathtt{t}}i} \varepsilon_{{\mathtt{t}}j})= \left \{
\begin{array}{rl}
\sigma^2_i , & \ \text{ for } i=j \ , \\
0 , & \  \text{ for } i \not = j \ ,
\end{array}
\right.
\end{eqnarray}
for $i,$ $j=1, \ldots, M$. Based on this, (\ref{m2}) can be written as
\begin{eqnarray}
\mathbb{E}(y_{{\mathtt{t}}i}y_{{\mathtt{t}}j})=\Phi_i \Phi_j + \left (\Phi_i \vPsi_j' + \Phi_j \vPsi_i' \right ) \mathbb{E}(\tilde{\mathbf{X}}_{{\mathtt{t}},1:d}) + (\mathbf{m}_2^{ij})^{\prime} \mathbb{E}(\tilde{\mathbf{X}}_{{\mathtt{t}},d+1:d+d_2}) + \mathbb{E} (\varepsilon_{{\mathtt{t}}i} \varepsilon_{{\mathtt{t}}j}) ,  \label{mm2}
\end{eqnarray}
\noindent where for $d=3$ we define $\mathbf{m}_2^{ij} = ( \Psi_{i1} \Psi_{j1}, \Psi_{i1} \Psi_{j2}+\Psi_{i2} \Psi_{j1}, $$ \Psi_{i1}\Psi_{j3}+\Psi_{i3}\Psi_{j1},$$ \Psi_{i2}\Psi_{j2}, $$ \Psi_{i2}\Psi_{j3} $$+ \Psi_{i3}\Psi_{j2}, $$ \Psi_{i3}\Psi_{j3} )^{\prime}
=\mathbf{m}_2^{ji} \in \mathbb{R}^{d_2}$  and thus $(\mathbf{m}_2^{ij})'\mathbb{E}(\tilde{\mathbf{X}}_{{\mathtt{t}},4:9})=\vPsi_i' \mathbb{E}(\mathbf{X}_{\mathtt{t}} \mathbf{X}_{\mathtt{t}}')\vPsi_j$.
Regarding the third moments we observe:
{\small
\begin{eqnarray}
\mathbb{E}(y_{{\mathtt{t}}i}^2y_{{\mathtt{t}}j})&=&  \mathbb{E}\left ( \left (\Phi_i + \vPsi_i'\mathbf{X}_{\mathtt{t}} + \varepsilon_{{\mathtt{t}}i} \right )^2 \left (\Phi_j + \vPsi_j'\mathbf{X}_{\mathtt{t}} + \varepsilon_{{\mathtt{t}}j} \right ) \right ) \nonumber \\
&=&  \mathbb{E}\left ( (\Phi_i + \vPsi_i'\mathbf{X}_{\mathtt{t}})^2 (\Phi_j + \vPsi_j'\mathbf{X}_{\mathtt{t}})+(\Phi_i + \vPsi_i'\mathbf{X}_{\mathtt{t}})^2 \varepsilon_{{\mathtt{t}}j}+2(\Phi_i + \vPsi_i'\mathbf{X}_{\mathtt{t}})(\Phi_j + \vPsi_j'\mathbf{X}_{\mathtt{t}})\varepsilon_{{\mathtt{t}}i} \right ) \nonumber \\
&& + \mathbb{E}\left ( 2(\Phi_i + \vPsi_i'\mathbf{X}_{\mathtt{t}})\varepsilon_{{\mathtt{t}}i}\varepsilon_{{\mathtt{t}}j}+(\Phi_j + \vPsi_j' \mathbf{X}_{\mathtt{t}})\varepsilon_{{\mathtt{t}}i}^2+\varepsilon_{{\mathtt{t}}i}^2\varepsilon_{{\mathtt{t}}j} \right )\nonumber \\
&=& \mathbb{E}\left ( \Phi_i^2 \Phi_j+ \Phi_i^2 (\vPsi_j'\mathbf{X}_{\mathtt{t}})+2\Phi_i\Phi_j(\vPsi_i'\mathbf{X}_{\mathtt{t}})+2\Phi_i(\vPsi_i'\mathbf{X}_{\mathtt{t}})(\vPsi_j'\mathbf{X}_{\mathtt{t}})+\Phi_j (\vPsi_i'\mathbf{X}_{\mathtt{t}})^2+(\vPsi_i'\mathbf{X}_{\mathtt{t}})^2(\vPsi_j'\mathbf{X}_{\mathtt{t}}) \right )\nonumber \\
&& +2(\Phi_i + \vPsi_i' \mathbb{E} (\mathbf{X}_{\mathtt{t}})) \sigma_i^2 \mathbb{I}_{\left ( {i=j}
 \right )}
+(\Phi_j + \vPsi_j' \mathbb{E} (\mathbf{X}_{\mathtt{t}})) \sigma_i^2 \nonumber \\
&=&  \Phi_i^2 \Phi_j+(\Phi_i^2\vPsi_j'+2\Phi_i\Phi_j\vPsi_i')\mathbb{E}(\mathbf{X}_{\mathtt{t}})+2\Phi_i\vPsi_i' \mathbb{E}(\mathbf{X}_{\mathtt{t}} \mathbf{X}_{\mathtt{t}}')\vPsi_j+\Phi_j\vPsi_i' \mathbb{E}(\mathbf{X}_{\mathtt{t}} \mathbf{X}_{\mathtt{t}}')\vPsi_i +  \mathbb{E}((\vPsi_i'\mathbf{X}_{\mathtt{t}})^2\vPsi_j'\mathbf{X}_{\mathtt{t}}) \nonumber \\
&& +2(\Phi_i + \vPsi_i' \mathbb{E} (\mathbf{X}_{\mathtt{t}})) \sigma_i^2
\mathbb{I}_{\left ( {i=j}
 \right )}
+(\Phi_j + \vPsi_j' \mathbb{E} (\mathbf{X}_{\mathtt{t}})) \sigma_i^2 \nonumber \\
&=& \left(\Phi_i^2 + \sigma_i^2\right) \Phi_j+\left(\Phi_i^2\vPsi_j'+2\Phi_i\Phi_j\vPsi_i' +  \sigma_i^2 \vPsi_j'\right)\mathbb{E} (\mathbf{X}_{\mathtt{t}})+\left (2\Phi_i (\mathbf{m}_2^{ij})'+\Phi_j(\mathbf{m}_2^{ii})'\right )\mathbb{E}(\tilde{\mathbf{X}}_{{\mathtt{t}},d+1:d+d_2}) \nonumber \\
&& +  (\mathbf{m}_3^{i^2j})'\mathbb{E}\left(\tilde{\mathbf{X}}_{{\mathtt{t}},d+d_2+1:d+d_2+d_3}\right) +2(\Phi_i + \vPsi_i' \mathbb{E} (\mathbf{X}_{\mathtt{t}})) \sigma_i^2 \mathbb{I}_{\left ( {i=j}
 \right )} \ , \label{m3i2j}
\end{eqnarray}
}
\noindent where $\mathbf{m}_3^{i^2j}  = \left(\Psi_{i1}^2\Psi_{j1}, \right.$ $\Psi_{i1}^2\Psi_{j2}+2\Psi_{i1}\Psi_{i2}\Psi_{j1},$ $\Psi_{i1}^2\Psi_{j3}+2\Psi_{i1}\Psi_{i3}\Psi_{j1},$ $\Psi_{i2}^2\Psi_{j1}+2\Psi_{i1}\Psi_{i2}\Psi_{j2},$ $\Psi_{i3}^2\Psi_{j1}+2\Psi_{i1}\Psi_{i3}\Psi_{j3},$ $2(\Psi_{i1}\Psi_{i2}\Psi_{j3}+\Psi_{i1}\Psi_{i3}\Psi_{j2}+\Psi_{i2}\Psi_{i3}\Psi_{j1}),$ $\Psi_{i2}^2\Psi_{j2},$ $\Psi_{i2}^2\Psi_{j3}+2\Psi_{i2}\Psi_{i3}\Psi_{j2},$ $\Psi_{i3}^2\Psi_{j2}+2\Psi_{i2}\Psi_{i3}\Psi_{j3},$ $\left. \Psi_{i3}^2\Psi_{j3}\right )^{\prime} \in \mathbb{R}^{d_3}$ and $\mathbb{I}_{(\cdot)}$ stands for an indicator function. By Assumption~\ref{ass:noise} we get  $\mathbb{E}(\mathbf{X}_{{\mathtt{t}}l}\varepsilon_{{\mathtt{t}}i})=0$, $\mathbb{E}(X_{{\mathtt{t}}l}^2\varepsilon_{{\mathtt{t}}i})=0$, $\mathbb{E}(X_{{\mathtt{t}}l}\varepsilon_{{\mathtt{t}}i}\varepsilon_{{\mathtt{t}}j})=0$, for $i \not = j$, and $\mathbb{E}(\varepsilon_{{\mathtt{t}}i}^2\varepsilon_{{\mathtt{t}}j})=0$ for $l=1, \ldots, d$ and $i,j=1, \ldots, M$. In a similar way and under the same assumptions, it can be shown that
{\small
\begin{eqnarray}
\mathbb{E}\left(y_{{\mathtt{t}}i}y_{{\mathtt{t}}j}^2 \right) &=& \left(\Phi_j^2 +\sigma_j^2 \right) \Phi_i +\left(\Phi_j^2\vPsi_i'+2\Phi_i\Phi_j\vPsi_j' + \sigma_j^2 \vPsi_i'\right)\mathbb{E}(\mathbf{X}_{\mathtt{t}})+\left (2\Phi_j (m_2^{ij})'+\Phi_i(\mathbf{m}_2^{jj})'\right )\mathbb{E}(\tilde{\mathbf{X}}_{{\mathtt{t}},d+1:d+d_2}) \nonumber \\
&& +  (\mathbf{m}_3^{ij^2})'\mathbb{E}\left(\tilde{\mathbf{X}}_{{\mathtt{t}},d+d_2+1:d+d_2+d_3}\right) +2(\Phi_i + \vPsi_i' \mathbb{E} (\mathbf{X}_{\mathtt{t}})) \sigma_i^2
\mathbb{I}_{\left ( {i=j}
 \right )} \ ,
\label{m3ij2}
\end{eqnarray}
}
\noindent
where $\mathbf{m}_3^{ij^2} = \left(\Psi_{i1}\Psi_{j1}^2, \right. $ $ \Psi_{i2}\Psi_{j1}^2+2\Psi_{i1}\Psi_{j1}\Psi_{j2},$ $\Psi_{i3}\Psi_{j1}^2+2\Psi_{i1}\Psi_{j1}\Psi_{j3},$ $\Psi_{i1}\Psi_{j2}^2+2\Psi_{i2}\Psi_{j1}\Psi_{j2},$ $\Psi_{i1}\Psi_{j3}^2+2\Psi_{i3}\Psi_{j1}\Psi_{j3},$ $2(\Psi_{i1}\Psi_{j2}\Psi_{j3}+\Psi_{i2}\Psi_{j1}\Psi_{j3}+\Psi_{i3}\Psi_{j1}\Psi_{j2}),$ $\Psi_{i2}\Psi_{j2}^2,$ $\Psi_{i3}\Psi_{j2}^2+2\Psi_{i2}\Psi_{j2}\Psi_{j3},$ $\Psi_{i2}\Psi_{j3}^2+2\Psi_{i3}\Psi_{j2}\Psi_{j3},$ $\left.\Psi_{i3}\Psi_{j3}^2\right)^{\prime} \in \mathbb{R}^{d_3}$ for $d=3$. %
For the fourth moment we obtain
\begin{eqnarray}
\mathbb{E}\left(y_{{\mathtt{t}}i}^2y_{{\mathtt{t}}j}^2 \right)&=&  \mathbb{E}\left ( \left (\Phi_i + \vPsi_i^{\prime} \mathbf{X}_{\mathtt{t}} + \varepsilon_{{\mathtt{t}}i} \right )^2 \left (\Phi_j + \vPsi_j^{\prime} \mathbf{X}_{\mathtt{t}} + \varepsilon_{{\mathtt{t}}j} \right )^2 \right ) \nonumber \\
&=&
\mE \left ( \left ( \Phi_i^2 + (\vPsi_i^{\prime} \mathbf{X}_{\mathtt{t}})^2 + \varepsilon_{{\mathtt{t}}i}^2 + 2\Phi_i \vPsi_i^{\prime} \mathbf{X}_{\mathtt{t}} + 2\Phi_i \varepsilon_{{\mathtt{t}}i} + 2\vPsi_i^{\prime} \mathbf{X}_{\mathtt{t}} \varepsilon_{{\mathtt{t}}i} \right ) \right. \nonumber \\
&& \times \ \left. \left ( \Phi_j^2 + (\vPsi_j^{\prime} \mathbf{X}_{\mathtt{t}})^2 + \varepsilon_{{\mathtt{t}}j}^2 + 2\Phi_j \vPsi_j^{\prime} \mathbf{X}_{\mathtt{t}} + 2\Phi_j \varepsilon_{{\mathtt{t}}j} + 2\vPsi_j^{\prime} \mathbf{X}_{\mathtt{t}} \varepsilon_{{\mathtt{t}}j} \right ) \right )  \nonumber \\
&=&\Phi_i^2 \Phi_j^2 +  \sigma_i^2 \Phi_j^2 + \sigma_j^2 \Phi_i^2 +  \mathbb{E}(\varepsilon_{{\mathtt{t}}i}^2 \varepsilon_{{\mathtt{t}}j}^2 ) \nonumber \\
 &&+ 2 \Phi_i \vPsi_i^{\prime} \mathbb{E}(\mathbf{X}_{\mathtt{t}}) ( \Phi_j^2 +  \sigma_j^2) + 2 \Phi_j \vPsi_j^{\prime} \mathbb{E}(\mathbf{X}_{\mathtt{t}}) (\Phi_i^2 + \sigma_i^2) \nonumber \\
 &&+ (\mathbf{m}_2^{ii})^{\prime} \mathbb{E}(\tilde{\mathbf{X}}_{{\mathtt{t}},d+1:d+d_2} ) (\Phi_j^2+\sigma_j^2)
 \nonumber \\
 &&
 +   (\mathbf{m}_2^{jj})^{\prime} \mathbb{E}(
 \tilde{\mathbf{X}}_{{\mathtt{t}},d+1:d+d_2} ) (\Phi_i^2+\sigma_i^2) + 4 \Phi_i \Phi_j (\mathbf{m}_2^{ij})' \mathbb{E}(\tilde{\mathbf{X}}_{{\mathtt{t}},d+1:d+d_2} ) \nonumber \\
 &&+ 2 \Phi_i (\mathbf{m}_3^{i j^2})' \mathbb{E}(\tilde{\mathbf{X}}_{{\mathtt{t}},d+d_2+1:d+d_2+d_3} ) + 2 \Phi_j (\mathbf{m}_3^{i^2 j})' \mathbb{E}(
 \tilde{\mathbf{X}}_{{\mathtt{t}},d+d_2+1:d+d_2+d_3} )
 \nonumber \\
 &&+  \mathbf{m}_4^{i^2 j^2} \mathbb{E}(\tilde{\mathbf{X}}_{{\mathtt{t}},d+d_2+d_3+1:d+d_2+d_3+d_4} ) \nonumber \\
 &&+ 4 \sigma_i^2 \left [ \Phi_i^2 + 2 \Phi_i \vPsi_i' \mathbb{E}(\mathbf{X}_{\mathtt{t}}) + (\mathbf{m}_2^{ii})' \tilde{\mathbf{X}}_{{\mathtt{t}},d+1:d+d_2} \right ] \mathbb{I}_{\left ( {i=j}
 \right )} \ ,
\label{eq:momy4}
\end{eqnarray}
\noindent where $\mathbf{m}_4^{i^2j^2} \mathbb{E}\left(\tilde{\mathbf{X}}_{{\mathtt{t}},d+d_2+d_3+1:d+d_2+d_3+d_4} \right) = \mathbb{E}\left((\vPsi_i' \mathbf{X}_{\mathtt{t}})^2(\vPsi_j' \mathbf{X}_{\mathtt{t}})^2 \right)$. By sticking to Assumption~\ref{ass:noise} the expectation $\mathbb{E}(\varepsilon_{{\mathtt{t}}i}^2 \varepsilon_{{\mathtt{t}}i}^2 )=\mathbb{E}(\varepsilon_{{\mathtt{t}}i}^4 )$ and $\mathbb{E}(\varepsilon_{{\mathtt{t}}i}^2 \varepsilon_{{\mathtt{t}}j}^2 )= \sigma_i^2 \sigma_j^2$ for $j \not= i$. Moreover,
$\mathbf{m}_4^{i^2j^2} = \left( \Psi_{i1}^2 \Psi _{j1}^2, \right.$ $2\Psi_{i1} \Psi_{j1} ( \Psi_{i2} \Psi_{j1}+ \Psi _{i1} \Psi _{j2}),$ $2\Psi_{i1} \Psi_{j1}(\Psi_{i1} \Psi_{j3}+ \Psi_{i3} \Psi_{j1}),$ $\Psi _{i1}^2 \Psi_{j2}^2 + \Psi _{i2}^2 \Psi_{j1}^2 + 4 \Psi_{i1} \Psi_{i2} \Psi_{j1} \Psi_{j2},$ $4\Psi_{i1} \Psi_{j1}( \Psi_{i2} \Psi_{j3} + \Psi_{i3} \Psi_{j2}) + 2 (\Psi_{i1}^2 \Psi_{j2} \Psi_{j3}+ \Psi _{i2} \Psi_{i3} \Psi_{j1}^2),$ $\Psi_{i1}^2 \Psi_{j3}^2 +\Psi_{i3}^2 \Psi_{j1}^2 + 4 \Psi_{i1} \Psi_{i3} \Psi_{j1} \Psi_{j3},$ $2 \Psi_{i2} \Psi_{j2}(\Psi_{i2} \Psi_{j1} + \Psi_{i1} \Psi_{j2}),$ $4 \Psi_{i2} \Psi_{j2}( \Psi_{i3} \Psi_{j1} + \Psi_{i1} \Psi_{j3})+ 2 (\Psi_{i1} \Psi _{i3} \Psi_{j2}^2 + \Psi_{i2}^2 \Psi_{j1} \Psi_{j3}),$ $4 \Psi_{i3} \Psi_{j3}(\Psi_{i1} \Psi_{j2} + \Psi _{i2} \Psi _{j1} )+ 2 (\Psi_{i1} \Psi_{i2} \Psi_{j3}^2 + \Psi_{i3}^2 \Psi_{j1} \Psi_{j2}),$ $2 \Psi_{i3} \Psi_{j3}(\Psi_{i3} \Psi_{j1} + \Psi_{i1} \Psi_{j3}),$ $\Psi_{i2}^2 \Psi_{j2}^2 ,$ $2 \Psi_{i2} \Psi_{j2}( \Psi_{i3} \Psi_{j2} +  \Psi_{i2} \Psi_{j3}),$ $\Psi_{i2}^2 \Psi_{j3}^2 + \Psi_{i3}^2 \Psi_{j2}^2 + 4 \Psi_{i2} \Psi_{i3} \Psi_{j2} \Psi _{j3},$ $2 \Psi_{i3} \Psi_{j3}(\Psi_{i2} \Psi_{j3} + \Psi_{i3} \Psi_{j2} ),$ $ \left. \Psi_{i3}^2 \Psi_{j3}^2 \right)^{\prime} \in \mathbb{R}^{d_4}$.

For the auto-covariance of the yields and the auto-covariance of the squared yields we have to calculate $\mE(\mathbf{X}_{\mathtt{t}}^v(\mathbf{X}_{{\mathtt{s}}}^w)')$ (which will become clear later). Before we proceed with these moments we obtain the result that only $\mathbf{x}_{{\mathtt{s}}}^{\iota}$ with exponents $\iota \leq v$ enter into the calculation of the conditional moment
$\mathbb{E}\left(\mathbf{X}_{{\mathtt{t}}}^v | \mathbf{X}_{{\mathtt{s}}}=\mathbf{x}_{{\mathtt{s}}} \right)$. In addition we derive a result on the structure of $\exp((t-s)  \mathbf{A})$, which is presented in the following lemma:

\begin{lemma}\label{structure} {\it
Let $\mathbf{D}$ and $\mathbf{B}$ be $n \times n$ lower-block triangular matrices such that: $\mathbf{D}_{m_i:n_i, n_i+1:n}=\mathbf{B}_{m_i:n_i, n_i+1:n}=0$ where $m_i \le n_i$ for $i=1, \ldots, k$, $k \le n$, $n_k=n$ and $n_i<n_{i+1}$ for $i=1, \ldots, k-1$. Then the matrix $\mathbf{C}= \mathbf{D}\mathbf{B}$ is of the same structure, namely $\mathbf{C}_{m_i:n_i, n_i+1:n}=0$ where $m_i \le n_i$ and $n_i<n_{i+1}$ for $i=1, \ldots, k-1$.}
\end{lemma}
\textit{Proof:} Let $j$ and $l$ be such that there exists $i \in \{1, \ldots, k \}$ such that $m_i \le j \le n_i$, $n_i+1 \le l \le n$. Then $C_{jl}=(D_{j1}, \ldots, D_{jn_i}, 0, \ldots, 0)(0, \ldots,  0, B_{n_i+1, l}, \ldots, B_{nl})'=0$. \hfill $\Box$

Note that for a square matrix $\mathbf{B}$, $\exp(\mathbf{B})=\sum_{i=0}^{+\infty} \frac{\mathbf{B}^i}{i!}.$ Thus, if $\mathbf{B}$ is a matrix of the structure described in the lemma then $\exp(\mathbf{B})$ has the same structure as well.
As the matrix $(t-s) \mathbf{A}$ is of the structure described in Lemma \ref{structure}, this and the definition of $\exp((t-s) \mathbf{A})$ imply that also the matrix $\exp((t-s) \mathbf{A})$ is of that same structure. Thus,
$$
\exp((t-s) \mathbf{A})_{N_v-1:N_v, N_v+1:N}=\mathbf{0},
$$
which gives
\begin{eqnarray}
&& \exp((t-s) \mathbf{A})_{N_{v-1}+1:N_v,:} \left[1, (\mathbf{x}^1)', (\mathbf{x}^2)', \ldots, (\mathbf{x}^p)'\right]' \nonumber \\
&& \ \ = [\exp((t-s) \mathbf{A})_{N_{v-1}+1:N_v,1:N_v}, \exp((t-s) \mathbf{A})_{N_{v-1}+1:N_v,N_v+1 : N}]\left[1, (\mathbf{x}^1)', \ldots, (\mathbf{x}^v)', (\mathbf{x}^{v+1})', \ldots,  (\mathbf{x}^p)'\right]' \nonumber \\
&& \ \ = [\exp((t-s) \mathbf{A})_{N_{v-1}+1:N_v,1:N_v},\mathbf{0}_{d_v \times N-N_v}]\left[1, (\mathbf{x}^1)', \ldots, (\mathbf{x}^v)', (\mathbf{x}^{v+1})', \ldots,  (\mathbf{x}^p)'\right]' \nonumber \\
&& \ \ = \exp((t-s) \mathbf{A})_{N_{v-1}+1:N_v,1:N_v} \left[1, (\mathbf{x}^1)', \ldots, (\mathbf{x}^v)'\right ]' + \mathbf{0}_{d_v \times N-N_v} \times \left [(\mathbf{x}^{v+1})', \ldots,  (\mathbf{x}^p)'\right]' \nonumber \\
&& \ \ = \exp((t-s) \mathbf{A})_{N_{v-1}+1:N_v,1:N_v} \left[1, (\mathbf{x}^1)', \ldots, (\mathbf{x}^v)'\right ]' \nonumber  .
\end{eqnarray}
(\ref{semigroup7}) and the above calculations show that:
Only $\mathbf{x}^{\iota}$ with $\iota \leq v$ enter into the calculation of the conditional moment
$\mathbb{E}(\mathbf{X}_{{\mathtt{t}}}^v | \mathbf{X}_{\mathtt{s}}=\mathbf{x})$. The conditional expectation of the $v$-th moment of $\mathbf{X}_{\mathtt{t}}$ with respect to $\mathbf{X}_{\mathtt{s}}=\mathbf{x}$ is
\begin{eqnarray}
&&\mathbb{E}(\mathbf{X}_{{\mathtt{t}}}^v|\mathbf{X}_{\mathtt{s}}=\mathbf{x})=\mathbb{E}\left(\tilde{\mathbf{X}}_{{{\mathtt{t}}}, N_{v-1}+1:N_v}|\mathbf{X}_{\mathtt{s}}=\mathbf{x} \right)= \exp((\mathtt{t}-\mathtt{s}) \Delta \mathbf{A})_{N_{v-1}+1:N_v,1:N_v} \left[1, (\mathbf{x}^1)', \ldots, (\mathbf{x}^v)'\right ]' \nonumber \\
&&= \exp((\mathtt{t}-\mathtt{s}) \Delta \mathbf{A})_{N_{v-1}+1:N_v,1}  + \exp((\mathtt{t}-\mathtt{s}) \Delta \mathbf{A})_{N_{v-1}+1:N_v,2:1+d_1} \mathbf{x}^1 + \cdots \nonumber \\
&& \cdots + \exp((\mathtt{t}-\mathtt{s}) \Delta \mathbf{A})_{N_{v-1}+1:N_v,N_{v-1}+1:N_v} \mathbf{x}^v \ ,
\end{eqnarray}
\noindent
which is of dimension $d_v\times 1$; $\Delta \in \mathbb{R}_{++}$ is the step-width already defined in Section~\ref{sect:moments}. This implies:%
{\footnote{A step width $\Delta=1$ was already assumed in the main text. To derive the following moments with a different step-width if necessary, $\Delta$ will be included in the following expressions.}}%
{\small
\begin{eqnarray}
\mathbb{E}(\mathbf{X}_{{\mathtt{t}}}^v (\mathbf{X}_{{\mathtt{s}}}^w)')&=&\mathbb{E}\left(\mathbb{E}(\mathbf{X}_{{\mathtt{t}}}^v|\mathbf{X}_{{\mathtt{s}}})(\mathbf{X}_{{\mathtt{s}}}^{w})' \right)\nonumber \\
&=&\exp((\mathtt{t}-\mathtt{s}) \Delta  \mathbf{A})_{N_{v-1}+1:N_v,1} \mathbb{E}((\mathbf{X}_{{\mathtt{s}}}^{w})') + \exp(({{\mathtt{t}}}-{{\mathtt{s}}}) \Delta  \mathbf{A})_{N_{v-1}+1:N_v,2:1+d_1} \mathbb{E}(\mathbf{X}_{{\mathtt{s}}} (\mathbf{X}_{{\mathtt{s}}}^{w})') \nonumber \\
&& + \cdots + \exp((\mathtt{t}-\mathtt{s}) \Delta  \mathbf{A})_{N_{v-1}+1:N_v,N_{v-1}+1:N_v} \mathbb{E}(\mathbf{X}_{{\mathtt{s}}}^v (\mathbf{X}_{{\mathtt{s}}}^{w})')  .
\end{eqnarray}
}
\noindent
Then for ${{\mathtt{t}}}>{{\mathtt{s}}}$ we obtain
{\footnotesize
\begin{eqnarray}
\mathbb{E}(y_{{{\mathtt{t}}}i} y_{{{\mathtt{s}}}i})&=&\mathbb{E}\left( \left( \Phi_i+\vPsi_i'\mathbf{X}_{{\mathtt{t}}}+\varepsilon_{{{\mathtt{t}}}i} \right)\left( \Phi_i+\vPsi_i'\mathbf{X}_{{\mathtt{s}}}+\varepsilon_{{{\mathtt{s}}}i} \right) \right) \nonumber \\
&=& \Phi_i^2 + 2 \Phi_i \vPsi_i'\mathbb{E}(X_{{\mathtt{t}}}) + \vPsi_i' \mathbb{E}(\mathbf{X}_{{\mathtt{t}}} \mathbf{X}_{{\mathtt{s}}}') \vPsi_i \nonumber \\
&=&  \Phi_i^2 + 2 \Phi_i \vPsi_i'\mathbb{E}(X_{{\mathtt{t}}}) + \vPsi_i' \exp(({{\mathtt{t}}}-{{\mathtt{s}}})\Delta \cdot \mathbf{A})_{2:1+d,1} \mathbb{E}(\mathbf{X}_{{\mathtt{t}}}') \vPsi_i + \vPsi_i' \exp(({{\mathtt{t}}}-{{\mathtt{s}}}) \Delta \cdot \mathbf{A})_{2:1+d,2:1+d} \mathbb{E}(\mathbf{X}_{{\mathtt{t}}} \mathbf{X}_{{\mathtt{t}}}') \vPsi_i \nonumber \\
&=&  \Phi_i^2 + 2 \Phi_i \vPsi_i'\mathbb{E}(\tilde{\mathbf{X}}_{{{\mathtt{t}}}, 1:d}) + \vPsi_i' \exp(({{\mathtt{t}}}-{{\mathtt{s}}})\Delta \cdot \mathbf{A})_{2:1+d,1} \mathbb{E}(\tilde{X}_{{{\mathtt{t}}}, 1:d}') \vPsi_i \nonumber \\
&& + \vPsi_i' \exp(({{\mathtt{t}}}-{{\mathtt{s}}}) \Delta \cdot \mathbf{A})_{2:1+d,2:1+d} \mathbb{E}(vech^{-1}(\tilde{\mathbf{X}}_{{{\mathtt{t}}}, 1+d:d+d_2})) \vPsi_i \ , \ \text{ and } \label{mst21} \\
\mE(y_{{{\mathtt{t}}}i}^2 y_{{{\mathtt{s}}}i}^2)&=&\mE \left ( \left ( \Phi_i + \vPsi_i' \mathbf{X}_{{\mathtt{t}}}+\varepsilon_{{{\mathtt{t}}}i}  \right )^2 \left ( \Phi_i + \vPsi_i' \mathbf{X}_{{\mathtt{s}}}+\varepsilon_{{{\mathtt{s}}}i}  \right )^2 \right ) \nonumber \\
&=& \mE \left ( \left ( \Phi_i^2 + (\vPsi_i' \mathbf{X}_{{\mathtt{t}}})^2 + \varepsilon_{ti}^2 + 2\Phi_i \vPsi_i'\mathbf{X}_{{\mathtt{t}}} + 2\Phi_i \varepsilon_{{{\mathtt{t}}}i} + 2\vPsi_i' \mathbf{X}_{{\mathtt{t}}} \varepsilon_{{{\mathtt{t}}}i} \right ) \right. \nonumber \\
&& \ \ \left. \left ( \Phi_i^2 + (\vPsi_i' \mathbf{X}_{{\mathtt{s}}})^2 + \varepsilon_{{{\mathtt{s}}}i}^2 + 2\Phi_i \vPsi_i' \mathbf{X}_{{\mathtt{s}}} + 2\Phi_i \varepsilon_{{{\mathtt{s}}}i} + 2\vPsi_i' \mathbf{X}_{{\mathtt{s}}} \varepsilon_{{{\mathtt{s}}}i} \right ) \right )  \nonumber \\
&=& \Phi_i^4 + 2\Phi_i^2 \mE((\vPsi_i' \mathbf{X}_{{\mathtt{s}}})^2)+2\Phi_i^2 \sigma_i^2 + 4 \Phi_i^3 \vPsi_i' \mE(\mathbf{X}_t)+ \mE((\vPsi_i' \mathbf{X}_{{\mathtt{t}}})^2(\vPsi_i' \mathbf{X}_{{\mathtt{s}}})^2) \nonumber \\
&& +2\sigma_i^2 \mE((\vPsi_i'\mathbf{X}_{{\mathtt{t}}})^2)+2\Phi_i \mE\left((\vPsi_i'\mathbf{X}_{{\mathtt{t}}})^2 \vPsi_i' \mathbf{X}_{{\mathtt{s}}} + \vPsi_i' \mathbf{X}_{{\mathtt{t}}} (\vPsi_i' \mathbf{X}_{{\mathtt{s}}})^2 \right) + 4\Phi_i \sigma_i^2 \vPsi_i' \mE(\mathbf{X}_{{\mathtt{t}}}) + 4 \Phi_i^2 \mE\left((\vPsi_i' \mathbf{X}_{{\mathtt{t}}})(\vPsi_i' \mathbf{X}_{{\mathtt{s}}}) \right) \nonumber \\
&=& \Phi_i^4 + 2(\Phi_i^2+\sigma_i^2) \mE\left((\vPsi_i' \mathbf{X}_{{\mathtt{t}}})^2 \right)+2\Phi_i^2 \sigma_i^2 + \mE\left((\vPsi_i' \mathbf{X}_{{\mathtt{t}}})^2(\vPsi_i' \mathbf{X}_{{\mathtt{s}}})^2 \right) \nonumber \\
&& +2\Phi_i \mE\left((\vPsi_i' \mathbf{X}_{{\mathtt{t}}})^2 \vPsi_i' \mathbf{X}_{{\mathtt{s}}} + \vPsi_i' \mathbf{X}_{{\mathtt{t}}} (\vPsi_i'X_{{\mathtt{s}}})^2 \right) + 4 ( \Phi_i^2+\sigma_i^2) \Phi_i \vPsi_i' \mE(X_{{\mathtt{t}}}) + 4 \Phi_i^2 \mE\left((\vPsi_i' \mathbf{X}_{{\mathtt{t}}})(\vPsi_i' \mathbf{X}_{{\mathtt{s}}}) \right)   .
\label{mst22}
\end{eqnarray}
}
%%%
\noindent To complete the calculation of these moments,  the quantities $\mE\left((\vPsi' \mathbf{X}_{{\mathtt{t}}})^2 \right)$, $\mE \left((\vPsi' \mathbf{X}_{{\mathtt{t}}})(\vPsi' \mathbf{X}_{{\mathtt{s}}}) \right)$, $\mE\left((\vPsi' \mathbf{X}_{{\mathtt{t}}})^2 \vPsi' \mathbf{X}_{{\mathtt{s}}} \right)$, $\mE\left(\vPsi' \mathbf{X}_{{\mathtt{t}}} (\vPsi' \mathbf{X}_{{\mathtt{s}}})^2 \right)$ and $\mE\left((\vPsi' \mathbf{X}_{{\mathtt{t}}})^2(\vPsi' \mathbf{X}_{{\mathtt{s}}})^2 \right)$ have to be derived. To simplify the notation, we omit the index $i$ in $\vPsi_i$ in the following expressions; $\Psi_l \in \mathbb{R}$ is the element $l$ of the $\vPsi \in \mathbb{R}^{d}$ (when the index $i$ is still included this would be $\Psi_{il}$). Note that
\begin{eqnarray}
\mE\left((\vPsi' \mathbf{X}_{{\mathtt{t}}})^2 \right) &=& \vPsi' \mE(\mathbf{X}_{{\mathtt{t}}} \mathbf{X}_{{\mathtt{t}}}') \vPsi = \vPsi' \mE(vech^{-1}(\mathbf{X}_{{\mathtt{t}}}^2)) \vPsi \text{ and  } \nonumber \\
\mE\left((\vPsi' \mathbf{X}_{{\mathtt{t}}})(\vPsi' \mathbf{X}_{{\mathtt{s}}}) \right)&=&\vPsi' \mE(\mathbf{X}_{{\mathtt{t}}} \mathbf{X}_{{\mathtt{s}}}') \vPsi = \vPsi' \left(\mE\left(\mE(\mathbf{X}_{{\mathtt{t}}}|\mathbf{X}_{{\mathtt{s}}})\mathbf{X}_{{\mathtt{s}}}' \right)\right) \vPsi  \\
&=& \vPsi' \left[ \exp(({{\mathtt{t}}}-{{\mathtt{s}}})\Delta  \mathbf{A})_{2:1+d,1} \mE(\mathbf{X}_{{\mathtt{t}}}') + \exp(({{\mathtt{t}}}-{{\mathtt{s}}})\Delta  \mathbf{A})_{2:1+d,2:1+d} \mE(\mathbf{X}_{{\mathtt{t}}} \mathbf{X}_{{\mathtt{t}}}') \right] \vPsi \nonumber \\
&=& \vPsi' \left[ \exp(({{\mathtt{t}}}-{{\mathtt{s}}})\Delta  \mathbf{A})_{2:1+d,1} \mE(\mathbf{X}_{{\mathtt{t}}}') + \exp(({{\mathtt{t}}}-{{\mathtt{s}}})\Delta  \mathbf{A})_{2:1+d,2:1+d} \mE(vech^{-1}(\mathbf{X}_{{\mathtt{t}}}^2)) \right] \vPsi .  \nonumber \label{ts}
\end{eqnarray}
In addition, we obtain
\begin{eqnarray}
\label{eq21}
\mE\left((\vPsi' \mathbf{X}_{{\mathtt{t}}})^2 \vPsi' \mathbf{X}_{{\mathtt{s}}} \right) &=& \vPsi' \mE(\mathbf{X}_{{\mathtt{t}}} \vPsi' \mathbf{X}_{{\mathtt{t}}} \mathbf{X}_{{\mathtt{s}}}')  \vPsi = \vPsi' \mE\left((\mathbf{X}_{{\mathtt{t}}} \mathbf{X}_{{\mathtt{s}}}') (\vPsi' \mathbf{X}_{{\mathtt{t}}}) \right)  \vPsi \ ,
\end{eqnarray}
\noindent for ${{\mathtt{t}}}>{{\mathtt{s}}}$. Here we observe the following equality:
\begin{eqnarray}
(\mathbf{X}_{{\mathtt{t}}} \mathbf{X}_{{\mathtt{s}}}') (\vPsi' \mathbf{X}_{{\mathtt{t}}})=
\left [
\begin{array}{cccc}
X_{{{\mathtt{t}}}1} X_{{{\mathtt{s}}}1} & X_{{{\mathtt{t}}}1} X_{{{\mathtt{s}}}2} & \cdots & X_{{{\mathtt{t}}}1} X_{{{\mathtt{s}}}d} \\
X_{{{\mathtt{t}}}2} X_{{{\mathtt{s}}}1} & X_{{{\mathtt{t}}}2} X_{{{\mathtt{s}}}2} & \cdots & X_{{{\mathtt{t}}}2} X_{{{\mathtt{s}}}d} \\
\cdots & \cdots & \cdots & \cdots \\
X_{{{\mathtt{t}}}d} X_{{{\mathtt{s}}}1} & X_{{{\mathtt{t}}}d} X_{{{\mathtt{s}}}2} & \cdots & X_{{{\mathtt{t}}}d} X_{{{\mathtt{s}}}d}
\end{array}
\right] \sum_{l=1}^d \Psi_{l} X_{{{\mathtt{t}}}l}  .
\end{eqnarray}
Equation (\ref{eq21}) requires the derivation of $\mE(X_{{{\mathtt{t}}}i} X_{{{\mathtt{s}}}j} X_{{{\mathtt{t}}}l} )$ where $i,$ $j$, $l \in \{ 1, \ldots, d \}.$ To simplify the notation, the following functions $g_i(\cdot)$, $i=2,3,4$, are introduced to facilitate tracking specific elements of the moments vectors. We obtain
\begin{equation}
\label{eq:gij}
g_2(i,j) = (i-1)\left(d-\frac{i}{2}\right)+j \  ,
\end{equation}
\noindent
for $i,j \in \mathbb{N}$ and $i \le j \le d$. Moreover we derive
{\small
\begin{eqnarray}\label{eq:g3ij}
g_3(i,j,m)&=&\sum_{k=1}^{i-1}d_{d-k}+ \frac{j-i}{2} \left ( 2d -i -j +3 \right ) + m-j+1 \\
\label{eq:g4ij}
g_4(i,j,m,n) &=& \sum_{k=1}^{i-1} d_{d-k}+\sum_{k=1}^{j-i} d_{(d-i+1-k,d-1)} + \left ( d+1 - \frac{m+j-1}{2} \right ) (m-j)  +n-m+1 \ ,
\end{eqnarray}
}
\noindent
for $i,j,m \in \mathbb{N}$ and $i \le j \le m \le d$ and
for $i,j,m \in \mathbb{N}$ and $i \le j \le m \le n \le d$, respectively. While $d$ was the dimension of the process $(\mathbf{X}(t))$, $d_{(.,.)}$ is the function already defined in (\ref{eq:defdv2}). For $d=3$ this yields
\begin{eqnarray}\label{g2}
g_2(i,j)=
\left \{
\begin{array}{rl}
1, \ \ & {\rm if} \ \ i=1, \ j=1 \\
2, \ \ & {\rm if} \ \ i=1, \ j=2 \\
3, \ \ & {\rm if} \ \ i=1, \ j=3 \\
4, \ \ & {\rm if} \ \ i=2, \ j=2 \\
5, \ \ & {\rm if} \ \ i=2, \ j=3 \\
6, \ \ & {\rm if} \ \ i=3, \ j=3 \ , \\
\end{array}
\right.
\end{eqnarray}

\begin{eqnarray}\label{g3}
g_3(i,j,m)=
\left \{
\begin{array}{rl}
1, \ \ & {\rm if} \ \ i=1, \ j=1, \ m=1 \\
2, \ \ & {\rm if} \ \ i=1, \ j=1, \ m=2 \\
3, \ \ & {\rm if} \ \ i=1, \ j=1, \ m=3 \\
4, \ \ & {\rm if} \ \ i=1, \ j=2, \ m=2 \\
5, \ \ & {\rm if} \ \ i=1, \ j=2, \ m=3 \\
6, \ \ & {\rm if} \ \ i=1, \ j=3, \ m=3 \\
7, \ \ & {\rm if} \ \ i=2, \ j=2, \ m=2 \\
8, \ \ & {\rm if} \ \ i=2, \ j=2, \ m=3 \\
9, \ \ & {\rm if} \ \ i=2, \ j=3, \ m=3 \\
10, \ \ & {\rm if} \ \ i=3, \ j=3, \ m=3 \ ,
\end{array}
\right.
\end{eqnarray}
\noindent
and
\begin{eqnarray}\label{g4}
g_4(i,j,m,n)=
\left \{
\begin{array}{rl}
1, \ \  & {\rm if} \ \ i=1, \ j=1, \ m=1, \ n=1 \\
2, \ \  & {\rm if} \ \ i=1, \ j=1, \ m=1, \ n=2 \\
3, \ \  & {\rm if} \ \ i=1, \ j=1, \ m=1, \ n=3 \\
4, \ \  & {\rm if} \ \ i=1, \ j=1, \ m=2, \ n=2 \\
5, \ \  & {\rm if} \ \ i=1, \ j=1, \ m=2, \ n=3 \\
6, \ \  & {\rm if} \ \ i=1, \ j=1, \ m=3, \ n=3 \\
7, \ \  & {\rm if} \ \ i=1, \ j=2, \ m=2, \ n=2 \\
8, \ \  & {\rm if} \ \ i=1, \ j=2, \ m=2, \ n=3 \\
9, \ \  & {\rm if} \ \ i=1, \ j=2, \ m=3, \ n=3 \\
10, \ \ & {\rm if} \ \ i=1, \ j=3, \ m=3, \ n=3 \\
11, \ \ & {\rm if} \ \ i=2, \ j=2, \ m=2, \ n=2 \\
12, \ \ & {\rm if} \ \ i=2, \ j=2, \ m=2, \ n=3 \\
13, \ \ & {\rm if} \ \ i=2, \ j=2, \ m=3, \ n=3 \\
14, \ \ & {\rm if} \ \ i=2, \ j=3, \ m=3, \ n=3 \\
15, \ \ & {\rm if} \ \ i=3, \ j=3, \ m=3, \ n=3  .
\end{array}
\right.
\end{eqnarray}
Let $\mathbf{e }$ be a vector of ones, $\mathbf{e} =(1, \ldots, 1)'$,\footnote{It's dimension is not specified on purpose as it will vary and will be clear from the context.} and ${\tilde{\mathbf{e}}} =(1, 2, \ldots, d)'$. Then $\mathbf{M}$, $\mathbf{M}^j$ and $\mathbf{M}^{j,l}$ are the following ${d_2 \times 2}$, ${d_2 \times 3}$ and ${d_2 \times 4}$ matrices:
\begin{eqnarray}
\mathbf{M} = \left(
\begin{array}{cc}
\mathbf{e} & \tilde{\mathbf{e}} \\
2 \mathbf{e} & \tilde{\mathbf{e}}_{2:d} \\
\cdots & \cdots \\
i \mathbf{e} & \tilde{\mathbf{e}}_{i:d} \\
\cdots & \cdots \\
d & d
\end{array}
\right) \ ,  \ {\text{ which for}} \ d=3 \ {\text{ is }} \ \ \
\mathbf{M} = \left(
\begin{array}{cc}
1 & 1 \\
1 & 2 \\
1 & 3 \\
2 & 2 \\
2 & 3 \\
3 & 3
\end{array}
\right) \ ,
\end{eqnarray}
\begin{eqnarray}
\mathbf{M}^j = (\mathbf{M}, j\mathbf{e}) \ \ \  {\rm and} \ \ \ \mathbf{M}^{j,l} = (\mathbf{M}^j,l\mathbf{e})= (\mathbf{M}, j \mathbf{e}, l \mathbf{e}) \ ,
\end{eqnarray}
\noindent
where $\mathbf{e}$ is here a vector of ones of the dimension $d_2 \times 1$. Thus, for ${{\mathtt{t}}}>{{\mathtt{s}}}$
\begin{eqnarray}
\mE\left(X_{{{\mathtt{t}}}i} X_{{{\mathtt{s}}}j} X_{{{\mathtt{t}}}l} \right) &=& \mE\left(\mE\left(X_{{{\mathtt{t}}}i} X_{{{\mathtt{t}}}l}|\mathbf{X}_{{\mathtt{s}}} \right)X_{{{\mathtt{s}}}j} \right) \nonumber \\
&=& \exp(({{\mathtt{t}}}-{{\mathtt{s}}})\Delta  \mathbf{A})_{k,1} \mE(X_{{{\mathtt{s}}}j})  + \exp(({{\mathtt{t}}}-{{\mathtt{s}}})\Delta  \mathbf{A})_{k,2:1+d}\mE(\mathbf{X}_{{\mathtt{s}}} X_{{{\mathtt{s}}}j}) \nonumber \\
&& + \exp(({{\mathtt{t}}}-{{\mathtt{s}}})\Delta  \mathbf{A})_{k,2+d:2+d+d_2}\mE(\mathbf{X}_{{\mathtt{s}}}^2 X_{{{\mathtt{s}}}j}) \nonumber \\
&=& \exp(({{\mathtt{t}}}-{{\mathtt{s}}})\Delta  \mathbf{A})_{k,1} \mE(X_{{{\mathtt{t}}}j}) + \exp(({{\mathtt{t}}}-{{\mathtt{s}}})\Delta  \mathbf{A})_{k,2:1+d}\mE\left(\mathbf{X}^2_{g_2([\tilde{\mathbf{e}}, j\mathbf{e}]),{{\mathtt{t}}}} \right)\nonumber \\
&&+ \exp(({{\mathtt{t}}}-{{\mathtt{s}}})\Delta  \mathbf{A})_{k,2+d:2+d+d_2}\mE\left(\mathbf{X}^3_{g_3(\mathbf{M}^j),{{\mathtt{t}}}}\right) \ ,  \label{eeq21}
\end{eqnarray}
\noindent
where $k=1+d+g_2(i,l)$. Thus, the $(i,j)$ element, $i,j=1, \ldots, d $, of matrix in (\ref{eq21}) is
$$
\left [ \mE\left((\mathbf{X}_{{\mathtt{t}}} \mathbf{X}_{{\mathtt{s}}}')\vPsi' \mathbf{X}_{{\mathtt{t}}}\right) \right ]_{ij}
= \sum_{l=1}^d \Psi_l \mE(X_{{{\mathtt{t}}}i} X_{{{\mathtt{s}}}j} X_{{{\mathtt{t}}}l}) \ ,
$$
where $\mE(X_{{{\mathtt{t}}}i} X_{{{\mathtt{s}}}j} X_{{{\mathtt{t}}}l})$ is given by (\ref{eeq21}).
Let ${{\mathtt{t}}}>{{\mathtt{s}}}$, then
\begin{eqnarray}\label{12}
\mE\left(\vPsi' \mathbf{X}_{{\mathtt{t}}} (\vPsi' \mathbf{X}_{{\mathtt{s}}})^2 \right)=\vPsi' \mE(\mathbf{X}_t \vPsi' \mathbf{X}_{{\mathtt{s}}} \mathbf{X}_{{\mathtt{s}}}')  \vPsi = \vPsi' \mE\left((\mathbf{X}_{{\mathtt{t}}} \mathbf{X}_{{\mathtt{s}}}') (\vPsi' \mathbf{X}_{{\mathtt{s}}}) \right)  \vPsi \ ,
\end{eqnarray}
\noindent where
\begin{eqnarray}
(\mathbf{X}_{{\mathtt{t}}} \mathbf{X}_{{\mathtt{s}}}') (\vPsi' \mathbf{X}_{{\mathtt{s}}})&=&
\left[
\begin{array}{cccc}
X_{{{\mathtt{t}}}1} X_{{{\mathtt{s}}}1} & X_{{{\mathtt{t}}}1} X_{{{\mathtt{s}}}2} & \cdots & X_{{{\mathtt{t}}}1} X_{{{\mathtt{s}}}d} \\
X_{{{\mathtt{t}}}2} X_{{{\mathtt{s}}}1} & X_{{{\mathtt{t}}}2} X_{{{\mathtt{s}}}2} & \cdots & X_{{{\mathtt{t}}}2} X_{{{\mathtt{s}}}d} \\
\cdots & \cdots & \cdots & \cdots \\
X_{{{\mathtt{t}}}d} X_{{{\mathtt{s}}}1} & X_{{{\mathtt{t}}}d} X_{{{\mathtt{s}}}2} & \cdots & X_{{{\mathtt{t}}}d} X_{{{\mathtt{s}}}d}
\end{array}
\right] \sum_{i=l}^d \Psi_l X_{{{\mathtt{s}}}l} \nonumber \\ \nonumber \\
&=& \sum_{i=l}^d \Psi_l
\left[
\begin{array}{cccc}
X_{{{\mathtt{t}}}1} X_{{{\mathtt{s}}}1} X_{{{\mathtt{s}}}l} & X_{{{\mathtt{t}}}1} X_{{{\mathtt{s}}}2} X_{{{\mathtt{s}}}l} & \cdots & X_{{{\mathtt{t}}}1} X_{{{\mathtt{s}}}d} X_{{{\mathtt{s}}}l} \\
X_{{{\mathtt{t}}}2} X_{{{\mathtt{s}}}1} X_{{{\mathtt{s}}}l} & X_{{{\mathtt{t}}}2} X_{{{\mathtt{s}}}2} X_{{{\mathtt{s}}}l} & \cdots & X_{{{\mathtt{t}}}2} X_{{{\mathtt{s}}}d} X_{{{\mathtt{s}}}l} \\
\cdots & \cdots & \cdots & \cdots \\
X_{{{\mathtt{t}}}d} X_{{{\mathtt{s}}}1} X_{{{\mathtt{s}}}l} & X_{{{\mathtt{t}}}d} X_{{{\mathtt{s}}}2} X_{{{\mathtt{s}}}l} & \cdots & X_{{{\mathtt{t}}}d} X_{{{\mathtt{s}}}d} X_{{{\mathtt{s}}}l}
\end{array}
\right] . \label{21}
\end{eqnarray}
\noindent
For expression (\ref{21}) one needs to know $\mE(X_{{{\mathtt{t}}}i} X_{{{\mathtt{s}}}j} X_{{{\mathtt{s}}}l})$ where $i,$ $j$, $l \in \{ 1, \ldots, d \}$. Thus, for ${{\mathtt{t}}}>{{\mathtt{s}}}$
\begin{eqnarray}
\mE(X_{{{\mathtt{t}}}i} X_{{{\mathtt{s}}}j} X_{{{\mathtt{s}}}l}) &=& \mE\left(\mE(X_{{{\mathtt{t}}}i}|X_{{\mathtt{s}}})X_{{{\mathtt{s}}}j} X_{{{\mathtt{s}}}l} \right)  \\
&=& \exp(({{\mathtt{t}}}-{{\mathtt{s}}})\Delta \mathbf{A})_{i+1,1} \mE( X_{{{\mathtt{s}}}j} X_{{{\mathtt{s}}}l}) + \exp(({{\mathtt{t}}}-{{\mathtt{s}}})\Delta \mathbf{A})_{i+1,2:1+d}\mE(X_{{\mathtt{s}}} X_{{{\mathtt{s}}}j} X_{{{\mathtt{s}}}l})  \nonumber \\
&=& \exp(({{\mathtt{t}}}-{{\mathtt{s}}})\Delta \mathbf{A})_{i+1,1} \mE \left(\mathbf{X}^2_{g_2(j,l),{{\mathtt{t}}}}\right) + \exp(({{\mathtt{t}}}-{{\mathtt{s}}})\Delta \mathbf{A})_{i+1,2:1+d}\mE\left(\mathbf{X}^3_{g_3([\tilde{e},je, le]),{{\mathtt{t}}}} \right)   . \nonumber \label{eq12}
\end{eqnarray}
The $(i,j)$ element, $i,j=1, \ldots, d $, of matrix in (\ref{21}) is
$$
\left [ \mE\left((\mathbf{X}_{{\mathtt{t}}} \mathbf{X}_{{\mathtt{s}}}')\vPsi' \mathbf{X}_{{\mathtt{s}}} \right) \right ]_{ij}
= \sum_{l=1}^d \Psi_l \mE(X_{{{\mathtt{t}}}i} X_{{{\mathtt{s}}}j} X_{{{\mathtt{s}}}l}) \ ,
$$
where $\mE(X_{{{\mathtt{t}}}i} X_{{{\mathtt{s}}}j} X_{{{\mathtt{s}}}l})$ is given by (\ref{eq12}).
Finally
\begin{eqnarray}
\label{eq:EPXPX}
\mE\left((\vPsi' \mathbf{X}_{{\mathtt{t}}})^2(\vPsi' \mathbf{X}_{{\mathtt{s}}})^2 \right) &=&\vPsi' \mE\left(\mathbf{X}_{{\mathtt{t}}} (\vPsi' \mathbf{X}_{{\mathtt{t}}}) (\vPsi' \mathbf{X}_{{\mathtt{s}}}) \mathbf{X}_{{\mathtt{s}}}' \right) \vPsi=\vPsi' \mE \left((\mathbf{X}_{{\mathtt{t}}} \mathbf{X}_{{\mathtt{s}}}') (\vPsi' \mathbf{X}_{{\mathtt{t}}}) (\vPsi' \mathbf{X}_{{\mathtt{s}}}) \right) \vPsi
\end{eqnarray}
and
\begin{eqnarray}
\label{eq:XXX}
(\mathbf{X}_{{\mathtt{t}}} \mathbf{X}_{{\mathtt{s}}}') (\vPsi' \mathbf{X}_{{\mathtt{t}}}) (\vPsi' \mathbf{X}_{{\mathtt{s}}})&=&
\left [
\begin{array}{cccc}
X_{{{\mathtt{t}}}1} X_{{{\mathtt{s}}}1} & X_{{{\mathtt{t}}}1} X_{{{\mathtt{s}}}2} & \cdots & X_{{{\mathtt{t}}}1} X_{{{\mathtt{s}}}d} \\
X_{{{\mathtt{t}}}2} X_{{{\mathtt{s}}}1} & X_{{{\mathtt{t}}}2} X_{{{\mathtt{s}}}2} & \cdots & X_{{{\mathtt{t}}}2} X_{{{\mathtt{s}}}d} \\
\cdots & \cdots & \cdots & \cdots \\
X_{{{\mathtt{t}}}d} X_{{{\mathtt{s}}}1} & X_{{{\mathtt{t}}}d} X_{{{\mathtt{s}}}2} & \cdots & X_{{{\mathtt{t}}}d} X_{{{\mathtt{s}}}d}
\end{array}
\right ] \sum_{i=1}^d  \sum_{j=1}^d \Psi_i \Psi_j  X_{{{\mathtt{t}}}i} X_{{{\mathtt{s}}}j}   \\
%&& \times \left ( \sum_{i=1}^d \vPsi_i^2 X_{{{\mathtt{t}}}i} X_{{{\mathtt{s}}}i}+\sum_{i=1}^d  \sum_{j=i+1}^d \vPsi_i \vPsi_j ( X_{{{\mathtt{t}}}i} X_{{{\mathtt{s}}}j}+ X_{{{\mathtt{t}}}j} X_{{{\mathtt{s}}}i}) \right) . \nonumber
&=& \sum_{i=1}^d  \sum_{j=1}^d \Psi_i \Psi_j
\left [
\begin{array}{cccc}
X_{{{\mathtt{t}}}1} X_{{{\mathtt{s}}}1} X_{{{\mathtt{t}}}i} X_{{{\mathtt{s}}}j} & X_{{{\mathtt{t}}}1} X_{{{\mathtt{s}}}2} X_{{{\mathtt{t}}}i} X_{{{\mathtt{s}}}j} & \cdots & X_{{{\mathtt{t}}}1} X_{{{\mathtt{s}}}d} X_{{{\mathtt{t}}}i} X_{{{\mathtt{s}}}j} \\
X_{{{\mathtt{t}}}2} X_{{{\mathtt{s}}}1} X_{{{\mathtt{t}}}i} X_{{{\mathtt{s}}}j} & X_{{{\mathtt{t}}}2} X_{{{\mathtt{s}}}2} X_{{{\mathtt{t}}}i} X_{{{\mathtt{s}}}j} & \cdots & X_{{{\mathtt{t}}}2} X_{{{\mathtt{s}}}d} X_{{{\mathtt{t}}}i} X_{{{\mathtt{s}}}j} \\
\cdots & \cdots & \cdots & \cdots \\
X_{{{\mathtt{t}}}d} X_{{{\mathtt{s}}}1} X_{{{\mathtt{t}}}i} X_{{{\mathtt{s}}}j} & X_{{{\mathtt{t}}}d} X_{{{\mathtt{s}}}2} X_{{{\mathtt{t}}}i} X_{{{\mathtt{s}}}j} & \cdots & X_{{{\mathtt{t}}}d} X_{{{\mathtt{s}}}d} X_{{{\mathtt{t}}}i} X_{{{\mathtt{s}}}j}
\end{array}
\right ]   . \nonumber
\end{eqnarray}
\noindent
Then for ${{\mathtt{t}}}>{{\mathtt{s}}}$ we have
\begin{eqnarray}
\label{eq:EXXX2}
\mE(X_{{{\mathtt{t}}}i} X_{{{\mathtt{s}}}j} X_{{{\mathtt{t}}}m} X_{{{\mathtt{s}}}n})&=&\mE \left(\mE( X_{{{\mathtt{t}}}i} X_{{{\mathtt{t}}}m}| \mathbf{X}_{{\mathtt{s}}})X_{{{\mathtt{s}}}j} X_{{{\mathtt{s}}}n}\right)\nonumber \\
&=& \exp(({{\mathtt{t}}}-{{\mathtt{s}}}) \Delta \cdot \mathbf{A})_{k,1} \mE \left(X_{{{\mathtt{s}}}j} X_{{{\mathtt{s}}}n} \right)+\exp(({{\mathtt{t}}}-{{\mathtt{s}}})\Delta \cdot \mathbf{A})_{k,2:1+d} \mE \left(\mathbf{X}_{{\mathtt{s}}} X_{{{\mathtt{s}}}j} X_{{{\mathtt{s}}}n}\right) \nonumber \\
&& +\exp(({{\mathtt{t}}}-{{\mathtt{s}}})\Delta \cdot \mathbf{A})_{k,2+d:1+d+d_2} \mE \left(X_{{\mathtt{s}}}^2 X_{{{\mathtt{s}}}j} X_{{{\mathtt{s}}}n}\right) \nonumber \\
&=& \exp(({{\mathtt{t}}}-{{\mathtt{s}}})\Delta \cdot \mathbf{A})_{k,1} \mE \left(\mathbf{X}^2_{g_2(j,n),{{\mathtt{t}}}} \right)+\exp(({{\mathtt{t}}}-{{\mathtt{s}}}) \Delta \cdot \mathbf{A})_{k,2:1+d} \mE \left(\mathbf{X}^3_{g_3(\tilde{e},je,ne),{{\mathtt{t}}}}\right) \nonumber \\
&& +\exp(({{\mathtt{t}}}-{{\mathtt{s}}})\Delta \cdot \mathbf{A})_{k,2+d:1+d+d_2} \mE \left(\mathbf{X}^4_{g_4(\mathbf{M}^{j,n}),{{\mathtt{t}}}}\right) \ ,
\end{eqnarray}
where $k=1+d+g_2(i,m).$ The $(i,j)$ element, $i,j=1, \ldots, d$, of the matrix in (\ref{eq:EPXPX}) is
$$
\left[ \mE\left((\mathbf{X}_{{\mathtt{t}}} \mathbf{X}_{{\mathtt{s}}}') (\vPsi'\mathbf{X}_{{\mathtt{t}}}) (\vPsi'\mathbf{X}_{{\mathtt{s}}}) \right) \right]_{ij}=\sum_{k=1}^d \sum_{l=1}^d  \Psi_k \Psi_l \mE(X_{{{\mathtt{t}}}i} X_{{{\mathtt{s}}}j} X_{{{\mathtt{t}}}k} X_{{{\mathtt{s}}}l}) \ ,
$$
\noindent
with expectations being given by (\ref{eq:EXXX2}).

\section{Solving for $\Phi(t,\mathbf{u})$ and $\vPsi(t,\mathbf{u})$}
\label{appa:psiphi1}

This section derives the functions $\Phi(t,\mathbf{u})$ and $\vPsi(t,\mathbf{u})$ of the Riccati differential equations described by
(\ref{transformeddk1}) for an $\mathbb{A}_m(d)$ model with diagonal $\vbeta_{II}$. By equation (\ref{eq:1}), which is based on \citet{filipovicbook2009}[Theorem~10.4 and Corollary~10.2], $\vPsi(t,\mathbf{u})$ and $\Phi(t,\mathbf{u})$ evaluated at $t=\tau_i$, $i=1,\dots,M$ and $\mathbf{u}=\mathbf{0}_{d}$ are necessary to compute the zero coupon prices $\pi^0(t,\tau_i)$ and corresponding model yields. For the \citet{vasicek77} and the \citet{cir85} model the solutions are presented e.g. in \citet{filipovicbook2009}[p.~162-163].

Now we apply the results obtained in \citet{GrasselliTebaldi2008}[Section~3.4.1] for $\mathbb{A}_{m}(d)$ models with diagonal $m \times m$ matrix $\vbeta_{II}$. In the first step we have to solve the linear ODE for the $J$ components. I.e. we consider\footnote{Note that the dimension of $\Psi_J$ is $n$.}
\begin{eqnarray}
\label{ode:eq1}
\partial_t \vPsi_J (t,\mathbf{u}) &=&  \left(\vbeta_{JJ}^Q \right)^{\prime} \vPsi_J(t,\mathbf{u}) - \vgamma_{xJ} ; \nonumber \\
\vPsi_J(0,\mathbf{u}) &=& \mathbf{u}_J \ , \ \  \vgamma_{xJ}=\mathbf{e}_{n \times 1}  .
\end{eqnarray}
\noindent A particular solution of (\ref{ode:eq1}) is of the structure
\begin{eqnarray}
\vPsi_J(t,\mathbf{u}) = \exp \left (t \left(\vbeta_{JJ}^{Q }\right)' \right ) \mathbf{c}_1 + \mathbf{c}_2  \ , \label{part_sol}
\end{eqnarray}
with
\begin{eqnarray}
\vPsi_J(0,\mathbf{u}) =\mathbf{c}_1+\mathbf{c}_2 =\mathbf{u}_J   . \label{initial_cond}
\end{eqnarray}
Then (\ref{part_sol}) implies
\begin{eqnarray}
\partial_t \vPsi_J (t,\mathbf{u}) &=& \left(\vbeta_{JJ}^{Q }\right)' \exp \left (t \left(\vbeta_{JJ}^{Q }\right)' \right ) \mathbf{c}_1 .
 \label{initial_solJ1}
\end{eqnarray}
\noindent
Plugging (\ref{part_sol}) and (\ref{initial_solJ1}) into (\ref{ode:eq1}) yields
$$
\left(\vbeta_{JJ}^{Q }\right)' \exp \left (t \left(\vbeta_{JJ}^{Q }\right)' \right ) \mathbf{c}_1 = \left(\vbeta_{JJ}^{Q }\right)' \left ( \exp \left (t \left(\vbeta_{JJ}^{Q }\right)' \right ) \mathbf{c}_1 + \mathbf{c}_2 \right ) - \vgamma_{xJ} \ ,
$$
which gives $\vgamma_{xJ}=\left(\vbeta_{JJ}^{Q }\right)' \mathbf{c}_2$ and thus $\mathbf{c}_2= \left ( \left(\vbeta_{JJ}^{Q }\right)' \right )^{-1} \vgamma_{xJ}$. This and (\ref{initial_cond}) imply that $\mathbf{c}_1=\mathbf{u}_J-\left ( \left(\vbeta_{JJ}^{Q }\right)' \right )^{-1} \vgamma_{xJ}$. Plugging the last expression and $\mathbf{c}_2$ into (\ref{part_sol}) gives%
{\footnote{Equation~(\ref{ode:eq5}) also follows from \citet{Perko1991}[Theorem 1, p.~60]. The matrix product in the last expression of (\ref{ode:eq5}) can be exchanged by the properties of the matrix exponential. I.e. $(\vbeta_{JJ}^{Q'})^{-1}\exp( t \vbeta_{JJ}^{Q'}) = \exp(t \vbeta_{JJ}^{Q'})(\vbeta_{JJ}^{Q'})^{-1}$ follows from $\exp( \mathbf{Y} \mathbf{X} \mathbf{Y}^{-1}) =  \mathbf{Y} \exp(\mathbf{X}) \mathbf{Y}^{-1}$.}}
\begin{eqnarray}
\vPsi_J (t,\mathbf{u}) &=& \exp \left (t \left(\vbeta_{JJ}^{Q }\right)' \right ) \mathbf{u}_J - \left ( \exp \left (t \left(\vbeta_{JJ}^{Q }\right)' \right ) - \mathbf{I}_n \right ) \left ( \left(\vbeta_{JJ}^{Q }\right)' \right )^{-1} \vgamma_{xJ} \nonumber \\
&=& \exp \left (t \left(\vbeta_{JJ}^{Q }\right)' \right ) \mathbf{u}_J - \left ( \left(\vbeta_{JJ}^{Q }\right)' \right )^{-1} \left ( \exp \left (t \left(\vbeta_{JJ}^{Q }\right)' \right ) - \mathbf{I}_n \right ) \vgamma_{xJ}  .
\label{ode:eq5}
\end{eqnarray}
In a second step the solution of the subsystem $\vPsi_J (t,\mathbf{u})$ is plugged in into the ODEs for the square root terms $\vPsi_I$.
Thus, the Riccati equations for the $I$ components are
\begin{eqnarray}
\label{ode:eq7}
\partial_t \Psi_i (t,\mathbf{u}) &=& \frac{1}{2} \Sigma_i^2 \Psi_i^2 (t,\mathbf{u})  + \beta_{ii}^Q  \Psi_i (t,\mathbf{u}) - \tilde \gamma_{xi} \   \nonumber \\
&& \tilde \gamma_{xi }(t,\mathbf{u}) = \gamma_{xi} - \sum_{j=1}^{n} \beta^{Q }_{m+j,i}  (\vPsi_J (t,\mathbf{u}))_j - \frac{1}{2} \sum_{j=1}^{n} \Sigma^2_{m+j} \mathcal{B}^x_{i,m+j} \left[ \vPsi_J(t,\mathbf{u})) \right]_j^2 \ , \nonumber \\
\Psi_i(0,\mathbf{u}) &=& u_i \ ,  \ \ \gamma_{xi}=1 \ , i=1, \dots, m.
\end{eqnarray}
\noindent As (\ref{ode:eq7}) is a time inhomogeneous Riccati equation, it can be solved in the following way: The ODE of interest is $\partial_t \Psi_i = \frac{1}{2} \Sigma_i^2 \Psi_i^2 + \beta_{ii}^Q \Psi_i - \tilde{\gamma}_{xi}$ for $i=1,\dots,m$. After the substitution $\nu_i=\Sigma_i^2 \Psi_i$, $i=1, \ldots, m,$ we get $\partial_t \nu_i = \frac{1}{2} \nu_i^2 + \beta_{ii}^Q \nu_i - \Sigma_i^2 \tilde{\gamma}_{xi}$. A solution for an inhomogenous Riccati ODE of this structure is provided in \citet{GrasselliTebaldi2008}[Section~3.4.1]. The solution for
 $\nu_i$ is
\begin{eqnarray}
\nu_i (t,\mathbf{u}) &=& \frac{M_1^{(i)} (t,\mathbf{u}) \, u_i + M_2^{(i)}(t,\mathbf{u}) }{M_3^{(i)}(t,\mathbf{u}) \, u_i + M_4^{(i)}(t,\mathbf{u})  } \ , \ \text{ where }  \nonumber \\
&&
\mathbf{M}^{(i)}(t,\mathbf{u})=
\left(
\begin{array}{cc}
M_1^{(i)}(t,\mathbf{u}) & M_2^{(i)}(t,\mathbf{u}) \\
M_3^{(i)}(t,\mathbf{u}) & M_4^{(i)}(t,\mathbf{u}) \\
\end{array}
\right) = \exp \left(
\begin{array}{cc}
t \beta^Q_{ii} & - \Sigma_{i}^2 \int_0^t \tilde \gamma_{xi} (s,\mathbf{u}) ds \\
-t/2 & 0 \\
\end{array}
\right)   .
 \end{eqnarray}
\noindent
At the end $\Psi_i=\frac{\nu_i}{\Sigma_i^2}$ for $i=1,\dots,m$. With $\mathbf{u}=\mathbf{0}_{d \times 1}$ we get
\begin{eqnarray}
\label{ode:eq9}
\Psi_i (t,\mathbf{0}) &=& \frac{1}{\Sigma_{i}^2} \, \frac{M_2^{(i)}(t,\mathbf{0}) }{M_4^{(i)}(t,\mathbf{0})  } \ , \ \text{ for $i=1,\dots,m$}  .
\end{eqnarray}

\noindent
To derive $\mathbf{M}^{(i)}(t,\mathbf{u})$ the integral $\int_0^t \tilde \gamma_{xi}(s,\mathbf{u}) ds$ has to be solved, where
\begin{eqnarray}
\label{ode:eq11}
\int_0^t \tilde \gamma_{xi }(s,\mathbf{u}) ds
&=& \gamma_{xi} t - \sum_{j=1}^{n}  \beta^{Q }_{m+j,i}  \int_0^t \left[ \vPsi_J (s,u) \right]_j ds - \frac{1}{2} \sum_{j=1}^{n} \Sigma^2_{m+j} \mathcal{B}^x_{i,m+j} \int_0^t \left[\vPsi^2_J(t,\mathbf{u}) \right]_{j}  ds  .
\end{eqnarray}
\noindent
The second term in (\ref{ode:eq11}) can be derived by means of
\begin{eqnarray}
\label{ode:eq13}
\int_0^t \vPsi_J (s,0) ds &=& -\int_0^t  \left[
\left( \left( \vbeta_{JJ}^{Q}\right)' \right)^{-1}  \left( \exp \left( s \left( \vbeta_{JJ}^{Q}\right)' \right) - \mathbf{I}_n \right) \vgamma_{xJ} \right] ds \,
\nonumber \\
&=& %\left( \vbeta_{JJ}^{Q \prime} \right)^{-1}  \left( \exp \left( t \vbeta_{JJ}^{Q \prime} \right) - I_n \right) u_J
-\left( \left( \vbeta_{JJ}^{Q}\right)' \right)^{-1} \left[ \left( \left( \vbeta_{JJ}^{Q}\right)' \right)^{-1}  \left( \exp \left( t \left( \vbeta_{JJ}^{Q}\right)' \right) - \mathbf{I}_n \right)  - t \mathbf{I}_n  \right] \gamma_{xJ}
\end{eqnarray}
using (\ref{ode:eq5}). The third term in (\ref{ode:eq11}) can be derived numerically as well as the whole expression (\ref{ode:eq11}).
It remains to calculate $\Phi(t, \mathbf{0})$, where by (\ref{transformeddk1})
\begin{eqnarray}
\label{transformeddk11}
\partial_t \Phi(t,\mathbf{u}) &=& \frac{1}{2} \vPsi(t,\mathbf{u})^{\prime}  \mathbf{a} \vPsi(t,\mathbf{u}) + \left(\mathbf{b}^Q\right)' \vPsi(t,\mathbf{u}) \ ,  \qquad \Phi(0,\mathbf{u})=0 \ ,  \nonumber \\
&=& \frac{1}{2}diag(\mathbf{\Sigma}^2_J )\vPsi_J(t,\mathbf{u})^2 + \left(\mathbf{b}^Q\right)' \vPsi(t,\mathbf{u})  . \nonumber
\end{eqnarray}
\noindent We can express $\Phi(t,\mathbf{0})$ by means of $\Phi(t,\mathbf{0})=\Phi_I(t,\mathbf{0})+\Phi_J(t,\mathbf{0})$. The $J$ components of the $d \times d$ matrix $ \mathbf{a}$ are equal to a $n \times n$ diagonal matrix having $\Sigma_{m+1}^2,\dots,\Sigma_{d}^2$ along the main diagonal such that
\begin{eqnarray}
\label{transformeddk13}
\Phi_J(t,\mathbf{0}) &=& \frac{1}{2} \int_0^t \vPsi_J (s,\mathbf{0})^{\prime} \left(
\begin{array}{ccc}
\Sigma_{m+1}^2 & 0 & 0 \\
0&\ddots &0 \\
0& 0 & \Sigma_{d}^2
\end{array}
\right)
 \vPsi_J (s ,\mathbf{0}) \ ds + \int_0^t (b_{m+1}^Q, \dots ,b_{d}^Q)^{} \vPsi_J(s,\mathbf{0}) \ ds    . \nonumber
\end{eqnarray}
\noindent
For $\Phi_I(t,\mathbf{0})$ we obtain
\begin{eqnarray}
\label{transformeddk15}
\Phi_I(t,\mathbf{0}) &=&  \left(\mathbf{b}_I^Q \right)^{\prime} \int_{0}^{t}  \vPsi_I(s,\mathbf{0}) \ ds  .
\end{eqnarray}
\noindent
$\Phi_I(t,\mathbf{0})$, $\Phi_J(t,\mathbf{0})$ and $\vPsi(t,\mathbf{0})$ can be easily obtained by means of numerical integration. To do this we generate a grid $\Gamma = \{t_0,t_1,\dots,t_G \}$ with $G+1$ grid points. We set $t_0=0$ and $t_G = \max(\tau_l)=\tau_{M}$. By including the maturities $\tau_l$, $l=1,\dots,M$, in $\Gamma$ we know that for each maturity  we have $t_{k_l} = \tau_l$ for some $k_l\in \{1,\dots,G+1\}$.\footnote{This has been implemented as follows: (i) generate an equally spaced grid, (ii) include the $M$ maturities, (iii) sort all these points in ascending order.} The step-widths are given by $\Delta_k=t_k-t_{k-1}$, $k=1,\dots,G+1$ (if some elements of $\Gamma$ coincide with $\tau_l$ this does not cause any problems since $\Delta_k=0$ for such grid-points). Then we evaluate $\Phi_J(t,\mathbf{0})$ at each $t=t_k$, $k=1,\dots,G+1$. By calculating the sums
$\frac{1}{2} \sum_{k=1}^{k_l-1} \vPsi_J (t_k,\mathbf{0})^{\prime} \left(
\begin{array}{cc}
\Sigma_{m+1}^2 & 0 \\
0 & \Sigma_{d}^2
\end{array}
\right)
 \vPsi_J (t_k ,\mathbf{0})  \Delta_k + \sum_{k=1}^{k_l-1} (b_{m+1}^Q,\dots,b_{d}^Q)^{} \vPsi_J(t_k,\mathbf{0}) \Delta_k$ (left Riemann sums),
$\frac{1}{2} \sum_{k=2}^{k_l} \vPsi_J (t_{k},\mathbf{0})^{\prime} \left(
\begin{array}{cc}
\Sigma_{m+1}^2 & 0 \\
0 & \Sigma_{d}^2
\end{array}
\right)
 \vPsi_J (t_{k} ,0)  \Delta_k + \sum_{k=2}^{k_l} (b_{m+1}^Q,\dots,b_{d}^Q)^{} \vPsi_J(t_{k},\mathbf{0}) \Delta_k$ (right Riemann sums),
 or
 $\frac{1}{2} \sum_{k=2}^{k_l} \frac{\vPsi_J (t_{k-1},\mathbf{0})^{\prime}+\vPsi_J (t_{k},\mathbf{0})^{\prime}}{2} \left(
\begin{array}{cc}
\Sigma_{m+1}^2 & 0 \\
0 & \Sigma_{d}^2
\end{array}
\right)
 \frac{\vPsi_J (t_{k-1},\mathbf{0})^{\prime}+\vPsi_J (t_{k},\mathbf{0})^{\prime}}{2}  \Delta_k + \sum_{k=2}^{k_l} (b_{m+1}^Q,\dots,b_{d}^Q) \frac{\vPsi_J (t_{k-1},\mathbf{0})^{\prime}+\vPsi_J (t_{k},\mathbf{0})^{\prime}}{2} \Delta_k$ (trapeze-rule) we get a numerical approximation of $\Phi_J(\tau_l,\mathbf{0})$, $k_l=G(+1)$ for $\tau_l=\tau_M$.  In our code right sums were implemented.
 Since integrals of
$\vPsi_J$ and $\vPsi_J^2$ are necessary to obtain $\int_{0}^{\tau_l} \tilde \gamma_{xi }(t,\mathbf{0}) dt$, we use numerical integration also to obtain $\int_{0}^{\tau_l} \tilde \gamma_{xi }(t,\mathbf{0}) dt$. These proxies are then used in
(\ref{ode:eq9}) to calculate $\Psi_i (\tau_l,\mathbf{0})$, $i=1,\dots,m$. Equipped with $\vPsi_I(t_k,\mathbf{0})$, $k=1,\dots,G+1$ we are also able to obtain a numerical approximation of $\Phi_I (\tau_l,\mathbf{0})$.

\section{Restrictions on the Parameters}
\label{appa:restrict1}

First we present the conditions for admissibility which guarantee that $(\mathbf{X}(t))$ remains with in the state space $\mathscr{S}$. All these restriction are applied in both measures, $\Pmeas$ and $\Qmeas$, respectively.

{\em Admissibility conditions} \citep[see][Theorem~10.2]{filipovicbook2009}: $\mathbf{a}$, $\valpha_i$ are symmetric and positive semidefinite. $\mathbf{a}_{II}=\mathbf{0}_{m \times m}$,
$\mathbf{a}_{IJ}= \mathbf{a}_{JI}'=\mathbf{0}_{m \times n}$, $\valpha_j=\mathbf{0}_{n \times n}$ for all $j=m+1,\dots,m+n$.
$\alpha_{i,kl}=\alpha_{i,lk}=0$ for $k \in I \setminus \{ i \}$ for all $1 \leq i,l \leq d$, $\mathbf{b}^{\cdot} \in \mathscr{S}$, $\vbeta_{IJ}^{\cdot} =\mathbf{0}_{m \times n}$ and
$\vbeta_{II}^{\cdot}$ has non-negative off-diagonal elements. In a model with diagonal diffusion matrix the admissibility restrictions are met
if the \citet{daisingleton00} conditions presented in Definition~\ref{def5} are met. To keep the process $(\mathbf{X}(t))$ off the boundaries of the state space $\mathscr{S}$ we can impose the
{\em Boundary conditions/Feller conditions} \citep[see][Eq. 15-17]{aitsahakiakimmel2009}:
$b_i^{\cdot} \geq \frac{1}{2} \Sigma_{i}^2$ for $i=1,\dots,m$. (The conditions $\vbeta_{IJ}^{\cdot} =\mathbf{0}_{m \times n}$ and
$\vbeta_{II}^{\cdot}$ having non-negative off-diagonal elements are already included in the admissibility conditions.)

Last but not least, we have some further restrictions for stationarity:

\noindent
{\em Stationarity conditions} \citep[see][Table~1]{aitsahakiakimmel2009}: The real part of the eigenvalues of $\vbeta$ is smaller than zero. A more general treatment regarding stationarity is provided in \citet{GlassermanKim2009}.

\section{$GMM$-Estimation}
\label{sect:gmmest1}
For our model it turned out that minimizing the $GMM$ distance function (\ref{def:Qtgmm1}) is non-trivial.
By using a standard minimization routine, as the MATLAB minimization routine $\mathtt{fminsearch}$ based on
the Nelder-Mead algorithm,\footnote{See $\mathtt{http://www.mathworks.de/de/help/matlab/ref/fminsearch.html}$} we observed that the estimation procedure preforms poorly.%
{\footnote{Detailed results of these simulation experiments can be obtained from the authors on request.}} %
Therefore, as being described in {\tt Step~1} below, we include multistart random search methods in our minimization procedure \citep[see, e.g.,][]{ToernZilinskas1989}. Compared to working with the above minimization routine only, this procedure improves parameter estimation, especially when looking at the means and the absolute deviation from the mean in percentage terms. Some results are presented in Table~\ref{tab:res1}.

In addition, we apply classical tests such as the {\em Wald} and the {\em distance difference} test \citep[see, e.g.,][]{Ruudbook2000,neweyMcfadden94}.
We observe that these tests do not perform well. Some results for tests of the null hypothesis $\theta^P = \theta^Q$ against the alternative $\theta^P \not= \theta^Q$ are presented in Table~\ref{tab:res3}, where it can be seen that power and size of these tests do not fulfill ``the usual quality standards". We explain this behavior by the problem of estimating a relatively large $(23 \times 23)$ covariance matrix and a matrix of gradients with the Wald test (see also equation~(\ref{eq:VNstand})). Regarding the distance difference test, we observe that the $(\mathfrak{q}\times \mathfrak{q})$ weighting matrix $\mathbf{C}_T= \hat{\vLambda}^{-1}$ has a strong impact on the results of the tests, which in turn introduces potential inaccuracies in case  $\hat{\vLambda}$ was not estimated accurately enough.

\begin{table}[h]
\begin{center}
\begin{tabular}{cr|rrrrrrrr}%
\hline \hline
\multicolumn{2}{c|}{$\vvartheta$}	&	$mean$	&	$median$	&	$min$	&	$max$	&	$std$	& $skew$	&	$kurt$ &	$|\vvartheta- \widehat{\vvartheta}|$		 \\
&& \multicolumn{1}{|c}{$\widehat{\vvartheta}$}  &&&&&&& \\
\hline
$\theta^Q$	&	10	&	10.3593	&	8.9022	&	1.0527	&	69.9629	&	6.0544	&	3.4352	&	23.6299	&	0.3593		 \\
$\theta^P$	&	1.5	&	1.5046	&	1.2986	&	0.0676	&	6.4437	&	1.0296	&	1.3471	&	5.1909	&	0.0046		 \\
$\beta_Q^{11}$	&	-1	&	-1.2823	&	-1.0328	&	-7.6430	&	-0.1108	&	0.9617	&	-2.0173	&	9.4853	&	0.2823		 \\
$\beta^Q_{21}$	&	0.2	&	0.2523	&	0.1729	&	0.0099	&	2.8282	&	0.2549	&	3.2707	&	22.3280	&	0.0523		 \\
$\beta^Q_{31}$	&	0.02	&	0.0326	&	0.0204	&	0.0009	&	0.5962	&	0.0416	&	5.2132	&	49.7281	&	0.0126		 \\
$ \beta^Q_{22}$	&	-1	&	-1.5493	&	-1.4686	&	-4.2679	&	-0.1046	&	0.7283	&	-0.5842	&	3.2132	&	0.5493		 \\
$\beta_Q^{32}$	&	0.04	&	0.0375	&	0.0354	&	-0.0734	&	0.1586	&	0.0404	&	0.1497	&	2.8928	&	0.0025		 \\
$\beta^Q_{23}$	&	0	&	-0.0005	&	-0.0002	&	-0.0343	&	0.0303	&	0.0097	&	0.0013	&	3.1280	&	0.0005		 \\
$\beta^Q_{33}$	&	-1	&	-1.5042	&	-1.4266	&	-4.7165	&	-0.0664	&	0.7906	&	-0.5289	&	3.0549	&	0.5042		 \\
$\beta_P^{11}$	&	-0.8	&	-1.6204	&	-0.8868	&	-43.8618	&	-0.0503	&	2.9373	&	-8.3139	&	99.3434	&	0.8204		 \\
$\beta^P_{21}$	&	0.02	&	0.0330	&	0.0210	&	0.0013	&	0.3927	&	0.0378	&	3.0352	&	17.4456	&	0.0130		 \\
$\beta^P_{31}$	&	0.01	&	0.0168	&	0.0102	&	0.0004	&	0.2022	&	0.0212	&	4.1593	&	27.5666	&	0.0068		 \\
$ \beta^P_{22}$	&	-0.7	&	-0.9193	&	-0.8646	&	-3.1251	&	0.2598	&	0.5646	&	-0.5446	&	2.8395	&	0.2193		 \\
$\beta_P^{32}$	&	0.01	&	0.0094	&	0.0094	&	-0.0182	&	0.0433	&	0.0099	&	-0.5446	&	2.8395	&	0.0006		 \\
$\beta^P_{23}$	&	0	&	0.0000	&	-0.0003	&	-0.0316	&	0.0305	&	0.0099	&	0.0824	&	2.8274	&	0.0000		 \\
$\beta^P_{33}$	&	-0.7	&	-0.9199	&	-0.8383	&	-3.0770	&	0.2518	&	0.5418	&	-0.5914	&	3.0650	&	0.2199		 \\
%$\mathbf{b}^Q_{1}$	&	10	&	11.4397	&	9.6903	&	0.5587	&	67.3939	&	7.7110	&	1.7613	&	8.4956	&	1.4397		 \\
%$\mathbf{b}^Q_{2}$	&	-2	&	-2.6418	&	-1.6222	&	-48.5571	&	-0.0572	&	3.5035	&	-5.3069	&	49.5371	&	0.6418		 \\
%$\mathbf{b}^Q_{3}$	&	-0.2	&	-0.3613	&	-0.1807	&	-16.2267	&	-0.0073	&	0.7110	&	-12.5923	&	255.4394	&	0.1613		 \\
%$\mathbf{b}^P_{1}$	&	1.2	&	2.4559	&	1.0640	&	0.0502	&	187.4110	&	7.6919	&	17.1564	&	374.3937	&	1.2559		 \\
%$\mathbf{b}^P_{2}$	&	-0.03	&	-0.0472	&	-0.0242	&	-0.7205	&	-0.0004	&	0.0711	&	-4.5160	&	31.4344	&	0.0172		 \\
%$\mathbf{b}^P_{3}$	&	-0.015	&	-0.0256	&	-0.0127	&	-0.3595	&	-0.0001	&	0.0386	&	-3.9035	&	23.3545	&	0.0106		 \\
$\mathcal{B}^x_{12}$	&	0.05	&	0.0791	&	0.0496	&	0.0029	&	1.2802	&	0.0964	&	4.3127	&	35.8452	&	0.0291		 \\
$\mathcal{B}^x_{13}$	&	0.1	&	0.1590	&	0.0978	&	0.0025	&	2.1414	&	0.1969	&	4.3398	&	32.1066	&	0.0590		 \\
$\gamma_0$	&	2	&	2.1224	&	2.1483	&	-4.1375	&	6.2455	&	1.6726	&	-0.2254	&	2.9938	&	0.1224		 \\
$\Sigma_{1}$	&	0.7	&	0.5450	&	0.4636	&	0.0176	&	2.7764	&	0.3771	&	1.6212	&	7.1421	&	0.1550		 \\
$\Sigma_{2}$	&	1	&	1.0037	&	0.7162	&	0.0267	&	5.8461	&	0.8681	&	1.9397	&	7.7289	&	0.0037		 \\
$\Sigma_{3}$	&	0.8	&	0.8538	&	0.6104	&	0.0253	&	8.2162	&	0.8849	&	3.6164	&	22.8961	&	0.0538		 \\
$\sigma_{\varepsilon}^2$	&	0.0067	&	0.0119	&	0.0068	&	0.0003	&	0.2908	&	0.0180	&	7.8621	&	102.5576	&	0.0051		 \\
\hline																	
\hline												
\end{tabular}
\caption{Parameter estimates for the $\mathbb{A}_{1}(3)$. Data simulated with $M=10$ and $T=500$. Estimation based on
using $\mathtt{fminsearch}$. $c_{\vartheta}=1$ is controlling for the noise in the generation of the starting value of the optimization routine.  Statistics are obtained from $1,000$ simulation runs. $mean$, $median$, $min$, $max$, $std$, $skew$ and $kurt$ stand for the sample mean, median, minimum, maximum, standard deviation, skewness and kurtosis of the point estimates $\widehat{\vvartheta}_{\ell}$, $\ell=1,\dots,1,000$. $|\vvartheta- \widehat{\vvartheta}|$ stands for absolute value of the mean deviation from the true parameter. The true parameter values  $\vvartheta$ are reported in the second column.}
\label{tab:res1}
\end{center}
\end{table}

\begin{table}[h]
\begin{center}
\begin{tabular}{c|cc|cc}%
\hline \hline
& \multicolumn{2}{c|}{$\theta^Q=10 \not=1.5= \theta^P$} & \multicolumn{2}{c}{$\theta^Q=\theta^P=1.5$}\\
\hline
$\alpha_{S}$	&	Wald	&	DD &	Wald	&	DD	\\
\hline
0.01	&	0.018	&	0.545	&	0.015	&	0.057		\\
0.05	&	0.028	&	0.583	&	0.021	&	0.062		\\
0.10	&	0.043	&	0.623	&	0.025	&	0.065		\\
\hline																	
\hline												
\end{tabular}
\caption{Parameter tests: Data are simulated with $M=10$, $T=500$ and $c_{\vartheta} =1$. $\left[ \vartheta \right]_1=\theta^Q$ and $\left[ \vartheta \right]_2=\theta^P$ and the remaining elements of $\vvartheta$ are equal to those of the second column in Table~\ref{tab:res1}.  $\alpha_{S}$ stands for the significance level. $c_{\vartheta} $ controls for the noise in the generation of the starting value of the optimization routine. The null hypothesis is $\theta^Q=\theta^P$ against the two sided alternative $\theta^Q \not= \theta^P$. The parameters $\vartheta$ estimated by combining multistart random search methods and a standard minimization procedure. The Wald test as well as the distance difference test (DD) are implemented as described in Chapter~22 in \citet{Ruudbook2000}.
Equation (\ref{eq:VNstand}) is used to estimate the asymptotic variance of $\sqrt{T} \left(\widehat{\vvartheta} - \vvartheta \right)$ with the Wald test,
while $\hat{\mathbf{\Lambda}}_T$, as presented in (\ref{eq:VNstand}), is used with the distance difference test.
The numbers in the table are rejection rates of the null hypothesis given the significance level $\alpha_{S}$, when using a Wald test and a distance difference test. Statistics are obtained from $1,000$ simulation runs. }
\label{tab:res3}
\end{center}
\end{table}

To further improve the properties of the estimation routine, we combine multistart random search methods with Quasi-Bayesian methods
\citep[see][]{chernozhukovhong03}. To apply Bayesian tools a prior $\tilde \pi(\vvartheta)$ has to be specified. The parameter space $\Theta$ is a subset of $\mathbb{R}^{\mathfrak{p}}$. It is a proper subset, since some parameters are strictly positive, nonnegative, etc. by the model assumptions. In addition, admissibility and stationarity further restrict the parameter space.
Hence, the prior $\tilde \pi(\vvartheta)=0$ for all $\vvartheta \not \in \Theta$.
In addition, to implement a random search method on a computer and to add ``prior information'' we restrict $\Theta$ to $\Theta_0 \subset \Theta$, where
$\tilde \pi(\vvartheta)=0$ for all $\vvartheta$ not contained in $\Theta_0$.

The subset $\Theta_0$ is constructed as follows:
For $\Sigma_{i}$ the lower bound is set to $0.1$, while the upper bound is set to $2$. The upper bound follows from variances of the yields observed, the lower bound from the assumption that the variance of each component is not too small. For the unrestricted $\mathcal{B}_{ij}^{{x}}$ we assume that $\mathcal{B}_{ij}^{{x}} \in [0,2]$, where $\mathcal{B}_{ij}^{{x}} \geq 0$ follows from the models assumptions, while $\mathcal{B}_{ij}^{{x}} \leq 2$ is used to keep the impact of the square root term on the other volatilities bounded.
In addition, $\sigma^2_{\varepsilon} \in [0.005,0.025]$. This is motivated by the argument that the observation error is small compared to the variance of the yields. The observation error can be due to  market-microstructure noise \citep[see, e.g.,][]{campbelllomackinlay97,chenetal2007}. The lower bound is based on the assumption that at least 10 basis points can be attributed to {the} noise.  To ensure that the matrices $\vbeta^Q$ and $\vbeta^P$ {are} sufficiently far away from a singular matrix, we assume $\beta_{ii}^{} \leq -0.1$. To cope with the high degree of serial correlation of the yields, we demand for $\beta_{ii}^{} \geq -50$. For $\beta_{ij}$, $i \not= j$ we apply a lower bound of $-10$ and an upper bound of $10$. The differences in the matrix exponential of $\vbeta$ become small, when values outside these intervals are used.

Since $\theta^P$ and $\gamma_0$ determine the mean of the instantaneous spot rate $\mathbb{E}(r_{\mathtt{t}}) = \gamma_0 + \theta^P$ defined by a stationary $(\mathbf{X}(t))$ (see equation~(\ref{eq:short1})), we assume that
$\frac{1}{c} \left[\tilde{\mathbf{m}}_T  (\mathbf{y}_{1:T}) \right]_1 \leq \gamma_0 + \theta^P \leq c  \left[\tilde{\mathbf{m}}_T (\mathbf{y}_{1:T}) \right]_1$, where $c=1.45$ is applied in the Bayesian sampler.
%\footnote{In a Bayesian framework the data used in the estimation should not be part of the prior.}
Since the sample mean of the instantaneous short rate cannot be observed, we use the sample mean of the shortest maturity, which in terms of our notation is $\left[ \tilde{\mathbf{m}}_T (\mathbf{y}_{1:T})\right]_{1}$.

In addition, the conditions on stationarity, identification and admissibility have to be met. Given these restrictions and the uniform prior on the components of $\Theta_0$, the prior $\tilde \pi(\vvartheta)$ is proportional to
$\mathbb{I}_{(\text{Stationarity,Identification,Admissibility})}$$\mathbb{I}_{\left(\frac{1}{c} \left[\tilde{\mathbf{m}}_T  (\mathbf{y}_{1:T})\right]_{1} \leq \gamma_0+\theta^P   \leq  c \left[\tilde{\mathbf{m}}_T (\mathbf{y}_{1:T})\right]_{1}\right)}$, where the term $\mathbb{I}_{(\cdot)}$ stands for an indicator function.
Summing up, all the above restrictions result in the set $\Theta_0$. For all elements $\vvartheta$ contained in $\Theta_0$ we use a uniform prior and for
all $\vvartheta \not \in \Theta_0$ we set $\tilde \pi(\vvartheta)=0$.

After the prior has been specified, parameter estimates are obtained as follows.

 \renewcommand{\arraystretch}{1.4}
\begin{tabular}{c l l }
\hline \hline
%{\tt Step 0}: & Choose $ \Theta_0 \subset \Theta$ and a prior $\tilde \pi(\vvartheta)$.  \\
%Perform {\tt Step 1} and {\tt Step 2} $\mathtt{L}-$times ($\mathtt{L}=200$)
{\tt Step 1}: & Run multistart random search methods, generate $\vvartheta^{(\mathtt{n})}$, where $\mathtt{n}=1,\dots, \mathtt{N}=2,000$. \\
{\tt Step 2}: & Run $MCMC$: \\[3pt]
& For each $MCMC$-step $\mathtt{m}$, where $\mathtt{m}=1,\dots,\mathtt{M}=20,000$, \\
& update $\vvartheta^{(\mathtt{m})}$ block-wise by means of the Metropolis-Hastings algorithm: \\
& $MCMC$ {\tt Sub-Step 1}: update block $\mathbb{J}_1$ \\
& $\vdots$ &  \\
& $MCMC$ {\tt Sub-Step $\mathtt{K}$}:  update  block $\mathbb{J}_{\mathtt{K}}$ \\
& $MCMC$ {\tt Sub-Step}: reversible jump step (with a probability of 90\%). \\[1pt]
\hline
& Obtain an estimate $\widehat{\vvartheta}$ from the draws  $\vvartheta^{(\mathtt{m})}$, where $\mathtt{m}=\mathtt{M}_b+1 ,\dots, \mathtt{M}=20,000$.\\ [1pt]
\hline \hline
\end{tabular}
\renewcommand{\arraystretch}{1.0}
\vskip3mm

Ad \underline{\tt Step 1}:
Given the set $\Theta_0$, we randomly generate initial points $\vvartheta^{(\mathtt{n})}$, $\mathtt{n}=1,\dots,\mathtt{N}{=2,000}$, which are independently drawn by means of $[\vvartheta^{(\mathtt{n})}]_j = [\vvartheta]_j + c_{\vartheta} [|\vvartheta|]_j \varepsilon_j$ for elements $j$, $j \in \{ 1, \ldots, \mathfrak{p} \}$, when the support is the real axis and $\log [|\vvartheta^{(\mathtt{n})}|]_j = \log [|\vvartheta|]_j + c_{\vartheta} \varepsilon_j$ such that $ [\vvartheta^{(\mathtt{n})}]_j = \exp \left( \log \left( [|\vvartheta|]_j \right) + c_{\vartheta} \varepsilon_j \right)$$\text{sgn} \left( [\vvartheta]_j \right)$ for elements $j$ from non-positive or non-negative part of the real axis. The random variables $\varepsilon_j$ are $iid$ standard normal and only $\vvartheta^{(\mathtt{n})}$ with $\tilde \pi \left( \vvartheta^{(\mathtt{n})} \right)>0$ are used.
In addition, as already stated in Section~\ref{sect:mc1}, our random search routine also generates samples, where $\left(\theta^P\right)^{(\mathtt{n})} =  \left(\theta^Q \right)^{(\mathtt{n})}$.
This is done by setting $\left(\theta^P\right)^{(\mathtt{n})}$ equal to the sampled $\left(\theta^Q \right)^{(\mathtt{n})}$ with a probability of 80\%. By sorting $\vvartheta^{(\mathtt{n})}$ according to ${Q}_{{T}}(\vvartheta^{(\mathtt{n})}; \mathbf{y}_{1:T})$ in ascending order, we are equipped with the sorted draws $\vvartheta_{[\mathtt{j}]}$  and distances $Q_{{T}}\left(\vvartheta_{[\mathtt{j}]} ; \mathbf{y}_{1:T}\right)$, where ${Q}_{T}\left(\vvartheta_{[\mathtt{1}]} ; \mathbf{y}_{1:T} \right)\leq Q_{T}\left(\vvartheta_{[\mathtt{2}]}; \mathbf{y}_{1:T} \right)\leq \dots \leq Q_{T}\left(\vvartheta_{[\mathtt{N}]} ; \mathbf{y}_{1:T}\right)$. The $GMM$ distance function    
${Q}_{{T}} \left( \vvartheta ; \mathbf{y}_{1:T} \right)$ is defined in (\ref{def:Qtgmm1}), where
$\mathbf{C}_T  = \mathbf{I}_{\mathfrak{q}}$ for all $\mathtt{n}=1,\dots,\mathtt{N}$.

Ad \underline{\tt Step 2}:
Based on the results in \citet{chernozhukovhong03}, the Metropolis-Hastings algorithm \citep[see, e.g.,][]{robertcasella04} can be used to minimize the $CUE-GMM$ criterion function $Q_T(\cdot)$. To do this we proceed as follows: Suppose that $\vvartheta^{(\mathtt{m}-1)}$ is available, where, just now, $\mathtt{m}$ stands for the index of the $MCMC$ step.
For $\mathtt{m}=1$ we start the Bayesian sampler at $\vvartheta_{[\mathtt{1}]}$, that is $\vvartheta^{(\mathtt{0})} $$=\vvartheta_{[\mathtt{1}]}$.

The parameter vector to be updated, $\vvartheta^{(\mathtt{m}-1)}$, is of dimension $\mathfrak{p}$, where the index set
 $\{1,\dots , \mathfrak{p} \}$ is covered by the blocks $\mathbb{J}_{\mathtt{k}} \subset \{1,\dots,\mathfrak{p} \}$, $\mathtt{k}=1,\dots,\mathtt{K}=5$.  The first block $\mathbb{J}_1$ consists of the first two parameters and the 19th parameter, which is $\gamma_0$, $\mathbb{J}_2=\{3,\dots,9\}$,
the third block $\mathbb{J}_{3}=\{10,\dots,15\}$, while $\mathbb{J}_4=\{16,17,18\}$. Finally, the fifth block $\mathbb{J}_5$ contains the volatility parameters. For the parameter odering see first column of Table~\ref{tab:res4}.

Within updating step $\mathtt{m}$, we consider the sub-steps $\mathtt{k}=1,\dots, \mathtt{K}$, where
$\vvartheta^{(\mathtt{m},\mathtt{k})}$ stands for the parameter vector in $MCMC$-step $\mathtt{m}$ at sub-step $\mathtt{k}$.
Let $\vvartheta^{old} = \vvartheta^{(\mathtt{m}-1)}= \vvartheta^{(\mathtt{m}-1,\mathtt{K})}$ for $\mathtt{k}=1$ and
$\vvartheta^{old} = \vvartheta^{(\mathtt{m},{\mathtt{k}-1})}$ for $\mathtt{k}=2,\dots,\mathtt{K}$.%
\footnote{The index of the sub-step $\mathtt{k}$ is not applied, when it is not essential.}
When the block $\mathbb{J}_{\mathtt{k}}$ is considered, $\left[ \vvartheta^{old} \right]_{i}$, $i \in \mathbb{J}_{\mathtt{k}}$, is updated.
To update $\left[ \vvartheta^{(\mathtt{m},{\mathtt{k}-1})} \right]_{i}$, $i \in \mathbb{J}_{\mathtt{k}}$, a random walk proposal, with {\em proposal density}
$q \left( \left[ \vvartheta^{new} \right]_{i}|\left[ \vvartheta^{old} \right]_{i} \right)$$=
f_{\mathcal{N}\left( \left[ \vvartheta^{old} \right]_{i}, \sigma_{RWi}^2  \right)}  \left( \left[ \vvartheta^{new} \right]_{i} \right)$, is used,
 where $f_{\mathcal{N}(\cdot)}(\cdot)$ stands for a normal density.
In the random walk proposals, we use small standard deviations of the noise in relative terms. In particular, $\sigma_{RWi}  = 0.01 [|\vvartheta^{old} |]_i$, with a probability of 90\%, for remaining 10\% we set the standard deviation of this noise term equal to $\sigma_{RWi} = 0.005 [|\vvartheta^{old} |]_i$. By applying these proposals to all elements $i \in \mathbb{J}_{\mathtt{k}}$, we get the parameter vector  $\left[\vvartheta^{new}\right]_{i}$
and the proposal density
$q \left( \vvartheta^{new }|\vvartheta^{old } \right) = \prod_{i \in \mathbb{J}_k}  q \left( \left[ \vvartheta^{new} \right]_{i}|\left[ \vvartheta^{old} \right]_{i} \right)$.
 For the remaining components $\left[\vvartheta^{new}\right]_{\ell}$$=\left[\vvartheta^{old}\right]_{\ell}$, where $\ell$ is not contained in the block $\mathbb{J}_k$.  Equipped with  ${Q}_{{T}}(\vvartheta^{new}; \mathbf{y}_{1:T})$ and ${Q}_{{T}}(\vvartheta^{old}; \mathbf{y}_{1:T})$, the prior $\tilde \pi(\cdot)$ and the proposal densities $q(\cdot)$, the Metropolis-Hastings algorithm can be used.
Let $\mathscr{L} ( \vvartheta) = \exp \left[ -\frac{1}{2} T {Q}_{{T}} \left( \vvartheta ; \mathbf{y}_{1:T} \right) \right]$. The $GMM$ distance function    
${Q}_{{T}} \left( \vvartheta ; \mathbf{y}_{1:T} \right)$ is defined in (\ref{def:Qtgmm1}), where
$\mathbf{C}_T  = \left(\hat{\vLambda}_T({\vvartheta}^{(\mathtt{m}-1)} )\right)^{-1}$ with $\hat{\vLambda}_T \left( \vvartheta^{(\mathtt{m}-1)} \right)$$ = \frac{1}{T-1} \sum_{ \mathtt{t}=2}^T \mathbf{h}_{(\mathtt{t})} ( {\vvartheta}^{(\mathtt{m}-1)} ;\mathbf{y}_{1:T} ) \, \mathbf{h}_{(\mathtt{t})} \, ( {\vvartheta}^{(\mathtt{m}-1)} ;\mathbf{y}_{1:T} )^{\prime}$.
Then, a transition from $\vvartheta^{old}$ to $\vvartheta^{new}$ is accepted with probability
\begin{eqnarray}
\label{EQ:MH1}
\varrho\left(  \vvartheta^{old},\vvartheta^{new} \right) &=& \min \left \{ 1, \frac{ \mathscr{L} ( \vvartheta^{new})}
{ \mathscr{L} ( \vvartheta^{old })}
\frac{ \tilde \pi( \vvartheta^{new})}
{ \tilde \pi( \vvartheta^{old })}
\frac{q \left( \vvartheta^{old }|\vvartheta^{new } \right)  }{q \left( \vvartheta^{new }|\vvartheta^{old } \right)} \right \}   .
\end{eqnarray}
\noindent
To implement this Metropolis-Hastings step, we draw a $[0,1]$ uniform random variable
and accept $\vvartheta^{new}$, i.e.
$\vvartheta^{(\mathtt{m},{\mathtt{k}})} = \vvartheta^{new}$, if this uniform random variable is smaller or equal to $\varrho\left(  \vvartheta^{old},\vvartheta^{new} \right)$, otherwise
$\vvartheta^{(\mathtt{m},{\mathtt{k}})} = \vvartheta^{old}$.
By our assumptions on the prior, it follows that ${ \tilde \pi( \vvartheta^{new})}
={ \tilde \pi( \vvartheta^{old })}$ as long as $\vvartheta^{new} \in \Theta_0$.
Whenever $\vvartheta^{new}  \notin \Theta_0$, then the probability $\varrho$ equals to zero. 
%$\vvartheta^{old} \in \Theta_0$ by the fact that 
%$\vvartheta^{(n)} \in \Theta_0$ in \texttt{Step~1} and only $\vvartheta^{new} \in \Theta_0$ can be accepted.
 Due the random walk proposal described above, we observe that $q \left( \vvartheta^{old }|\vvartheta^{new } \right) = q \left( \vvartheta^{new }|\vvartheta^{old } \right)$. The next block is then updated such that $\vvartheta^{old}$ becomes equal to the current $\vvartheta^{(\mathtt{m},{\mathtt{k}})}$. After having performed these updating steps for all blocks, $\mathtt{k}=1,\dots,\mathtt{K}$, we obtain $\vvartheta^{(\mathtt{m})} = \vvartheta^{(\mathtt{m} , {\mathtt{K}})}$.

To improve the properties of the Bayesian sampler in the case when $\theta^Q=\theta^P$ or when $\theta^Q \not= \theta^P$, a reversible jump move based on \citet{Green1995} and \citet{RichardsonGreen1997} has been implemented. Suppose that ${\vvartheta}^{({\mathtt{m}})}$ has been obtained by the above steps. Let $\vvartheta^{old}= {\vvartheta}_{}^{({\mathtt{m}})}$. With a probability of 90\%
we add the following step to sampling step $\mathtt{m}$:  Consider the state $s_1$, where $\theta_{}^Q=\theta^P$ and state $s_2$, where $\theta^Q \not= \theta^P$. The state $S$ is Bernoulli distributed random variable with prior probability $\mathbb{P}(S= s_1)$$=p_{s_1}=0.90$. By applying \citet{Green1995}, transitions from $\{ S=s_1 \}$ to  $\{ S=s_2 \}$ and vice versa can be performed by means of the Metropolis Hastings algorithm. In particular,  consider the uniformly distributed random variable $\eta$, as well as the normal $iid$ random variables $u$ and $u_{\gamma}$. The proposal densities are $f_{\mathcal{N}(0,\sigma_u^2)}(u)$ and $f_{\mathcal{N}(0,\sigma_{u_{\gamma}}^2)}(u_{\gamma})$.  Let $\{ S=s_1\}$, where $\theta^{old}=\theta^Q=\theta^P$.  A possible split transition from $\{ S=s_1\}$ to  $\{ S=s_2\}$ works as follows
\begin{eqnarray}
\label{greensplit1}
\theta^{P,new} &=&  \theta^{old} -  2 \eta u \ , \nonumber \\
\theta^{Q,new} &=&  \theta^{old} +  2 (1-\eta) u \ , \nonumber \\
 \gamma^{new}_0 &=& \gamma^{old}_0 - 2 \eta u +  u_{\gamma}  .
 \end{eqnarray}
\noindent
By replacing the corresponding elements in $\vvartheta^{old}$ by
$\theta^{P,new}$, $\theta^{Q,new}$ and $\gamma^{new}_0$,
 we get the new parameter vector $\vvartheta^{new}$.%
\footnote{Note that $\vvartheta^{old}$ contains the old parameters, where
$\theta^{P}=\theta^{Q}$. In the notation of \citet{Green1995}, the dimension of the parameter of interest with state $s_1$ is $n_1=2$, (consisting of $\theta_{}^{old}$ and $\gamma_{0}^{old}$), the dimension of the noise component is $m_1=3$ (due to $\eta$, $u$ and $u_{\gamma}$). With $s_2$ we get $n_2=3$ (consisting of $\theta^{P,new}$, $\theta^{Q,new}$ and $\gamma_{0}^{new}$) and  $m_2=2$ (due to $\eta$ and $u_{\gamma}$). This yields,
$n_1+m_1=n_2+m_2$. } %
Let $\vchi^{old}=(s_1;\theta^{old}, \gamma_{0}^{old})$ and
$\vchi^{new}=(s_2;\theta^{P,new}, \theta^{Q,new}  \gamma_{0}^{new})$.
By taking partial derivatives of the terms in
(\ref{greensplit1}), we obtain the Jacobian matrix

\begin{eqnarray}
\label{greensplit3}
{\mathbf{J}} = \frac{\partial (\theta^{P}, \theta^{Q}, \eta, \gamma_{0}, u_{\gamma})^{\prime}}
        {\partial (\theta, \eta, \gamma_{0}, u, u_{\gamma})^{\prime} }
&=&
\left(
\begin{array}{ccccc}
1 & - 2 u &  0 & -2 \eta & 0 \\
1 & - 2 u &  0 & 2 (1-\eta) & 0 \\
0 & 1 &  0 &  0 & 0 \\
0 & - 2 u &  1 & - 2 \eta & 1 \\
0 & 0 & 0 & 0 & 1\\
\end{array}
\right)  .
\end{eqnarray}
\noindent
The determinant of the matrix ${\mathbf{J}}$ is equal to 2. Given the proposal densities $q(u)=f_{\mathcal{N}(0,\sigma_u^2)}(u)$ and $q(u_{\gamma}) =f_{\mathcal{N}(0,\sigma_{u_{\gamma}}^2)}(u_{\gamma})$ for  $u$ and $u_{\gamma}$, a transition from $\vchi^{old}$ to $\vchi^{new}$ is accepted with probability \citep[see][equation (7)]{Green1995}
\begin{eqnarray}
\label{EQ:MH3a}
&&\varrho\left(  \vchi^{old} , \vchi^{new} \right)  \nonumber \\
&& \ \ \ =
\min \left( 1,
\frac{ \mathscr{L}( \vvartheta^{new})}
{ \mathscr{L} ( \vvartheta^{old })}
\frac{ \tilde \pi( \vvartheta^{new})}
{ \tilde \pi( \vvartheta^{old })}
\frac{1-p_{s_1}}{p_{s_1}  }
\frac{f_{\mathcal{N}(0, \sigma_{u_{\gamma}}^2)} \left( u_{\gamma}^{old} \right) }{ f_{\mathcal{N}(0,\sigma_{u_{\gamma}}^2)} \left( u_{\gamma}^{new} \right) f_{\mathcal{N}(0,\sigma_{u}^2)} ( u )  }
 |{\mathbf{J}}| \right)  \nonumber \\
&& \ \ \
= \min \left( 1,
\frac{ \mathscr{L}( \vvartheta^{new})}
{ \mathscr{L}( \vvartheta^{old })}
\frac{ \tilde \pi( \vvartheta^{new})}
{ \tilde \pi( \vvartheta^{old })}
\frac{1-p_{s_1}} {p_{s_1}  }
\frac{ 2 }{f_{\mathcal{N}(0,\sigma_u^2)} (u) }
 \right) .
\end{eqnarray}
\noindent
Since $u_{\gamma}^{old} = \gamma^{new}_0 - \gamma^{old}_0 + 2 \eta u $, the densities $f_{\mathcal{N}(0,\sigma_{\gamma}^2)}$ cancel out in
(\ref{EQ:MH3a}). An equivalent Metropolis-Hastings move can be performed without an update of $\gamma_0$.
A possible merge transition from $\{ S=s_2\}$ to  $\{ S=s_1\}$ works as follows
\begin{eqnarray}
\label{greensplit7}
\theta_{}^{new}  &=& \theta_{}^{Q,new} =\theta_{}^{P,new}= (1-\eta) \theta_{}^{Q,new} + \eta \theta_{}^{Q,old} \nonumber \\
 \gamma^{new}_0 &=& \gamma^{old}_0 - 2 \eta u +  u_{\gamma} \ , \text{ such that } \nonumber \\
 u &=& \frac{\theta_{}^{P} - \theta_{}^{new} }{ - 2 \eta  } = \frac{\theta_{}^{Q} + \theta_{}^{new} }{ 2(1- \eta ) }  .
 \end{eqnarray}
\noindent
By means of (\ref{greensplit7}) we get $\theta^{new}$  and $\gamma_0^{new}$.
Then, a transition from $\vchi^{old}= \left(s_2; \theta^{P,old},\theta^{Q,old},\gamma_0^{old} \right)$ to
$\vchi^{new}= \left(s_1;\theta^{P,new}=\theta^{Q,new}=\theta^{new}, \gamma_0^{new} \right)$
is accepted with probability \citep[][equation (7)]{Green1995}:

\begin{eqnarray}
\label{EQ:MH51}
&& \varrho\left(  \vchi^{old} , \vchi^{new} \right) =
\min \left \{ 1,
\frac{  \mathscr{L} ( \vvartheta^{new})}
{  \mathscr{L} ( \vvartheta^{old })}
\frac{p_{s_1}  }{1-p_{s_1}} f_{\mathcal{N}(0,\sigma_u^2)} (u) \frac{1}{2} \right \}  .
\end{eqnarray}
\noindent
If either a split or a merge transition is accepted we set $\vvartheta^{(m)}=\vvartheta^{new}$. After a merge move $\theta^P=\theta^Q$ in updating sub-step $\mathtt{k}=1$, until a split move takes place. %PS checked this on July 31, 2015; this has been implemented correctly

{\em Parameter Estimation:}
To obtain the parameter estimates $\widehat{\vvartheta}$, we consider the draws ${\vvartheta}^{(\mathtt{m})}$, where $\mathtt{m} = \mathtt{M}_b+1,\dots, \mathtt{M}$ of the convergent part of the  Markov chain. We work with
$\mathtt{M}_b=5,000$ and  $\mathtt{M}_{}=20,000$. Then $\widehat{\vvartheta}$ is provided by the sample mean. Tables~\ref{tab:res4} and \ref{tab:res5} show parameter estimates obtained by using the Bayesian algorithm described above.

In addition, as shown by \citet{chernozhukovhong03}, the draws after burn-in phase can also be used to estimate the asymptotic variance of the parameters. To do this, we can simply calculate the sample variance of
$\widehat{\vvartheta}^{(m)}$, where $\mathtt{m}= \mathtt{M}_b+1,\dots, \mathtt{M}$. To account for the serial correlation observed with the Markov chain, we follow Bayesian literature to estimate the variance of the components of
$\widehat{\vvartheta}^{}$ by means of the batch-means approach described in \citet{FlegalJones2010}[ in particular, Equation~(6) is used].

{\em Monte Carlo Study:} In the simulation studies described in Section~\ref{sect:mc1}, {\tt Steps 1} and {\tt 2} are performed for each Monte-Carlo replication $(\mathtt{l} = 1,\dots,\mathtt{L}=200)$.

\begin{remark}
\label{rem:time}
The implementation of the Quasi-Bayesian sampler based \citet{chernozhukovhong03} is not {``}free of cost{''}. Running
multistart random search methods and a standard minimization procedure and then performing the Wald test based on  (\ref{eq:VNstand}) takes approximately 20 minutes, while one full estimation step based on running a random search and then obtaining 20,000 draws from a Markov Chain lasts for approximately 24 hours on the same standard PC.
\end{remark}

\clearpage
%\bibliography{masteer}
\bibliographystyle{apalike}
%\bibliographystyle{named}
%\bibliography{/../../masterbib/trunk/master} %%% poldi office
\bibliography{/../../../texnew/masterbib/trunk/master} %%% poldi office

\begin{thebibliography}{}

\bibitem[A\"it-Sahalia, 1996a]{Ait1996}
A\"it-Sahalia, Y. (1996a).
\newblock Nonparametric pricing of interest rate derivative securities.
\newblock {\em Econometrica}, 64:527--560.

\bibitem[A\"it-Sahalia, 1996b]{aitsahalia96}
A\"it-Sahalia, Y. (1996b).
\newblock Testing continuous-time models of the spot interest rate.
\newblock {\em The Review of Financial Studies}, 9(2):385--426.

\bibitem[A\"it-Sahalia, 2002]{aitsahalia02}
A\"it-Sahalia, Y. (2002).
\newblock Maximum likelihood estimation of discretely-sampled diffusions: A
  closed-form approximation approach.
\newblock {\em Econometrica}, 70:223--262.

\bibitem[A\"it-Sahalia and Kimmel, 2010]{aitsahakiakimmel2009}
A\"it-Sahalia, Y. and Kimmel, R.~L. (2010).
\newblock {Estimating Affine Multifactor Term Structure Models Using
  Closed-Form Likelihood Expansions}.
\newblock {\em Journal of Financial Economics}, 98:113--144.

\bibitem[Altonji and Segal, 1996]{AltonjiSegal1996}
Altonji, J.~G. and Segal, L.~M. (1996).
\newblock {Small-Sample Bias in GMM Estimation of Covariance Structures}.
\newblock {\em Journal of Business \& Economic Statistics}, 14(3):353--66.

\bibitem[Andersen et~al., 1999]{Andersenetal1999}
Andersen, T.~G., Chung, H.-J., and Sorensen, B.~E. (1999).
\newblock Efficient method of moments estimation of a stochastic volatility
  model: A monte carlo study.
\newblock {\em Journal of Econometrics}, 91(1):61--87.

\bibitem[Campbell et~al., 1997]{campbelllomackinlay97}
Campbell, J.~Y., Lo, A.~W., and MacKinlay, A.~C. (1997).
\newblock {\em The Econometrics of Financial Markets}.
\newblock Princeton University Press, Princeton.

\bibitem[Chen and Joslin, 2012]{chenjoslin11}
Chen, H. and Joslin, S. (2012).
\newblock Generalized transform analysis of affine processes and applications
  in finance.
\newblock {\em The Review of Financial Studies}, 25(7):2225--2256.

\bibitem[Chen et~al., 2007]{chenetal2007}
Chen, L., Lesmond, D.~A., and Wei, J. (2007).
\newblock Corporate yield spreads and bond liquidity.
\newblock {\em The Journal of Finance}, 62(1):119--149.

\bibitem[Cheridito et~al., 2007]{cheriditofilipovickimmel03}
Cheridito, P., Filipovi\'c, D., and Kimmel, R.~L. (2007).
\newblock Market price of risk specifications for affine models: {T}heory and
  evidence.
\newblock {\em Journal of Financial Economics}, 83(1):123--170.

\bibitem[Cheridito et~al., 2008]{Cheriditoetal2008}
Cheridito, P., Filipovic, D., and Kimmel, R.~L. (2008).
\newblock {A Note on the Dai-Singleton Canonical Representation of Affine Term
  Structure Models}.
\newblock {\em SSRN eLibrary}.

\bibitem[Chernozhukov and Hong, 2003]{chernozhukovhong03}
Chernozhukov, V. and Hong, H. (2003).
\newblock An mcmc approach to classical estimation.
\newblock {\em Journal of Econometrics}, 115:293--346.

\bibitem[Chib and Ergashev, 2009]{ChibErgashev2008}
Chib, S. and Ergashev, B. (2009).
\newblock Analysis of multifactor affine yield curve models.
\newblock {\em Journal of the American Statistical Association},
  104(488):1324--1337.

\bibitem[Cochrane, 2005]{Cochranebook2005}
Cochrane, J. (2005).
\newblock {\em Asset Pricing}.
\newblock Princeton University Press, revised edition.

\bibitem[Cox et~al., 1985]{cir85}
Cox, J.~C., Ingersoll, J.~E., and Ross, S.~A. (1985).
\newblock A theory of the term structure of interest rates.
\newblock {\em Econometrica}, 53(2):385--407.

\bibitem[Cuchiero et~al., 2012]{cuchieroetal08}
Cuchiero, C., Teichmann, J., and Keller-Ressel, M. (2012).
\newblock Polynomial processes and their application to mathematical finance.
\newblock {\em Finance \& Stochastics}, 16(4):711--740.

\bibitem[Dai and Singleton, 2000]{daisingleton00}
Dai, Q. and Singleton, K.~J. (2000).
\newblock Specification analysis of affine term structure models.
\newblock {\em Journal of Finance}, 55(5):1943--1978.

\bibitem[Dai and Singleton, 2003]{dai03}
Dai, Q. and Singleton, K.~J. (2003).
\newblock Term structure dynamics in theory and reality.
\newblock {\em The Review of Financial Studies}, 16(3):631--678.

\bibitem[Diebold et~al., 2006]{Dieboldetal2006}
Diebold, F.~X., Rudebusch, G.~D., and Aruoba, S.~B. (2006).
\newblock The macroeconomy and the yield curve: A dynamic latent factor
  approach.
\newblock {\em Journal of Econometrics}, 131:309--338.

\bibitem[Duffee, 2011]{Duffee2011}
Duffee, G.~R. (2011).
\newblock Information in (and not in) the term structure.
\newblock {\em The Review of Financial Studies}, 24(9):2895--2934.

\bibitem[Duffie et~al., 2003]{duffiefilipovicschachermayer03}
Duffie, D., Filipovi\'c, D., and Schachermayer, W. (2003).
\newblock Affine processes and applications in finance.
\newblock {\em Annals of Applied Probability}, 13:984--1053.

\bibitem[Duffie and Kan, 1996]{duffiekan96}
Duffie, D. and Kan, R. (1996).
\newblock A yield-factor model of interest rates.
\newblock {\em Mathematical Finance}, 6(4):379--406.

\bibitem[Duffie et~al., 2000]{duffiepansingleton00}
Duffie, D., Pan, J., and Singleton, K.~J. (2000).
\newblock Transform analysis and asset pricing for affine jump-diffusions.
\newblock {\em Econometrica}, 68(6):1343--1376.

\bibitem[Egorov et~al., 2011]{Egorovetal2011}
Egorov, A.~V., Li, H., and Ng, D. (2011).
\newblock A tale of two yield curves: Modeling the joint term structure of
  dollar and euro interest rates.
\newblock {\em Journal of Econometrics}, 162(1):55--70.

\bibitem[Filipovi\'c, 2009]{filipovicbook2009}
Filipovi\'c, D. (2009).
\newblock {\em Term-Structure Models: A Graduate Course}.
\newblock Springer, Berlin.

\bibitem[Filipovi\'{c} et~al., 2013]{filipovicmayerhoferschneider13}
Filipovi\'{c}, D., Mayerhofer, E., and Schneider, P. (2013).
\newblock Transition density approximations for multivariate affine jump
  diffusion processes.
\newblock {\em Journal of Econometrics}.
\newblock forthcoming.

\bibitem[Flegal and Jones, 2010]{FlegalJones2010}
Flegal, J.~M. and Jones, G.~L. (2010).
\newblock Batch {M}eans and {S}pectral {V}ariance {E}stimators in {M}arkov
  {C}hain {M}onte {C}arlo.
\newblock {\em The Annals of Statistics}, 38(2):1034--1070.

\bibitem[Fr\"uhwirth-Schnatter and Geyer, 1996]{fruehwirthgeyer96}
Fr\"uhwirth-Schnatter, S. and Geyer, A. (1996).
\newblock Bayesian estimation of economemtric multi-factor
  cox-ingersoll-ross-models of the term structure of interest rates via {MCMC}
  methods.
\newblock Working paper, Vienna University of Economics and Business.

\bibitem[Glasserman and Kim, 2010]{GlassermanKim2009}
Glasserman, P. and Kim, K.-K. (2010).
\newblock {Moment Explosions and Stationary Distributions in Affine Diffusion
  Models}.
\newblock {\em Mathematical Finance}, 20(1):1--33.

\bibitem[Grasselli and Tebaldi, 2008]{GrasselliTebaldi2008}
Grasselli, M. and Tebaldi, C. (2008).
\newblock Solvable affine term structure models.
\newblock {\em Mathematical Finance}, 18(1):135--153.

\bibitem[Green, 1995]{Green1995}
Green, P. (1995).
\newblock Reversible jump markov chain monte carlo computation and bayesian
  model determination.
\newblock {\em Biometrika}, 82(4):711--732.

\bibitem[Guggenberger and Smith, 2005]{Guggenberger2005}
Guggenberger, P. and Smith, R.~J. (2005).
\newblock Generalized empirical likelihood estimators and tests under partial,
  weak, and strong identification.
\newblock {\em Econometric Theory}, null:667--709.

\bibitem[Hamilton and Wu, 2012]{HamiltonWu2010}
Hamilton, J.~D. and Wu, J.~C. (2012).
\newblock Identification and estimation of {G}aussian affine term structure
  models.
\newblock {\em Journal of Econometrics}, 168(2):315 -- 331.

\bibitem[Hansen, 1982]{hansen82}
Hansen, L.~P. (1982).
\newblock Large sample properties of generalized method of moments estimators.
\newblock {\em Econometrica}, 50(4):1029--1054.

\bibitem[Jones, 2003]{Jones2003b}
Jones, C.~S. (2003).
\newblock Nonlinear mean reversion in the short-term interest rate.
\newblock {\em The Review of Financial Studies}, 16(3):793--843.

\bibitem[Joslin et~al., 2010]{Joslinetal2009}
Joslin, S., Singleton, K.~J., and Zhu, H. (2010).
\newblock A new perspective on gaussian dynamic term structure models.
\newblock {\em The Review of Financial Studies}, 24:926--970.

\bibitem[Keller-Ressel and Mayerhofer, 2012]{KellerMayerhofer2011}
Keller-Ressel, M. and Mayerhofer, E. (2012).
\newblock Exponential moments of affine processes.
\newblock Technical report, Deutsche Bundesbank, Frankfurt.

\bibitem[Klenke, 2008]{Klenke2008}
Klenke, A. (2008).
\newblock {\em Probability Theory - A Comprehensive Course}.
\newblock Springer.

\bibitem[Mayerhofer et~al., 2010]{mayerhoferetal10}
Mayerhofer, E., Pfaffel, O., and Stelzer, R. (2010).
\newblock On strong solutions of matrix valued jump-diffusions.
\newblock {\em preprint}.

\bibitem[Newey and McFadden, 1994]{neweyMcfadden94}
Newey, W.~K. and McFadden, D. (1994).
\newblock Large sample estimation and hypothesis testing.
\newblock In {\em Handbook of econometrics, {V}ol.\ {IV}}, volume~2 of {\em
  Handbooks in Econom.}, pages 2111--2245. North-Holland, Amsterdam.

\bibitem[Newey and Windmeijer, 2009]{NeweyWindmeijer2009}
Newey, W.~K. and Windmeijer, F. (2009).
\newblock Generalized method of moments with many weak moment conditions.
\newblock {\em Econometrica}, 77(3):687--719.

\bibitem[Perko, 1991]{Perko1991}
Perko, L. (1991).
\newblock {\em Differential Equations and Dynamical Systems}.
\newblock Texts in Applied Mathematics, No.~7. Springer.

\bibitem[Piazzesi, 2010]{piazzesi03}
Piazzesi, M. (2010).
\newblock {\em Affine Term Structure Models}.
\newblock In Y. A\"it-Sahalia and L. Hansen (Eds.), Handbook of Financial
  Econometrics, North-Holland, Amsterdam.

\bibitem[Poirier, 1995]{poirier1995}
Poirier, D.~J. (1995).
\newblock {\em Intermediate Statistics and Econometrics: A Comparative
  Approach}.
\newblock MIT Press, Cambridge, Massachusetts.

\bibitem[P\"otscher and Prucha, 1997]{PoetscherPrucha1997}
P\"otscher, B.~M. and Prucha, I.~R. (1997).
\newblock {\em Dynamic Nonlinear Econometric Models, Asymptotic Theory}.
\newblock Springer, New York.

\bibitem[Richardson and Green, 1997]{RichardsonGreen1997}
Richardson, S. and Green, P.~J. (1997).
\newblock On bayesian analysis of mixtures with an unknown number of components
  (with discussion).
\newblock {\em Journal of the Royal Statistical Society: Series B (Statistical
  Methodology)}, 59(4):731--792.

\bibitem[Robert and Casella, 2004]{robertcasella04}
Robert, C. and Casella, G. (2004).
\newblock {\em Monte Carlo Statistical Methods}.
\newblock Springer, New York, 2nd edition.

\bibitem[Ruud, 2000]{Ruudbook2000}
Ruud, P.~A. (2000).
\newblock {\em An Introduction to Classical Econometric Theory}.
\newblock Oxford University Press, New York.

\bibitem[Stanton, 1997]{Stanton1997}
Stanton, R. (1997).
\newblock A nonparametric model of term structure dynamics and the market price
  of interest rate risk.
\newblock {\em Journal of Finance}, 52(5):1973--2002.

\bibitem[T{\"o}rn and Zilinskas, 1989]{ToernZilinskas1989}
T{\"o}rn, A. and Zilinskas, A. (1989).
\newblock {\em Global Optimization}.
\newblock Lecture Notes in Computer Science 350. Springer.

\bibitem[Vasicek, 1977]{vasicek77}
Vasicek, O. (1977).
\newblock An equilibrium characterization of the term structure.
\newblock {\em Journal of Financial Economics}, 5:177--188.

\bibitem[Windmeijer, 2005]{Windmeijer2005}
Windmeijer, F. (2005).
\newblock {A finite sample correction for the variance of linear efficient
  two-step GMM estimators}.
\newblock {\em Journal of Econometrics}, 126(1):25--51.

\bibitem[Zhou, 2001]{zhou01}
Zhou, H. (2001).
\newblock Finite sampler properties of {EMM}, {GMM}, {QMLE}, and {MLE} for a
  square-root interest rate diffusion model.
\newblock {\em Journal of Computational Finance}, 5:89--122.

\bibitem[Zhou, 2003]{zhou03}
Zhou, H. (2003).
\newblock It\^o conditional moment generator and the estimation of short-rate
  process.
\newblock {\em Journal of Financial Econometrics}, 1(2):250--271.

\end{thebibliography}

%\bibliography{/../../../../masterbib/trunk/master} %%% poldi home
%\bibliography{master}

\end{document}